\documentclass[12pt]{article}

\newif\ifnature
\naturetrue

\usepackage{cite}
\usepackage{graphicx} 
\usepackage{amssymb}
\usepackage{array}
\usepackage{amsmath}
\usepackage{dsfont}
\usepackage{physics}
\usepackage{amsfonts}
\usepackage{siunitx}
\usepackage{soul}
\usepackage{xr-hyper}
\usepackage{hyperref}
\usepackage{lineno}

\ifnature
\usepackage{caption}
\else
\fi

\newcommand{\Hxxz}{H_\mathrm{XXZ} }

\newenvironment{sciabstract}{%
\begin{quote} \bf}
{\end{quote}}

\newcommand{\figOne}{
 \begin{figure}[htbp]
\centering
\ifnature
\includegraphics[width=0.76\textwidth]{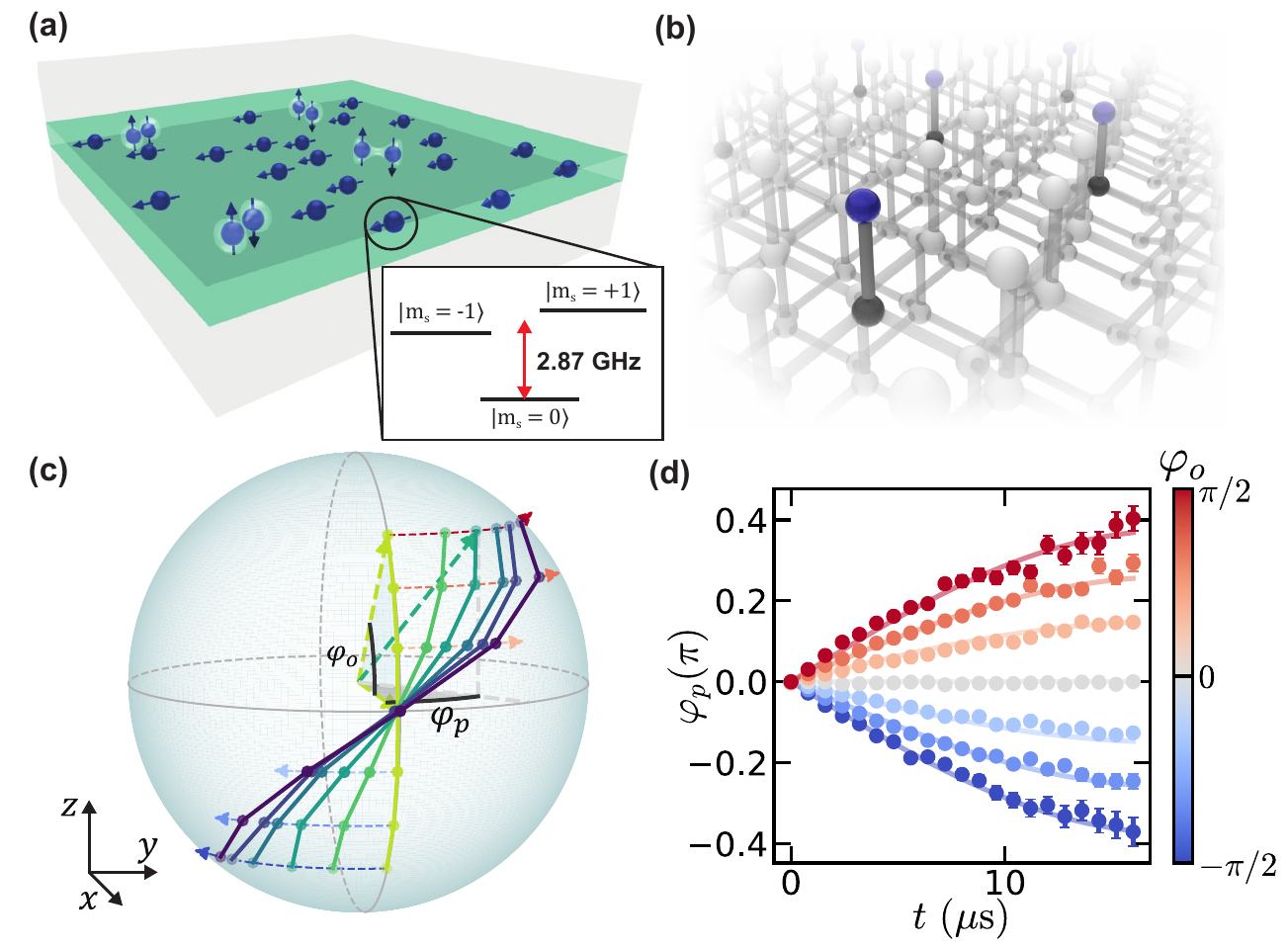}
\else
\includegraphics[width=0.5\textwidth]{Figures/Fig1_v5.pdf}
\fi
\caption{
\fontsize{7pt}{9pt}\selectfont 
\textbf{Twisting dynamics of a strongly-interacting, two-dimensional NV ensemble.} 
\textbf{(a)} Schematic depiction of a two-dimensional ensemble of NV centers. 
NVs are confined to a $\sim7$~nm layer (green) formed via nitrogen delta-doping during diamond growth. 
Each NV exhibits an electronic spin-1 ground state and  we encode an effective two-level system in the $\{\ket{m_s = 0}, \ket{m_s = -1}\}$ subspace. 
 NV centers are randomly positioned within the diamond lattice, leading to the presence of strongly coupled dimers. 
\textbf{(b)} Depicts the orientation of our diamond lattice with one subgroup of NV centers aligned in the [111] lattice direction, perpendicular to the delta-doped layer. 
Black, blue, and white spheres represent vacancies, nitrogen, and carbon atoms, respectively.
\textbf{(c)} Shows the dynamics of a spin polarized initial state under $H_\textrm{XXZ}$.
Initial states (represented by the yellow arrow on the Bloch sphere) are offset above (or below) the $x$-axis by an angle $\varphi_o$.
$H_\textrm{XXZ}$ induces effective  ``twisting'' dynamics where an initial state 
precesses about the $z$-axis at a rate proportional to  $\expval{S_z}$; the average precession angle in the equatorial plane is given by $\varphi_p$.
Colored lines connect data points for different initial states at the same evolution time.
\textbf{(d)} Analogous data from (c) showing the precession angle as a function of time for initial states with different  offsets above (or below)
the $x$-axis.
}
\label{fig:fig1}
\end{figure}
}

\newcommand{\figTwo}{
 \begin{figure*}[htbp]
    \centering
    \includegraphics[width=\textwidth]{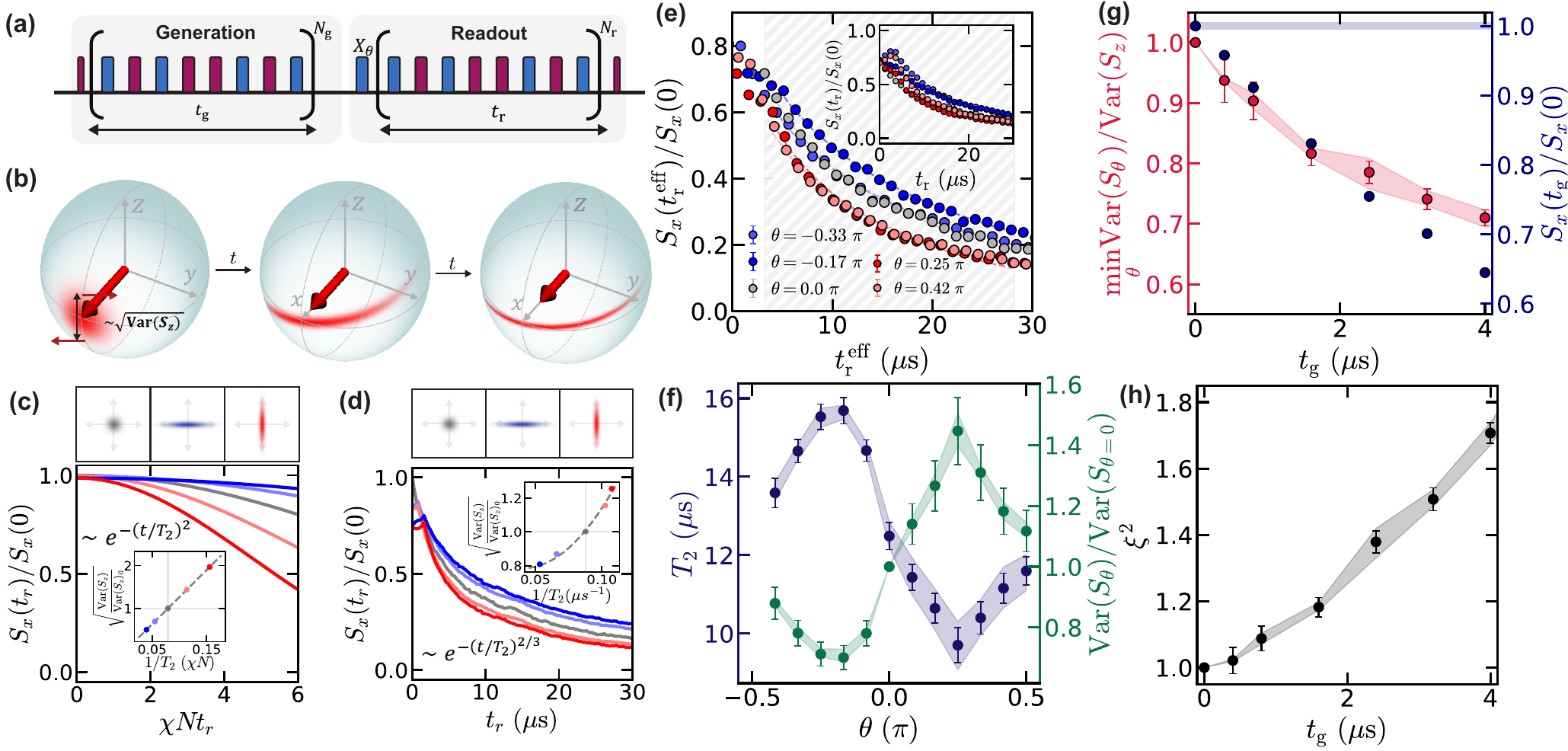}
    \caption{ \fontsize{7pt}{9pt}\selectfont 
    \textbf{Interaction-enabled readout of the quantum projection noise.} 
    \textbf{(a)} Our experimental sequence consists of two steps. 
    First, a spin squeezed state is generated via evolution under $H_\textrm{XXZ}$ for a time, $t_\textrm{g}$, starting from an initial spin-polarized state.
    After a variable global rotation by angle $\theta$ around the $x$-axis, the squeezed state’s
    anisotropic spin projection noise is read out via quench dynamics (for a time $t_\textrm{r}$) under $H_\textrm{XXZ}$.
    Throughout our experiments, we utilize an XY-8 dynamical decoupling sequence in order to isolate the NV ensemble from other paramagnetic defects in the diamond lattice (see Methods). 
    \textbf{(b)} Schematic depiction of the intuition underlying interaction-enabled spin projection noise readout. 
    The red patch represents the Wigner quasiprobability distribution (projection noise) of the collective spin in the plane spanned by $S_y$ and $S_z$.
    For a many-body spin system undergoing effective twisting dynamics, this distribution becomes sheared at a rate  $\sim \chi \sqrt{\text{Var}(S_z)}$.
    This shearing yields a concomitant shrinking of the total spin length, $S_x$, and enables a measurement of the state's quantum variance, $\text{Var}(S_z)$, via the decay timescale of $\langle S_x (t) \rangle$.
    \textbf{(c)} Interaction-enabled readout in the one-axis twisting model. 
    Initial states with different spin projection noise,  $\text{Var}(S_z)$ (top), lead to different decay timescales for $\langle S_x(t) \rangle$ (bottom).
    A state exhibiting a smaller projection noise (blue) exhibits a slower decay than the initial spin-polarized state (gray), while a state exhibiting a larger projection noise (red) exhibits a faster decay. 
    For the OAT model, the form of this decay can be analytically derived, 
    $\langle S_x(t) \rangle \sim e^{-2\chi^2 \text{Var}(S_z) t^2} \equiv e^{-(t/T_2)^2}$.
    (inset) Depicts the mapping between the decay timescale, $T_2$, and the quantum variance (normalized by the variance of the initial spin-polarized state, see Methods). 
    \textbf{(d)} Interaction-enabled readout in our disordered dipolar XXZ model. 
    Unlike the OAT model, there does not exist a simple analytic mapping between $\text{Var}(S_z)$ and the decay timescale of $\langle S_x(t) \rangle$.
    Thus, we numerically simulate the quench dynamics under $H_\mathrm{XXZ}$ for different initial states (top) using the discrete-cluster truncated Wigner approximation. 
    The dynamics of $\langle S_x(t) \rangle$ exhibit a similar qualitative dependence on the state's spin projection noise, $\mathrm{Var}(S_z)$, as in panel (c). 
    (inset) Depicts the one-to-one mapping between the fitted $T_2$ decay timescale and the spin projection noise, $\text{Var}(S_z)$, of the initial state. 
    \textbf{(e)} A spin-polarized initial state, $\ket{\mathbf{x}}$, is evolved under $H_\text{XXZ}$ for a time $t_\mathrm{g} = 3.2\ \mu\mathrm{s}$. 
    We measure the subsequent quench dynamics for different rotation angles $\theta$ (inset), and the $S_x(t_r)$ decay is normalized by $S_x(t_\mathrm{g} = 0)$.
    For each angle, the state's projection noise is squeezed along a different axis, and we redefine the time, $t_\mathrm{r} \rightarrow t_\mathrm{r}^\mathrm{eff}$, in order to account for this (see Methods). 
    Consistent with our numerical simulations (d), certain rotation angles yield a longer effective decay timescale (blue), while other rotation angles yield a shorter effective decay timescale (red).
    For each angle, a timescale $T_2$ is extracted from the decay in the striped fitting time window by fitting the data to $e^{-(t/T_2)^{2/3}}$.
    These timescales (purple data) are shown as a function of $\theta$ in \textbf{(f)}. 
    The shaded purple region surrounding the data correspond to changes in the extracted $T_2$ as a function of changing the fitting window (from $3.2-25$~$\mu$s to $3.2-35$~$\mu$s), and demonstrates the robustness of our approach. 
    By using the numerically determined mapping (d), one can immediately convert the decay timescale to the quantum variance of the state (green). 
    Note that the mapping we utilize accounts for experimental imperfections such as the polarization fidelity of the initial state (see Methods). 
    \textbf{(g)} In order to probe the dynamics of spin squeezing, we evolve the spin-polarized initial state for different times, $t_\textrm{g}$.
    At each time, we perform our interaction-enabled readout protocol, and extract the normalized quantum variance, 
     $\mathrm{Var}(S_\theta) / \mathrm{Var}(S_{\theta=0})$ as a function of $\theta$.  
     This allows us to determine the minimum variance as a function of $t_\textrm{g}$ (red data).
     We also directly measure the decay of the collective spin length as a function of $t_\textrm{g}$ (navy data).
    Taken together, this enables us to compute the dynamics of the squeezing parameter, $\xi^2$ as a function of the preparation time \textbf{(h)}. 
    We find that $\xi^2$ exhibits a monotonic increase as a function of time; thus, despite the presence of an anisotropic spin projection noise distribution, the prepared state does not exhibit metrologically useful spin squeezing. 
    }
    \label{fig:fig2}
\end{figure*}
}

\newcommand{\figThree}{
 \begin{figure*}[t]
    \centering
    \includegraphics[width=\textwidth]{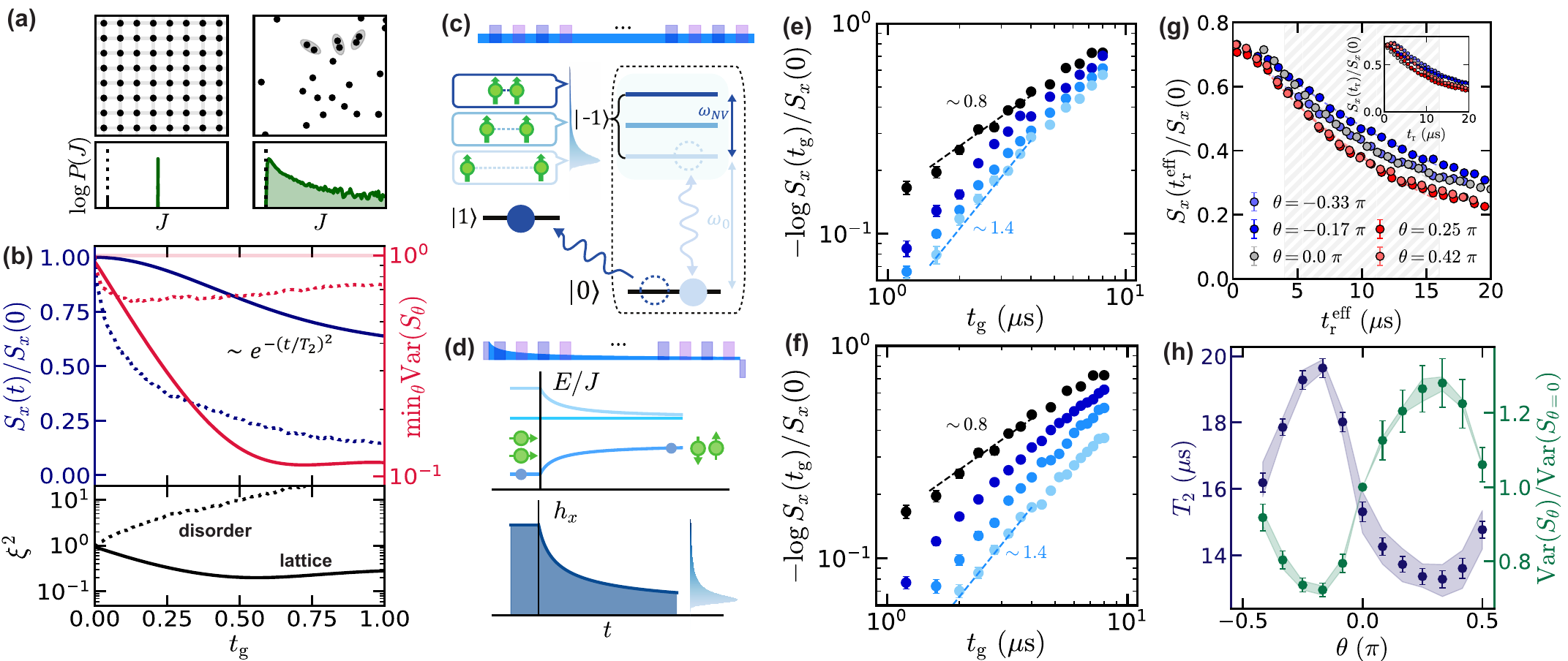}
    \caption{\fontsize{7pt}{9pt}\selectfont 
    \textbf{Reduction of positional disorder via lattice engineering.} 
    \textbf{(a)} Highlights the difference in the distribution of dipolar interaction strengths, $P(J)$, for an ordered two-dimensional array (left) and a positionally disordered ensemble (right). 
    For the disordered ensemble, $P(J)$ exhibits heavy tails corresponding to strongly-interacting clusters of spins, such as ``dimers'' (outlined in gray). $J = 0$ is indicated by the vertical dashed line.
    \textbf{(b)} Numerical (cluster DTWA) simulations of the squeezing dynamics of an ordered versus disordered NV ensemble with $N=100$. 
    (top) Shows the minimum quantum variance and the collective spin length as a function of the squeezing generation time, $t_\textrm{g}$.
    (bottom) Shows the squeezing parameter as a function of $t_\textrm{g}$.
    Consistent with our experimental observations [Fig.~\ref{fig:fig2}h], the disordered case (dotted lines) exhibits a monotonically increasing squeezing parameter, $\xi^2$.
    By contrast, for the lattice case (solid lines), the squeezing parameter is below unity.
    \textbf{(c,d)} Schematic illustrating two lattice-engineering approaches for reducing the amount of positional disorder in our NV ensemble. 
    Both strategies attempt to isolate NVs in the central region of the interaction spectrum, $P(J)$, thereby eliminating strongly-coupled dimers in the tails of $P(J)$.
    The first approach (c), leverages shelving to the $|m_s = +1\rangle$ state.
    In particular, after optical pumping to $|m_s = 0 \rangle$, a weak microwave pulse with high-frequency selectivity is used to drive NVs in the central region of the interaction spectrum from $|m_s = 0 \rangle \rightarrow |m_s = -1 \rangle$ (light blue).
    Next, a strong microwave pulse is used to shelve the remaining NVs from $|m_s = 0 \rangle \rightarrow |m_s = +1 \rangle$ (dark blue), where they are decoupled from the subsequent squeezing dynamics. 
    Finally, another microwave pulse is used to bring those NVs in $|m_s = -1 \rangle$ back to the $|m_s = 0 \rangle$ state. 
    The second approach (d), uses a form of adiabatic depolarization. 
    After optical pumping, we initialize the NV spins along the $x$-axis via a $\pi/2$-pulse.
    Next, we turn on a strong transverse field, $h_x S_x$, and slowly ramp this field down to a final value $h_x^f$. 
    NVs in the central region of the interaction spectrum (with $|J|  \lesssim h_x^f$) will maintain their initial $x$-polarization. 
    By contrast, NVs in the tails of the interaction spectrum (with $|J|  \gtrsim h_x^f$) will exhibit
    rapid, dipolar-induced depolarization, which effectively removes them from the subsequent squeezing dynamics (see Methods).
    \textbf{(e,f)}  The functional form of the decay of $\langle S_x(t) \rangle$ can act as a proxy for the amount of positional disorder present in our NV ensemble. 
    As previously shown in Fig.~\ref{fig:fig2}(d), in the fully disordered case, $\langle S_x(t) \rangle \sim e^{-(t/T_2)^{2/3}}$.
    In the lattice case [as shown in panel (b)], one finds that $\langle S_x(t) \rangle$ instead scales as $\sim e^{-(t/T_2)^{2}}$.
    Thus, the stretch exponent of the decay of $\langle S_x(t) \rangle$ provides a metric for characterizing our lattice engineering. 
    For both the shelving approach (e) and the adiabatic depolarization approach (f), we find that the stretch exponent can be improved from $\sim 0.8$ up to $\sim 1.4$, representing a significant reduction in the positional disorder. 
    \textbf{(g,h)} A spin-polarized initial state, $\ket{\mathbf{x}}$, is evolved under $H_\text{XXZ}$ in the presence of lattice engineering (via the shelving approach) for a time $t_\mathrm{g} = \SI{2.4}{\micro\second}$.
    Here, we demonstrate the read out of the quantum variance of the resulting state via the same interaction-based approach as before. 
    (g)  Displays the readout quench dynamics 
    for different rotation angles $\theta$ in direct analogy to Fig.~\ref{fig:fig2}(e). 
    For each angle, a timescale $T_2$ is extracted from the decay in the striped time window by fitting the data to $e^{-t/T_2}$.
    We note that although the early-time stretch exponent is larger than one (panels e,f), this stretch exponent crosses over to $2/3$ at late times; thus we pick a stretch exponent of one as a simple interpolation that accurately represents the decay timescale during the striped time window.
    These timescales (purple data) are shown as a function of $\theta$ in (h). 
    The shaded
purple region surrounding the data correspond to changes in the extracted $T_2$ as a function of changing the fitting window
(from $2.4-12$~$\mu$s to $2.4-16$~$\mu$s). 
By using the numerically determined
mapping (explicitly including the lattice engineering, see Methods),  one can immediately convert the decay timescale to the quantum variance of the state (green).
    }
    \label{fig:fig3}
\end{figure*}
}

\newcommand{\figFour}{
 \begin{figure*}[htbp]
\centering

\ifnature
\includegraphics[width=0.8\textwidth]{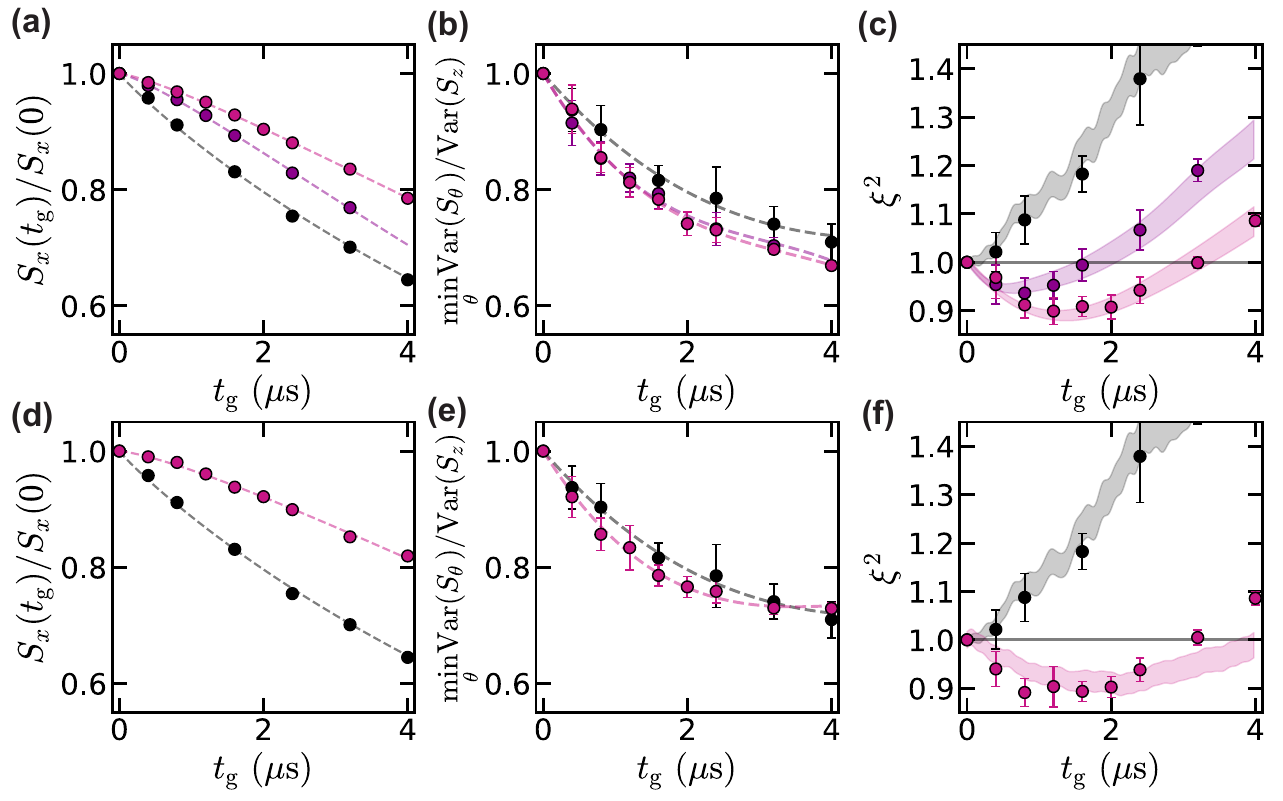}
\else
\includegraphics[width=0.7\textwidth]{Figures/Fig4.pdf}
\fi
\caption{
\fontsize{7pt}{9pt}\selectfont 
\textbf{Spin squeezing in a disordered NV ensemble}. 
\textbf{(a,d)} Depict the decay of the collective spin length as a function of the squeezing preparation time, $t_\textrm{g}$, in the presence of lattice engineering. 
For both the shelving approach (a) and the adiabatic depolarization approach (d), we observe a significantly slower decay, compared to the fully disordered case (black data).
For the shelving approach, colored data indicate different amounts of lattice engineering corresponding to the frequency selectivity (i.e.~Rabi frequency, $\Omega$) of the microwave pulse, with $\Omega = 310$~kHz (purple data) and $\Omega = 125$~kHz (pink data). 
For the adiabatic depolarization approach, the transverse field starts at $h_x = 380$~kHz and is ramped down to $h_x = 27$~kHz over $24$~$\mu$s. 
Dashed lines are a guide to the eye.
\textbf{(b,e)} 
 Depict the minimum variance as a function of $t_\textrm{g}$, in the presence of lattice engineering (see Methods).
 Experimental conditions for the two approaches are analogous to panels (a,d). 
 Dashed lines are a guide to the eye.
\textbf{(c,f)} 
Show the squeezing parameter as a function of $t_\textrm{g}$ for both lattice engineering strategies. 
In both cases, the squeezing parameter drops below unity for an extended period of evolution time. 
For the shelving method, we observe spin squeezing with an optimal squeezing parameter, $\xi^2 = 0.90(3)$, at a time $t_\mathrm{g} = 1.6\ \mu$s, while for the depolarization method, we observe spin squeezing with an optimal squeezing parameter, $\xi^2 = 0.89(2)$, at a time $t_\mathrm{g} = 1.6\ \mu$s. 
Shaded regions correspond to numerical predictions from cluster DTWA accounting for experimental imperfections,  lattice engineering, and the uncertainty due to the average over positional disorder. 
}
\label{fig:fig4}
\end{figure*}
}

\begin{document}

\title{Spin squeezing in an ensemble of nitrogen-vacancy centers in diamond}

% % 
\author
{
Weijie Wu$^{1*}$, 
Emily J. Davis$^{1,2*}$,
Lillian B. Hughes$^{3}$, 
Bingtian Ye$^{1}$, \\
Zilin Wang$^{1}$, 
Dominik Kufel$^{1}$,
Tasuku Ono$^{1}$,
Simon A. Meynell$^{4}$, \\
Maxwell Block$^{1}$, 
Che Liu$^{1}$, 
Haopu Yang$^{1}$,\\
Ania C. Bleszynski Jayich$^{4,\dagger}$, 
Norman Y. Yao$^{1,\dagger}$
\\
%
% 7
\normalsize{\hspace{-8mm}$^{1}$ Department of Physics, Harvard University, Cambridge, MA 02138, USA}\\
\normalsize{\hspace{-8mm}$^{2}$ Department of Physics, New York University, New York, NY 10012, USA}\\
\normalsize{\hspace{-8mm}$^{3}$ Materials Department, University of California, Santa Barbara, CA 93106, U.S.A}\\
\normalsize{\hspace{-8mm}$^{4}$ Department of Physics, University of California, Santa Barbara, CA 93106, U.S.A}\\
\normalsize{$^*$These authors contributed equally to this work.}\\
\normalsize{\hspace{-8mm}$^\dagger$To whom correspondence should be addressed:}\\
\normalsize{\hspace{-8mm}e-mail: ania@physics.ucsb.edu; \hspace{1mm}nyao@fas.harvard.edu}
}

\date{}

\ifnature
\baselineskip24pt
\maketitle
\vspace{5mm}
\else
\fi

\ifnature
\begin{sciabstract}
\else
\begin{abstract}
\fi
Spin squeezed states provide a seminal example of how the structure of quantum mechanical correlations can be controlled to produce metrologically useful entanglement~\cite{winelandSpinSqueezingReduced1992, kitagawaSqueezedSpinStates1993, meyer2001experimental, LouchetChauvet2010, gross2010nonlinear, sewell2012magnetic, ockeloen2013quantum, hostenMeasurementNoise1002016}.
%
%riedel2010atom
Such squeezed states have been demonstrated in a wide variety of artificial quantum systems ranging from atoms in optical cavities to trapped ion crystals~\cite{fernholz2008spin, lerouxImplementationCavitySqueezing2010,hamley2012spin,riedel2010atom,bohnet2014reduced,bohnetQuantumSpinDynamics2016,bornetScalableSpinSqueezing2023,frankeQuantumenhancedSensingOptical2023,ecknerRealizingSpinSqueezing2023, hines2023spin}. 
By contrast, despite their numerous advantages as practical sensors,  spin ensembles in solid-state materials have yet to be controlled with sufficient precision to generate targeted entanglement such as spin squeezing. 
% RT 
In this work, we present the first experimental demonstration of spin squeezing in a solid-state spin system.
Our experiments are performed on a strongly-interacting ensemble of nitrogen-vacancy (NV) color centers in diamond at room temperature, and squeezing (-0.5~$\pm$~0.1~dB) is generated by the native magnetic dipole-dipole interaction between NVs. 
In order to generate and detect squeezing in a solid-state spin system, we overcome a number of key challenges of broad experimental and theoretical interest. 
First, we develop a novel approach, using interaction-enabled noise spectroscopy, to characterize the quantum projection noise in our system without directly resolving the spin probability distribution. 
Second, noting that the random positioning of spin defects severely limits the generation of spin squeezing, we implement a pair of strategies aimed at isolating the dynamics of a relatively ordered sub-ensemble of NV centers. 
% can maybe still be made better???
%
Our results open the door to entanglement-enhanced metrology using macroscopic ensembles of optically-active spins in solids.
\ifnature
\end{sciabstract}
\else
\end{abstract}
\date{\today}
\fi

\ifnature
\else
\maketitle
\newpage
\fi

Controlling the structure of quantum mechanical correlations in a many particle system promises fundamental advances in quantum sensing, simulation, and computing. 
Engineered systems of atoms, ions and superconducting qubits have been leading this charge, demonstrating the remarkable ability to generate a diverse array of \emph{targeted} entangled states~\cite{grossQuantumSimulationsUltracold2017, browaeysManybodyPhysicsIndividually2020, kjaergaardSuperconductingQubitsCurrent2020, monroeProgrammableQuantumSimulations2021}. 
While disparate, these systems share a number of key ingredients: Their degrees of freedom are individually addressable, can be assembled with engineered couplings, and are well isolated from their environment. 
Extending the control of programmable entanglement to solid-state materials remains a central challenge of modern quantum science, one which is particularly acute because such systems typically lack the crucial ingredients above.

Despite this challenge, the last decade has seen tremendous efforts toward the creation of structured entanglement in ensembles of solid-state spins~\cite{auccaiseSpinSqueezingQuadrupolar2015,
wolfowiczQuantumGuidelinesSolidstate2021, randall2021many,kianiniaQuantumEmitters2D2022,burkardSemiconductorSpinQubits2023,kirsteinSqueezedDarkNuclear2023}.
One setting where many-body entanglement and solid-state spins naturally meet is quantum metrology.
% why? 
% what the entanglement does for you
%
Indeed, entanglement can enable enhanced measurement sensitivities in spin sensors, which are already used as powerful probes of biological and condensed matter systems~\cite{maze2008nanoscale, le2013optical, kucsko2013nanometre, kolkowitz2015probing, casolaProbingCondensedMatter2018, jenkins2022imaging, aslamQuantumSensorsBiomedical2023}.
The promise of achieving sensitivities beyond the limit imposed by classical correlations (i.e.~the standard quantum limit) motivates the following question: In the absence of single-particle control and with access to only a limited set of global observables, can one generate and detect
metrologically-useful, many-body entanglement in a solid-state spin ensemble?

Here, we take a crucial step towards answering this question and bridging the entanglement gap for solid-state systems. In particular, we demonstrate, for the first time, spin squeezing in a room-temperature, solid-state spin ensemble. The  structure of a spin-squeezed state is designed to suppress the quantum projection noise below that of an uncorrelated spin-polarized state and can be harnessed for computing, networking, and metrology~\cite{winelandSpinSqueezingReduced1992,kitagawaSqueezedSpinStates1993,furusawa1998unconditional,braunsteinQuantumInformationContinuous2005, pezzeQuantumMetrologyNonclassical2018,maliaDistributedQuantumSensing2022}. Our experimental platform consists of a two-dimensional ensemble of nitrogen-vacancy (NV) color centers delta-doped within diamond \cite{hughesStronglyInteractingTwodimensional2024}. The NV spins are randomly positioned within the 2D layer and interact via magnetic dipole-dipole interactions~[Fig.~\ref{fig:fig1}].

Our main results are threefold. First, we demonstrate that time-evolution under the disordered dipolar interaction yields mean-field \emph{twisting} dynamics: a spin-polarized state displaced along $\hat{z}$ undergoes a rotation proportional to its displacement [Fig.~\ref{fig:fig1}c]. Analogous dynamics generated by all-to-all interactions, i.e., the one-axis twisting (OAT) Hamiltonian, are known to produce spin squeezing of a large collective spin~\cite{kitagawaSqueezedSpinStates1993, lerouxImplementationCavitySqueezing2010, bohnetQuantumSpinDynamics2016, braverman2019near}. However, such interactions are neither natural nor easily engineered in the solid-state~\cite{cappellaro2009quantum,bennett2013phonon}.  %goldstein2011environment 
To this end, our work is motivated by and builds upon recent theoretical predictions that short-range interactions within two-dimensional arrays of quantum dipoles can also generate spin squeezing~\cite{perlinSpinSqueezingShortRange2020,comparinScalableSpinSqueezing2022, blockScalableSpinSqueezing2024,bornetScalableSpinSqueezing2023,ecknerRealizingSpinSqueezing2023, frankeQuantumenhancedSensingOptical2023}.

Second, we turn to the task of measuring the reduction in the spin projection noise generated by the dipolar twisting dynamics. Despite progress along this direction, projection-noise-limited readout has yet to be demonstrated in ensembles of NV centers~\cite{barrySensitivityOptimizationNVdiamond2020, arunkumarQuantumLogicEnhanced2023}. To this end, we utilize the decoherence dynamics of the NV centers themselves to measure the reshaped, elliptical spin projection noise [Fig.~\ref{fig:fig2}];  we note that this approach is quite generic and should enable the witnessing of entanglement in platforms without access to single-particle detection.%~\cite{millerTwoaxisTwistingUsing2024}.

Finally, we find that the strong positional disorder intrinsic to our NV ensemble prohibits squeezing by inducing a rapid decay of the total spin length. 
Intuitively, this rapid decay arises from the fast dynamics of strongly-coupled clusters  of nearby NV centers.
Thus, we introduce and implement two methods for tunably removing positional disorder from our system: (i) frequency-selective shelving of clusters, and (ii) adiabatic depolarization of clusters [Fig.~\ref{fig:fig3}(c,d)]. These techniques allow us to probe the squeezing dynamics of a relatively ordered subset of NV centers, which alleviates the rapid decay of the total spin length. Combining all of the above experimental innovations, we demonstrate spin squeezing in an NV ensemble (-0.5~$\pm$~0.1~dB) and probe its time evolution [Fig.~\ref{fig:fig4}].

\ifnature
\emph{Dipolar twisting dynamics} ---
\else
\section{Dipolar twisting dynamics}
\fi
Each NV center comprises an electronic spin-1  ground state that can be optically polarized and read out~\cite{dohertyNitrogenvacancyColourCentre2013}. 
We work with an effective two-level system encoded in the $\{\ket{m_s = 0}, \ket{m_s = -1}\}$ sublevels of the NV ground-state manifold~[Fig.~\ref{fig:fig1}a].
The native interaction between NV spins is a magnetic dipole-dipole coupling, which generically exhibits an angular dependence as a function of both the relative position and orientation of the dipoles. 
%
% quantized along the NV axis? or not?
Crucially, our (111)-oriented diamond sample contains a subgroup of NV centers whose quantization axis is perpendicular to the 2D plane~[Fig.~\ref{fig:fig1}b].
For these NVs, the angular dependence drops out and the system is governed by a uniform, long-range XXZ Hamiltonian: 
\begin{equation}\label{eq:HXXZ}
    \Hxxz = -\sum_{i,j}\frac{J_0}{r_{ij}^3}(s^i_x s^j_x + s^i_y s^j_y - s^i_z s^j_z),
\end{equation}
where $\hbar=1$, $J_0 = (2\pi)\times\SI{52}{\mega\hertz\cdot\nano\meter^3}$ 
characterizes the strength of the dipolar interaction, $\mathbf{s^i}$ is the effective spin-1/2 operator acting on our two-level system, and $r_{ij}$ is  the distance between NVs, with an average spacing of $\sim 20$~nm (corresponding to an areal density of $\rho = 8$~ppm$\cdot$nm).

\ifnature
\else
\figOne
\fi

In order to generate spin squeezing, the Hamiltonian dynamics must somehow reshape the spin projection noise.
In the case of one-axis twisting, this reshaping is achieved by a Hamiltonian,  $H_\text{OAT} = \chi S_z^2$, which causes a spin-polarized state with mean value $\expval{S_z}$, 
to precess about the $z$-axis at a rate proportional to  $\expval{S_z}$; here, $S_z = \sum_i s_z^i$ corresponds to the collective spin operator~\cite{kitagawaSqueezedSpinStates1993, gross2010nonlinear, lerouxImplementationCavitySqueezing2010, bohnetQuantumSpinDynamics2016, braverman2019near}.  
%,borishTransverseFieldIsingDynamics2020}.
%
On inspection, $\Hxxz$ is quite distinct from the OAT Hamiltonian: rather than all-to-all Ising interactions, it features an effectively short-range power-law as well as spin-exchange interactions.
Interestingly, recent works have shown that, for lattice systems, the quench dynamics of a spin-polarized state under $\Hxxz$ can be remarkably similar to those under  $H_\text{OAT}$~\cite{perlinSpinSqueezingShortRange2020,comparinScalableSpinSqueezing2022,blockScalableSpinSqueezing2024,bornetScalableSpinSqueezing2023}.

Whether this similarity extends to ensembles of optically addressable solid-state spin defects is a much more delicate question. 
Indeed, all such systems exhibit strong positional disorder owing to the random positioning of defects. 
Such positional disorder can yield a variety of subtle effects, ranging from strongly coupled dimers to the onset of spin glass physics~\cite{edwardsTheorySpinGlasses1975, burin2015many, abaninColloquiumManybodyLocalization2019,martinControllingLocalThermalization2023}.

To this end, we begin by measuring the quench dynamics of NV centers under
the disordered dipolar $\Hxxz$ Hamiltonian [Eqn.~\ref{eq:HXXZ}].
After optically pumping to the $\ket{m_s = 0}$ state, we initialize a spin-polarized state, $\ket{\mathbf{x_{\varphi}}}$, which is offset above (or below) the $x$-axis by an angle $\varphi_o$ [Fig.~\ref{fig:fig1}c].
We then allow this state to evolve under $\Hxxz$ and measure the time evolution of the collective spin operators $S_x = \sum_i s_x^i$ and $S_y = \sum_i s_y^i$, in order to calculate the average precession angle in the equatorial plane,  $\varphi_p = \arctan(S_y/S_x)$ [Fig.~\ref{fig:fig1}d].
Throughout the time evolution, we utilize an XY-8 pulse sequence (see Methods) to dynamically decouple the NV ensemble from quasi-static fields, generated for example,  by other paramagnetic impurities.

As depicted in Fig.~\ref{fig:fig1}c,d, we observe a linear increase in the precession angle $\varphi_p$ at early times, at a rate proportional to the expectation value of $\langle S_z \rangle \sim S\varphi_o$, where $S$ is the total spin length.
This is reminiscent of the aforementioned ``twisting'' dynamics associated with the OAT Hamiltonian.
From the early-time slope of $\varphi_p(t)$ [Fig.~\ref{fig:fig1}d],  we extract an average twisting strength, $\chi_\textrm{eff}=(2\pi) \times $150~kHz (see Methods). 
At late times, we observe that $\varphi_p$ begins to deviate from linear behavior, in contrast with the OAT model. The difference arises from an inhomogeneous distribution of twisting rates, and is a manifestation of the positional disorder in our system.  

\ifnature
\else
\figTwo
\fi

The fact that $\Hxxz$ induces early-time twisting dynamics analogous to $H_\text{OAT}$ seems promising for realizing spin squeezing: although the observed twisting of the mean spin vector [Fig.~\ref{fig:fig1}c,d] can be understood classically, the same dynamics also generate shearing of the quantum spin projection noise~\cite{borishTransverseFieldIsingDynamics2020, hines2023spin}.
In particular, this projection noise is often visualized as a Wigner quasiprobability distribution on the  Bloch sphere of the collective spin; as illustrated in Fig.~\ref{fig:fig2}b, the twisting rate of this distribution is effectively given by  $\sim \chi \sqrt{\text{Var}(S_z)}$, where $\sqrt{\text{Var}(S_z)}$ corresponds to the width of the spin projection noise along the $z$-axis.
For a fixed spin length, this twisting shears the projection noise and leads to  spin squeezing [Fig.~\ref{fig:fig2}b], 
characterized by the parameter,
\begin{equation}
    \xi^2 = N\frac{\min_\theta \mathrm{Var}(S_\theta)}{\expval{S_x}^2},
    \label{squeezing_parameter}
\end{equation}
where $N$ is the system size and $\mathrm{Var}(S_\theta)$ is the variance of the collective spin operator $S_\theta = \cos(\theta) S_z + \sin(\theta) S_y$; $\xi^2 < 1$ represents the improvement in the signal to noise ratio achieved when performing  Ramsey interferometry with a squeezed state versus a spin-polarized state~\cite{winelandSpinSqueezingReduced1992,sorensenEntanglementExtremeSpin2001, sorensenManyparticleEntanglementBose2001}.

\ifnature
\emph{Interaction-enabled readout of the quantum spin projection noise} --- 
\else
\section{Interaction-enabled readout of the quantum spin projection noise}
\fi
For quantum simulation platforms with spin-projection-noise-limited readout, the variance of the collective spin operator, $\mathrm{Var}(S_\theta)$, can be directly measured~\cite{degenQuantumSensing2017}. 
However, to date, such readout has yet to be demonstrated for ensembles of optically active, solid-state spin defects. 
In the specific case of NV centers, while higher-fidelity readout schemes exist, including nuclear-spin-enabled repetitive readout, resonant excitation at low temperatures, and spin-to-charge conversion~\cite{jiangRepetitiveReadoutSingle2009,robledoHighfidelityProjectiveReadout2011, shieldsEfficientReadoutSingle2015, arunkumarQuantumLogicEnhanced2023}, the best measurements of NV ensembles are still far from the quantum spin projection noise  limit~\cite{arunkumarQuantumLogicEnhanced2023}.

In order to demonstrate spin squeezing in our  NV ensemble without direct detection of the noise, our experiments will proceed in two conceptually separate steps [Fig.~\ref{fig:fig2}a]: 
(i) \emph{generation} of the spin squeezed state via Hamiltonian evolution from an initial spin-polarized state $\ket{\mathbf{x}}$ under $\Hxxz$ for a time $t_\textrm{g}$, and (ii) \emph{readout} of the state's 
anisotropic spin projection noise via a quench protocol for a time $t_\mathrm{r}$ involving \emph{only} measurements of the collective spin vector.
The latter can be understood as a form of interaction-enabled noise spectroscopy, where a measurement of the decay of the collective spin length, $\langle S_x (t_\mathrm{r}) \rangle$,  provides information about the quantum state of the system. 
In particular, from a mean-field perspective, each NV center experiences a net magnetic field originating from all of the other NV spins. 
As usual, the fluctuations of this field govern the decoherence dynamics of the system.
Crucially, when the many-body dynamics of the NV ensemble is dominated by the intrinsic interactions between the spins, then these
 fluctuations are effectively given by the spin projection noise of the state itself; a larger spin projection noise leads to faster decoherence, and vice versa [Fig.~\ref{fig:fig2}c,d].

% \emph{Protocol for spin-projection noise readout}---
The simplest setting to explicate our protocol is via the OAT model, where the connection between the decay of the mean spin length, $\langle S_x (t_\mathrm{r}) \rangle$, and the ensemble's quantum variance is particularly transparent. 
Specifically, consider the class of spin-squeezed states prepared via evolution under $H_\text{OAT}=\chi S_z^2$.
As the Wigner distribution twists, it begins to sample the curvature of the Bloch sphere, causing the total spin length to decrease (Fig.~\ref{fig:fig2}b);
since the twisting rate $\sim \chi \sqrt{\text{Var}(S_z)}$, a state with a larger (smaller) variance wraps around the Bloch sphere more quickly (slowly).
In the case of one-axis twisting, this intuition can be analytically formalized into an equation relating the decay of $\langle S_x (t_\mathrm{r}) \rangle$ to  $\mathrm{Var}(S_z)$.  
Specifically, for a squeezed state whose spin projection noise is stretched (or compressed) along the $z$-axis [Fig.~\ref{fig:fig2}c,e], one can immediately derive that (see Supplementary Information): 
\begin{equation}\label{eq:decaySx}
    \langle S_x(t_\mathrm{r}) \rangle \sim e^{-2\chi^2 \text{Var}(S_z) t_\mathrm{r}^2} \equiv e^{-(t_\mathrm{r}/T_2)^2}.
\end{equation}
For states where the projection noise is squeezed along a different axis, one simply needs to redefine the time, $t_\mathrm{r} \rightarrow t_\mathrm{r}^\mathrm{eff}$, that enters the $S_x(t_\mathrm{r})$ decay profile (see Methods).

From an operational perspective, Eqn.~\ref{eq:decaySx} enables one to directly characterize the spin projection noise of the squeezed state by measuring the characteristic decay timescale, $T_2$, of the collective spin.
To compute the squeezing parameter, $\xi^2$, one must measure $\mathrm{Var}(S_\theta)$ as a function of $\theta$.
This can be achieved by using a global spin rotation, to rotate the variance along an arbitrary direction $\theta$ to the $z$-axis, i.e.~$\mathrm{Var}(S_\theta) \rightarrow \mathrm{Var}(S_z)$.

While we have described our strategy in the context of the OAT model, our protocol is expected to apply for \emph{any} system undergoing  ``twisting'' dynamics.
However, unlike the OAT setting, for our disordered, dipolar NV ensemble, the mapping between the decay
of $\langle S_x(t_\mathrm{r}) \rangle$ and the spin projection noise, $\sqrt{\text{Var}(S_z)}$, cannot be analytically determined. 
To this end, as illustrated in Fig.~\ref{fig:fig2}d, we perform an extensive set of simulations using the discrete-cluster truncated Wigner approximation (cluster DTWA)\cite{schachenmayerManyBodyQuantumSpin2015,braemerClusterTruncatedWigner2024} in order to numerically determine the mapping (see Methods); we carefully benchmark this mapping via both exact Krylov subspace methods and a variational approach based upon neural quantum states. 

We end on a quick remark. Since ``twisting dynamics'' naturally occur for XXZ Hamiltonians, which arise in a wide-variety of quantum simulation platforms~\cite{wolfowiczQuantumGuidelinesSolidstate2021,chomazDipolarPhysicsReview2022a,cornishQuantumComputationQuantum2024},  we emphasize that our protocol provides a general strategy for measuring the quantum variance in such systems without the need for projection-noise-limited readout. 

\ifnature
\else
\figThree
\fi

\ifnature
\emph{Probing spin squeezing in a disordered, dipolar spin ensemble} ---
\else
\section{Probing spin squeezing in a disordered, dipolar spin ensemble}
\fi
Having established both twisting dynamics [Fig.~\ref{fig:fig1}c,d] and a global protocol for measuring the spin projection noise [Fig.~\ref{fig:fig2}d], we now turn to directly characterizing spin squeezing in our disordered NV ensemble.
Figure~\ref{fig:fig2}e depicts the results for $t_\mathrm{g} = 3.2\ \mu$s.
Compared to $\theta = 0$ (gray data points), for certain rotation angles (red data points), $\langle S_x(t_\mathrm{r}^\mathrm{eff}) \rangle$ exhibits a faster decay in time, while for others (blue data points), it exhibits a slower decay.
Quantitatively, for each rotation angle, $\theta$, we extract the characteristic decay time-scale, $T_2$ and, utilize the computed mapping to determine the corresponding quantum variance [Fig.~\ref{fig:fig2}f].
The sinusoidal shape of the variance as a function of $\theta$ precisely reflects the elliptical shape of the state's spin projection noise.
Moreover, it immediately allows one to determine the minimum variance, $\min_\theta[\mathrm{Var}(S_\theta)]$, and thus the squeezing parameter for $t_\mathrm{g} = 3.2\ \mu$s.

To characterize the time evolution of spin squeezing, we vary the preparation time, $t_\mathrm{g}$, and repeat the same experiment.
As evinced in Fig.~\ref{fig:fig2}g (red data), despite the fact that the minimum projection noise is decreasing as a function of time, the squeezing parameter (Fig.~\ref{fig:fig2}h) is monotonically increasing, implying that the state does not exhibit any enhanced metrological utility. The lack of squeezing arises because the state's mean spin length is decreasing too quickly relative to the reduction of the spin projection noise [Fig.~\ref{fig:fig2}g]. As is evident from Eqn.~\ref{squeezing_parameter}, in order to achieve spin squeezing ($\xi^2 < 1$), one must balance two (typically competing) processes: the twisting-induced reduction of the minimal quantum variance and the shrinking of the collective spin length.

Interestingly, this issue stems from the positional disorder of NV centers.
Indeed, as shown in Fig.~\ref{fig:fig3}b, numerical simulations of an ordered spin ensemble [left, Fig.~\ref{fig:fig3}a] predict that the optimal value of the squeezing parameter is well below unity; meanwhile, consistent with our experimental observations, the squeezing dynamics of a disordered spin ensemble [right, Fig.~\ref{fig:fig3}a] never yield $\xi^2(t_\mathrm{g}) < 1$.
This difference can be understood by examining the underlying distribution of interaction strengths, $P(J) = \sum_j J_0/r_{ij}^3$, in the two cases. Compared to the lattice, the disordered spin ensemble exhibits a heavy tail owing to the presence of strongly-coupled dimers [Fig.~\ref{fig:fig3}a].

These dimers have two effects. 
First, their large interaction strength leads to fast dynamics, which manifest as the rapid decay of their initial $x$-polarization. 
Thus, such dimers contribute an anomalously fast decay to the mean spin length. 
In particular, by contrast to the lattice case, where $\langle S_x (t_\mathrm{g}) \rangle \sim e^{-t_\mathrm{g}^{2}}$, the disordered case exhibits  $\langle S_x (t_\mathrm{g}) \rangle \sim e^{-t_\mathrm{g}^{2/3}}$ (Fig.~\ref{fig:fig2}d, Fig.~\ref{fig:fig3}b)~\cite{davisProbingManybodyDynamics2023}; importantly, the former results in an early-time derivative $dS_x/dt_\mathrm{g}|_{t_\mathrm{g}=0}$ of zero, whereas  the latter displays a rapid initial decrease in $\langle S_x \rangle$. 
Second, their large interaction strength implies that they are only weakly coupled to the rest of the many-body system, and thus, their dynamics do not significantly contribute to the reduction of the system's spin projection noise. 
In combination, these effects prevent spin squeezing in strongly-disordered spin ensembles [Fig.~\ref{fig:fig2}h, Fig.~\ref{fig:fig3}b bottom]. 

\ifnature
\emph{Spectral tailoring of positional disorder} ---
\else
\section{Spectral tailoring of positional disorder}
\fi
To this end, we develop and implement two \emph{lattice-engineering} protocols aimed at reducing the positional disorder in our NV ensemble
[Fig.~\ref{fig:fig3}(c,d)]. 
Conceptually, both protocols address the tail of $P(J)$ in order to spectrally isolate NV dimers and ``remove'' them from the many-body dynamics (Fig.~\ref{fig:fig3}a).
In our first lattice-engineering method, we take advantage of the unused  $\ket{m_s = 1}$ NV sub-level (Fig.~\ref{fig:fig1}a inset).
In particular, we utilize a weak microwave $\pi$-pulse with high frequency selectivity, to shelve dimers into the $\ket{m_s = 1}$ state (Fig.~\ref{fig:fig3}c), while NVs with more typical interaction strengths remain in the $\{\ket{m_s=0}, \ket{m_s = -1}\}$ subspace.
In our second method, after initializing the spin-polarized state, $\ket{\mathbf{x}}$, we turn on a strong, resonant microwave drive, in order to apply a large transverse field, $h_xS_x$, to the NV centers. 
We then ramp the strength of this transverse field to a final value, $h^f_x$, chosen based on the shape of the interaction spectrum $P(J)$ (Fig.~\ref{fig:fig3}d).
For NVs whose interaction is weaker than $h^f_x$, the transverse field maintains the state's initial $x$-polarization. 
By contrast, for NVs whose interaction is stronger than $h^f_x$, the dynamics remain dominated by $H_\textrm{XXZ}$ and the spins still exhibit rapid depolarization.
Crucially, depolarization of the strongly-interacting dimers separates them from the remaining spin-polarized NV-subensemble, whose squeezing dynamics can then be probed.

As a proxy for characterizing the effectiveness of our dimer removal protocols, as well as the reduction in the positional disorder, we investigate the early time decay of $\langle S_x (t) \rangle \sim e^{-t^{p}}$.
As aforementioned, in the fully disordered case, $p=2/3$, while in the lattice case, $p=2$. 
Remarkably, as depicted in Fig.~\ref{fig:fig3}(e,f), both protocols yield a dramatic increase in the early-time stretch exponent, $p$.
This suggests that the remnant NV sub-ensemble --- i.e. the un-shelved NVs in our first method, or the polarized NVs in our second method --- exhibits an interaction distribution, $P(J)$, that is significantly closer to the lattice case. 

\ifnature
\emph{Spin squeezing of an NV ensemble} ---
\else
\section{Spin squeezing of an NV ensemble}
\fi
We now explore the dynamics of spin squeezing in the presence of our lattice-engineering protocols. 
For both protocols, 
the decay of $\langle S_x(t_\mathrm{r}) \rangle$ depends on $\theta$ [Fig.~\ref{fig:fig3}g] and the extracted variance exhibits the sinusoidal shape characteristic of squeezing [Fig.~\ref{fig:fig3}h]. 
This is qualitatively similar to our observations in the absence of lattice engineering [Fig.~\ref{fig:fig2}e,f].
However, owing to the removal of dimers, the decay of the mean spin length is significantly slower, while the twisting-induced reduction of the system's quantum variance remains largely unchanged [Fig.~\ref{fig:fig4}(a,b) for the shelving method and Fig.~\ref{fig:fig4}(d,e) for the depolarization method].
For the shelving method, we observe spin squeezing with an optimal squeezing parameter, $\xi^2 = 0.90(3)$, at a time $t_\mathrm{g} = 1.6\ \mu$s, while for the depolarization method, we observe spin squeezing with an optimal squeezing parameter, $\xi^2 = 0.89(2)$, at a time $t_\mathrm{g} = 1.6\ \mu$s. 
These correspond, respectively to $-0.46\pm 0.14$ dB and $-0.50\pm0.10$ dB of squeezing, and represent the first observation of spin squeezing in the solid-state.

\ifnature
\else
\figFour
\fi

\ifnature
\emph{Discussion and outlook} ---
\else
\section{Discussion and outlook}
\fi
Our experimental demonstration of spin squeezing in a practical, solid-state quantum sensor opens up a wealth of possibilities.
Most important is the prospect of actually utilizing spin squeezing to measure signals that would be otherwise undetectable. 
In particular, combining our approach with 
recent efforts to push the detection fidelity of NV ensembles towards the spin projection noise limit \cite{arunkumarQuantumLogicEnhanced2023} could enable quantum-enhanced sensing of biological systems and materials~\cite{casolaProbingCondensedMatter2018, aslamQuantumSensorsBiomedical2023}.
In addition, to further increase the amount of spin squeezing, one can envision two natural directions.
First, improving the deterministic placement of spin defects via localized ion implantation or annealing \cite{chenLaserWritingIndividual2019, groot-berningFabrication15NVCenters2021}, and second, engineering the effective Hamiltonian to operate in a regime more favorable for scalable spin squeezing~\cite{blockScalableSpinSqueezing2024, bornetScalableSpinSqueezing2023}.

More broadly, our work sheds light on a number of fundamental challenges for the realization of spin squeezing in optically active, solid-state sensors. 
Perhaps most formidably, is the presence of strongly coupled clusters of spins, which contribute an anomalously fast decay to the collective spin length. 
Our approach to overcoming this challenge, namely, spectrally-resolved lattice engineering, is widely applicable to other 
disordered dipolar platforms, including magnetic atoms and polar molecules in optical lattices with low filling ~\cite{carrollObservationGeneralizedTJ2024}, as well as ultracold gases of Rydberg atoms~\cite{signoles2021glassy}.
Finally, our work also introduces and implements a novel technique for reading out the quantum variance of a state, without the need for projection-noise-limited measurements. 

\emph{Acknowledgments}---We gratefully acknowledge the insightful discussions with M. Aidelsburger, S. Chern, P. Crowley, H. Gao, B. Kobrin, N. Leitao, F. Machado, T. Schuster and B. Zhu. 
This work was supported by the 
U.S. Department of Energy via the Office of Science, National Quantum Information Science Research Centers, Quantum Systems Accelerator and via the BES grant no.~DE-SC0019241, as well as  the Army Research Office via grant no.~W911NF-24-1-0079 and through the MURI program (grant no.~W911NF-20-1-0136).
We acknowledge the use of shared facilities of the UCSB Quantum Foundry through Q-AMASE-i program (NSF DMR-1906325), the UCSB MRSEC (NSF DMR 1720256), and the Quantum Structures Facility within the UCSB California NanoSystems Institute.
A.B.J. acknowledges support from the NSF QLCI program through grant number OMA-2016245.
L.B.H. acknowledges support from the NSF Graduate Research Fellowship Program (DGE 2139319) and the UCSB Quantum Foundry.
D.K. acknowledges support from Generation-Q AWS and HQI fellowships.
T.O. acknowledges support from the Ezoe Memorial Recruit Foundation.
S.A.M. acknowledges support from the UCSB Quantum Foundry (NSF DMR-1906325) and support from the Canada NSERC (Grant No. AID 516704-2018).

\vspace{3mm}

\emph{Author Contributions}--- 
W.W., E.J.D., and Z.W. performed experiments with the help of T.O., C.L. and H.Y. 
L.B.H., S.A.M. and A.B.J. synthesized the diamond sample.
W.W. and E.J.D. developed the experimental protocols and performed data analysis.
E.J.D., B.Y. and M.B. developed the theoretical models and methodology.
W.W., D.K., B.Y. and Z.W. performed the numerical simulation.
A.B.J. and N.Y.Y. supervised the project. 
W.W., E.J.D., B.Y., Z.W. and N.Y.Y. wrote the manuscript with input from all authors.

\vspace{3mm}

\emph{Competing interests}--- The authors declare no competing interests..

\vspace{3mm}

\ifnature
\newpage
\figOne
\newpage
\figTwo
\newpage
\figThree
\newpage
\figFour
\clearpage
\else
\fi

\ifnature
\bibliographystyle{naturemag}
\else
\bibliographystyle{apsrev4-2}
\fi

\bibliography{main_arXiv.bib}

\begin{thebibliography}{10}
\expandafter\ifx\csname url\endcsname\relax
  \def\url#1{\texttt{#1}}\fi
\expandafter\ifx\csname urlprefix\endcsname\relax\def\urlprefix{URL }\fi
\providecommand{\bibinfo}[2]{#2}
\providecommand{\eprint}[2][]{\url{#2}}

\bibitem{winelandSpinSqueezingReduced1992}
\bibinfo{author}{Wineland, D.~J.}, \bibinfo{author}{Bollinger, J.~J.}, \bibinfo{author}{Itano, W.~M.}, \bibinfo{author}{Moore, F.~L.} \& \bibinfo{author}{Heinzen, D.~J.}
\newblock \bibinfo{title}{Spin squeezing and reduced quantum noise in spectroscopy}.
\newblock \emph{\bibinfo{journal}{Physical Review A}} \textbf{\bibinfo{volume}{46}}, \bibinfo{pages}{R6797--R6800} (\bibinfo{year}{1992}).

\bibitem{kitagawaSqueezedSpinStates1993}
\bibinfo{author}{Kitagawa, M.} \& \bibinfo{author}{Ueda, M.}
\newblock \bibinfo{title}{Squeezed spin states}.
\newblock \emph{\bibinfo{journal}{Physical Review A}} \textbf{\bibinfo{volume}{47}}, \bibinfo{pages}{5138--5143} (\bibinfo{year}{1993}).

\bibitem{meyer2001experimental}
\bibinfo{author}{Meyer, V.} \emph{et~al.}
\newblock \bibinfo{title}{Experimental demonstration of entanglement-enhanced rotation angle estimation using trapped ions}.
\newblock \emph{\bibinfo{journal}{Phys. Rev. Lett.}} \textbf{\bibinfo{volume}{86}}, \bibinfo{pages}{5870--5873} (\bibinfo{year}{2001}).
\newblock \urlprefix\url{https://link.aps.org/doi/10.1103/PhysRevLett.86.5870}.

\bibitem{LouchetChauvet2010}
\bibinfo{author}{Louchet-Chauvet, A.} \emph{et~al.}
\newblock \bibinfo{title}{Entanglement-assisted atomic clock beyond the projection noise limit}.
\newblock \emph{\bibinfo{journal}{New Journal of Physics}} \textbf{\bibinfo{volume}{12}}, \bibinfo{pages}{065032} (\bibinfo{year}{2010}).
\newblock \urlprefix\url{https://dx.doi.org/10.1088/1367-2630/12/6/065032}.

\bibitem{gross2010nonlinear}
\bibinfo{author}{Gross, C.}, \bibinfo{author}{Zibold, T.}, \bibinfo{author}{Nicklas, E.}, \bibinfo{author}{Esteve, J.} \& \bibinfo{author}{Oberthaler, M.~K.}
\newblock \bibinfo{title}{Nonlinear atom interferometer surpasses classical precision limit}.
\newblock \emph{\bibinfo{journal}{Nature}} \textbf{\bibinfo{volume}{464}}, \bibinfo{pages}{1165--1169} (\bibinfo{year}{2010}).

\bibitem{sewell2012magnetic}
\bibinfo{author}{Sewell, R.~J.} \emph{et~al.}
\newblock \bibinfo{title}{Magnetic sensitivity beyond the projection noise limit by spin squeezing}.
\newblock \emph{\bibinfo{journal}{Physical review letters}} \textbf{\bibinfo{volume}{109}}, \bibinfo{pages}{253605} (\bibinfo{year}{2012}).

\bibitem{ockeloen2013quantum}
\bibinfo{author}{Ockeloen, C.~F.}, \bibinfo{author}{Schmied, R.}, \bibinfo{author}{Riedel, M.~F.} \& \bibinfo{author}{Treutlein, P.}
\newblock \bibinfo{title}{Quantum metrology with a scanning probe atom interferometer}.
\newblock \emph{\bibinfo{journal}{Physical review letters}} \textbf{\bibinfo{volume}{111}}, \bibinfo{pages}{143001} (\bibinfo{year}{2013}).

\bibitem{hostenMeasurementNoise1002016}
\bibinfo{author}{Hosten, O.}, \bibinfo{author}{Engelsen, N.~J.}, \bibinfo{author}{Krishnakumar, R.} \& \bibinfo{author}{Kasevich, M.~A.}
\newblock \bibinfo{title}{Measurement noise 100 times lower than the quantum-projection limit using entangled atoms}.
\newblock \emph{\bibinfo{journal}{Nature}} \textbf{\bibinfo{volume}{529}}, \bibinfo{pages}{505--508} (\bibinfo{year}{2016}).

\bibitem{fernholz2008spin}
\bibinfo{author}{Fernholz, T.} \emph{et~al.}
\newblock \bibinfo{title}{Spin squeezing of atomic ensembles via nuclear-electronic spin entanglement}.
\newblock \emph{\bibinfo{journal}{Physical review letters}} \textbf{\bibinfo{volume}{101}}, \bibinfo{pages}{073601} (\bibinfo{year}{2008}).

\bibitem{lerouxImplementationCavitySqueezing2010}
\bibinfo{author}{Leroux, I.~D.}, \bibinfo{author}{{Schleier-Smith}, M.~H.} \& \bibinfo{author}{Vuleti{\'c}, V.}
\newblock \bibinfo{title}{Implementation of {{Cavity Squeezing}} of a {{Collective Atomic Spin}}}.
\newblock \emph{\bibinfo{journal}{Physical Review Letters}} \textbf{\bibinfo{volume}{104}}, \bibinfo{pages}{073602} (\bibinfo{year}{2010}).

\bibitem{hamley2012spin}
\bibinfo{author}{Hamley, C.~D.}, \bibinfo{author}{Gerving, C.}, \bibinfo{author}{Hoang, T.~M.}, \bibinfo{author}{Bookjans, E.~M.} \& \bibinfo{author}{Chapman, M.~S.}
\newblock \bibinfo{title}{Spin-nematic squeezed vacuum in a quantum gas}.
\newblock \emph{\bibinfo{journal}{Nature Physics}} \textbf{\bibinfo{volume}{8}}, \bibinfo{pages}{305--308} (\bibinfo{year}{2012}).

\bibitem{riedel2010atom}
\bibinfo{author}{Riedel, M.~F.} \emph{et~al.}
\newblock \bibinfo{title}{Atom-chip-based generation of entanglement for quantum metrology}.
\newblock \emph{\bibinfo{journal}{Nature}} \textbf{\bibinfo{volume}{464}}, \bibinfo{pages}{1170--1173} (\bibinfo{year}{2010}).

\bibitem{bohnet2014reduced}
\bibinfo{author}{Bohnet, J.~G.} \emph{et~al.}
\newblock \bibinfo{title}{Reduced spin measurement back-action for a phase sensitivity ten times beyond the standard quantum limit}.
\newblock \emph{\bibinfo{journal}{Nature Photonics}} \textbf{\bibinfo{volume}{8}}, \bibinfo{pages}{731--736} (\bibinfo{year}{2014}).

\bibitem{bohnetQuantumSpinDynamics2016}
\bibinfo{author}{Bohnet, J.~G.} \emph{et~al.}
\newblock \bibinfo{title}{Quantum spin dynamics and entanglement generation with hundreds of trapped ions}.
\newblock \emph{\bibinfo{journal}{Science}} \textbf{\bibinfo{volume}{352}}, \bibinfo{pages}{1297--1301} (\bibinfo{year}{2016}).

\bibitem{bornetScalableSpinSqueezing2023}
\bibinfo{author}{Bornet, G.} \emph{et~al.}
\newblock \bibinfo{title}{Scalable spin squeezing in a dipolar {{Rydberg}} atom array}.
\newblock \emph{\bibinfo{journal}{Nature}} \textbf{\bibinfo{volume}{621}}, \bibinfo{pages}{728--733} (\bibinfo{year}{2023}).

\bibitem{frankeQuantumenhancedSensingOptical2023}
\bibinfo{author}{Franke, J.} \emph{et~al.}
\newblock \bibinfo{title}{Quantum-enhanced sensing on optical transitions through finite-range interactions}.
\newblock \emph{\bibinfo{journal}{Nature}} \textbf{\bibinfo{volume}{621}}, \bibinfo{pages}{740--745} (\bibinfo{year}{2023}).

\bibitem{ecknerRealizingSpinSqueezing2023}
\bibinfo{author}{Eckner, W.~J.} \emph{et~al.}
\newblock \bibinfo{title}{Realizing spin squeezing with {{Rydberg}} interactions in an optical clock}.
\newblock \emph{\bibinfo{journal}{Nature}} \textbf{\bibinfo{volume}{621}}, \bibinfo{pages}{734--739} (\bibinfo{year}{2023}).

\bibitem{hines2023spin}
\bibinfo{author}{Hines, J.~A.} \emph{et~al.}
\newblock \bibinfo{title}{Spin squeezing by rydberg dressing in an array of atomic ensembles}.
\newblock \emph{\bibinfo{journal}{Physical Review Letters}} \textbf{\bibinfo{volume}{131}}, \bibinfo{pages}{063401} (\bibinfo{year}{2023}).

\bibitem{grossQuantumSimulationsUltracold2017}
\bibinfo{author}{Gross, C.} \& \bibinfo{author}{Bloch, I.}
\newblock \bibinfo{title}{Quantum simulations with ultracold atoms in optical lattices}.
\newblock \emph{\bibinfo{journal}{Science}} \textbf{\bibinfo{volume}{357}}, \bibinfo{pages}{995--1001} (\bibinfo{year}{2017}).

\bibitem{browaeysManybodyPhysicsIndividually2020}
\bibinfo{author}{Browaeys, A.} \& \bibinfo{author}{Lahaye, T.}
\newblock \bibinfo{title}{Many-body physics with individually controlled {{Rydberg}} atoms}.
\newblock \emph{\bibinfo{journal}{Nature Physics}} \textbf{\bibinfo{volume}{16}}, \bibinfo{pages}{132--142} (\bibinfo{year}{2020}).

\bibitem{kjaergaardSuperconductingQubitsCurrent2020}
\bibinfo{author}{Kjaergaard, M.} \emph{et~al.}
\newblock \bibinfo{title}{Superconducting {{Qubits}}: {{Current State}} of {{Play}}}.
\newblock \emph{\bibinfo{journal}{Annual Review of Condensed Matter Physics}} \textbf{\bibinfo{volume}{11}}, \bibinfo{pages}{369--395} (\bibinfo{year}{2020}).

\bibitem{monroeProgrammableQuantumSimulations2021}
\bibinfo{author}{Monroe, C.}
\newblock \bibinfo{title}{Programmable quantum simulations of spin systems with trapped ions}.
\newblock \emph{\bibinfo{journal}{Reviews of Modern Physics}} \textbf{\bibinfo{volume}{93}} (\bibinfo{year}{2021}).

\bibitem{auccaiseSpinSqueezingQuadrupolar2015}
\bibinfo{author}{Auccaise, R.} \emph{et~al.}
\newblock \bibinfo{title}{Spin {{Squeezing}} in a {{Quadrupolar Nuclei NMR System}}}.
\newblock \emph{\bibinfo{journal}{Physical Review Letters}} \textbf{\bibinfo{volume}{114}}, \bibinfo{pages}{043604} (\bibinfo{year}{2015}).

\bibitem{wolfowiczQuantumGuidelinesSolidstate2021}
\bibinfo{author}{Wolfowicz, G.} \emph{et~al.}
\newblock \bibinfo{title}{Quantum guidelines for solid-state spin defects}.
\newblock \emph{\bibinfo{journal}{Nature Reviews Materials}} \textbf{\bibinfo{volume}{6}}, \bibinfo{pages}{906--925} (\bibinfo{year}{2021}).

\bibitem{randall2021many}
\bibinfo{author}{Randall, J.} \emph{et~al.}
\newblock \bibinfo{title}{Many-body--localized discrete time crystal with a programmable spin-based quantum simulator}.
\newblock \emph{\bibinfo{journal}{Science}} \textbf{\bibinfo{volume}{374}}, \bibinfo{pages}{1474--1478} (\bibinfo{year}{2021}).

\bibitem{kianiniaQuantumEmitters2D2022}
\bibinfo{author}{Kianinia, M.}, \bibinfo{author}{Xu, Z.-Q.}, \bibinfo{author}{Toth, M.} \& \bibinfo{author}{Aharonovich, I.}
\newblock \bibinfo{title}{Quantum emitters in {{2D}} materials: {{Emitter}} engineering, photophysics, and integration in photonic nanostructures}.
\newblock \emph{\bibinfo{journal}{Applied Physics Reviews}} \textbf{\bibinfo{volume}{9}}, \bibinfo{pages}{011306} (\bibinfo{year}{2022}).

\bibitem{burkardSemiconductorSpinQubits2023}
\bibinfo{author}{Burkard, G.}, \bibinfo{author}{Ladd, T.~D.}, \bibinfo{author}{Pan, A.}, \bibinfo{author}{Nichol, J.~M.} \& \bibinfo{author}{Petta, J.~R.}
\newblock \bibinfo{title}{Semiconductor spin qubits}.
\newblock \emph{\bibinfo{journal}{Reviews of Modern Physics}} \textbf{\bibinfo{volume}{95}}, \bibinfo{pages}{025003} (\bibinfo{year}{2023}).

\bibitem{kirsteinSqueezedDarkNuclear2023}
\bibinfo{author}{Kirstein, E.} \emph{et~al.}
\newblock \bibinfo{title}{The squeezed dark nuclear spin state in lead halide perovskites}.
\newblock \emph{\bibinfo{journal}{Nature Communications}} \textbf{\bibinfo{volume}{14}}, \bibinfo{pages}{6683} (\bibinfo{year}{2023}).

\bibitem{maze2008nanoscale}
\bibinfo{author}{Maze, J.~R.} \emph{et~al.}
\newblock \bibinfo{title}{Nanoscale magnetic sensing with an individual electronic spin in diamond}.
\newblock \emph{\bibinfo{journal}{Nature}} \textbf{\bibinfo{volume}{455}}, \bibinfo{pages}{644--647} (\bibinfo{year}{2008}).

\bibitem{le2013optical}
\bibinfo{author}{Le~Sage, D.} \emph{et~al.}
\newblock \bibinfo{title}{Optical magnetic imaging of living cells}.
\newblock \emph{\bibinfo{journal}{Nature}} \textbf{\bibinfo{volume}{496}}, \bibinfo{pages}{486--489} (\bibinfo{year}{2013}).

\bibitem{kucsko2013nanometre}
\bibinfo{author}{Kucsko, G.} \emph{et~al.}
\newblock \bibinfo{title}{Nanometre-scale thermometry in a living cell}.
\newblock \emph{\bibinfo{journal}{Nature}} \textbf{\bibinfo{volume}{500}}, \bibinfo{pages}{54--58} (\bibinfo{year}{2013}).

\bibitem{kolkowitz2015probing}
\bibinfo{author}{Kolkowitz, S.} \emph{et~al.}
\newblock \bibinfo{title}{Probing johnson noise and ballistic transport in normal metals with a single-spin qubit}.
\newblock \emph{\bibinfo{journal}{Science}} \textbf{\bibinfo{volume}{347}}, \bibinfo{pages}{1129--1132} (\bibinfo{year}{2015}).

\bibitem{casolaProbingCondensedMatter2018}
\bibinfo{author}{Casola, F.}, \bibinfo{author}{{van der Sar}, T.} \& \bibinfo{author}{Yacoby, A.}
\newblock \bibinfo{title}{Probing condensed matter physics with magnetometry based on nitrogen-vacancy centres in diamond}.
\newblock \emph{\bibinfo{journal}{Nature Reviews Materials}} \textbf{\bibinfo{volume}{3}}, \bibinfo{pages}{17088} (\bibinfo{year}{2018}).

\bibitem{jenkins2022imaging}
\bibinfo{author}{Jenkins, A.} \emph{et~al.}
\newblock \bibinfo{title}{Imaging the breakdown of ohmic transport in graphene}.
\newblock \emph{\bibinfo{journal}{Physical Review Letters}} \textbf{\bibinfo{volume}{129}}, \bibinfo{pages}{087701} (\bibinfo{year}{2022}).

\bibitem{aslamQuantumSensorsBiomedical2023}
\bibinfo{author}{Aslam, N.} \emph{et~al.}
\newblock \bibinfo{title}{Quantum sensors for biomedical applications}.
\newblock \emph{\bibinfo{journal}{Nature Reviews Physics}} \textbf{\bibinfo{volume}{5}}, \bibinfo{pages}{157--169} (\bibinfo{year}{2023}).

\bibitem{furusawa1998unconditional}
\bibinfo{author}{Furusawa, A.} \emph{et~al.}
\newblock \bibinfo{title}{Unconditional quantum teleportation}.
\newblock \emph{\bibinfo{journal}{science}} \textbf{\bibinfo{volume}{282}}, \bibinfo{pages}{706--709} (\bibinfo{year}{1998}).

\bibitem{braunsteinQuantumInformationContinuous2005}
\bibinfo{author}{Braunstein, S.~L.} \& \bibinfo{author}{{van Loock}, P.}
\newblock \bibinfo{title}{Quantum information with continuous variables}.
\newblock \emph{\bibinfo{journal}{Reviews of Modern Physics}} \textbf{\bibinfo{volume}{77}}, \bibinfo{pages}{513--577} (\bibinfo{year}{2005}).

\bibitem{pezzeQuantumMetrologyNonclassical2018}
\bibinfo{author}{Pezz{\`e}, L.}, \bibinfo{author}{Smerzi, A.}, \bibinfo{author}{Oberthaler, M.~K.}, \bibinfo{author}{Schmied, R.} \& \bibinfo{author}{Treutlein, P.}
\newblock \bibinfo{title}{Quantum metrology with nonclassical states of atomic ensembles}.
\newblock \emph{\bibinfo{journal}{Reviews of Modern Physics}} \textbf{\bibinfo{volume}{90}}, \bibinfo{pages}{035005} (\bibinfo{year}{2018}).

\bibitem{maliaDistributedQuantumSensing2022}
\bibinfo{author}{Malia, B.~K.}, \bibinfo{author}{Wu, Y.}, \bibinfo{author}{{Mart{\'i}nez-Rinc{\'o}n}, J.} \& \bibinfo{author}{Kasevich, M.~A.}
\newblock \bibinfo{title}{Distributed quantum sensing with mode-entangled spin-squeezed atomic states}.
\newblock \emph{\bibinfo{journal}{Nature}} \textbf{\bibinfo{volume}{612}}, \bibinfo{pages}{661--665} (\bibinfo{year}{2022}).

\bibitem{hughesStronglyInteractingTwodimensional2024}
\bibinfo{author}{Hughes, L.~B.} \emph{et~al.}
\newblock \bibinfo{title}{A strongly interacting, two-dimensional, dipolar spin ensemble in (111)-oriented diamond} (\bibinfo{year}{2024}).
\newblock \eprint{2404.10075}.

\bibitem{braverman2019near}
\bibinfo{author}{Braverman, B.} \emph{et~al.}
\newblock \bibinfo{title}{Near-unitary spin squeezing in yb 171}.
\newblock \emph{\bibinfo{journal}{Physical review letters}} \textbf{\bibinfo{volume}{122}}, \bibinfo{pages}{223203} (\bibinfo{year}{2019}).

\bibitem{cappellaro2009quantum}
\bibinfo{author}{Cappellaro, P.} \& \bibinfo{author}{Lukin, M.~D.}
\newblock \bibinfo{title}{Quantum correlation in disordered spin systems: Applications to magnetic sensing}.
\newblock \emph{\bibinfo{journal}{Physical Review A—Atomic, Molecular, and Optical Physics}} \textbf{\bibinfo{volume}{80}}, \bibinfo{pages}{032311} (\bibinfo{year}{2009}).

\bibitem{bennett2013phonon}
\bibinfo{author}{Bennett, S.} \emph{et~al.}
\newblock \bibinfo{title}{Phonon-induced spin-spin interactions in diamond nanostructures: Application to spin squeezing}.
\newblock \emph{\bibinfo{journal}{Physical Review Letters}} \textbf{\bibinfo{volume}{110}}, \bibinfo{pages}{156402} (\bibinfo{year}{2013}).

\bibitem{perlinSpinSqueezingShortRange2020}
\bibinfo{author}{Perlin, M.~A.}, \bibinfo{author}{Qu, C.} \& \bibinfo{author}{Rey, A.~M.}
\newblock \bibinfo{title}{Spin {{Squeezing}} with {{Short-Range Spin-Exchange Interactions}}}.
\newblock \emph{\bibinfo{journal}{Physical Review Letters}} \textbf{\bibinfo{volume}{125}}, \bibinfo{pages}{223401} (\bibinfo{year}{2020}).

\bibitem{comparinScalableSpinSqueezing2022}
\bibinfo{author}{Comparin, T.}, \bibinfo{author}{Mezzacapo, F.}, \bibinfo{author}{{Robert-de-Saint-Vincent}, M.} \& \bibinfo{author}{Roscilde, T.}
\newblock \bibinfo{title}{Scalable {{Spin Squeezing}} from {{Spontaneous Breaking}} of a {{Continuous Symmetry}}}.
\newblock \emph{\bibinfo{journal}{Physical Review Letters}} \textbf{\bibinfo{volume}{129}}, \bibinfo{pages}{113201} (\bibinfo{year}{2022}).

\bibitem{blockScalableSpinSqueezing2024}
\bibinfo{author}{Block, M.} \emph{et~al.}
\newblock \bibinfo{title}{Scalable spin squeezing from finite-temperature easy-plane magnetism}.
\newblock \emph{\bibinfo{journal}{Nature Physics}} \textbf{\bibinfo{volume}{20}}, \bibinfo{pages}{1575--1581} (\bibinfo{year}{2024}).

\bibitem{barrySensitivityOptimizationNVdiamond2020}
\bibinfo{author}{Barry, J.~F.} \emph{et~al.}
\newblock \bibinfo{title}{Sensitivity optimization for {{NV-diamond}} magnetometry}.
\newblock \emph{\bibinfo{journal}{Reviews of Modern Physics}} \textbf{\bibinfo{volume}{92}}, \bibinfo{pages}{015004} (\bibinfo{year}{2020}).

\bibitem{arunkumarQuantumLogicEnhanced2023}
\bibinfo{author}{Arunkumar, N.} \emph{et~al.}
\newblock \bibinfo{title}{Quantum {{Logic Enhanced Sensing}} in {{Solid-State Spin Ensembles}}}.
\newblock \emph{\bibinfo{journal}{Physical Review Letters}} \textbf{\bibinfo{volume}{131}}, \bibinfo{pages}{100801} (\bibinfo{year}{2023}).

\bibitem{dohertyNitrogenvacancyColourCentre2013}
\bibinfo{author}{Doherty, M.~W.} \emph{et~al.}
\newblock \bibinfo{title}{The nitrogen-vacancy colour centre in diamond}.
\newblock \emph{\bibinfo{journal}{Physics Reports}} \textbf{\bibinfo{volume}{528}}, \bibinfo{pages}{1--45} (\bibinfo{year}{2013}).

\bibitem{edwardsTheorySpinGlasses1975}
\bibinfo{author}{Edwards, S.~F.} \& \bibinfo{author}{Anderson, P.~W.}
\newblock \bibinfo{title}{Theory of spin glasses}.
\newblock \emph{\bibinfo{journal}{Journal of Physics F: Metal Physics}} \textbf{\bibinfo{volume}{5}}, \bibinfo{pages}{965} (\bibinfo{year}{1975}).

\bibitem{burin2015many}
\bibinfo{author}{Burin, A.~L.}
\newblock \bibinfo{title}{Many-body delocalization in a strongly disordered system with long-range interactions: Finite-size scaling}.
\newblock \emph{\bibinfo{journal}{Physical Review B}} \textbf{\bibinfo{volume}{91}}, \bibinfo{pages}{094202} (\bibinfo{year}{2015}).

\bibitem{abaninColloquiumManybodyLocalization2019}
\bibinfo{author}{Abanin, D.~A.}, \bibinfo{author}{Altman, E.}, \bibinfo{author}{Bloch, I.} \& \bibinfo{author}{Serbyn, M.}
\newblock \bibinfo{title}{Colloquium : {{Many-body}} localization, thermalization, and entanglement}.
\newblock \emph{\bibinfo{journal}{Reviews of Modern Physics}} \textbf{\bibinfo{volume}{91}}, \bibinfo{pages}{021001} (\bibinfo{year}{2019}).

\bibitem{martinControllingLocalThermalization2023}
\bibinfo{author}{Martin, L.~S.} \emph{et~al.}
\newblock \bibinfo{title}{Controlling {{Local Thermalization Dynamics}} in a {{Floquet-Engineered Dipolar Ensemble}}}.
\newblock \emph{\bibinfo{journal}{Physical Review Letters}} \textbf{\bibinfo{volume}{130}}, \bibinfo{pages}{210403} (\bibinfo{year}{2023}).

\bibitem{borishTransverseFieldIsingDynamics2020}
\bibinfo{author}{Borish, V.}, \bibinfo{author}{Markovi{\'c}, O.}, \bibinfo{author}{Hines, J.~A.}, \bibinfo{author}{Rajagopal, S.~V.} \& \bibinfo{author}{{Schleier-Smith}, M.}
\newblock \bibinfo{title}{Transverse-{{Field Ising Dynamics}} in a {{Rydberg-Dressed Atomic Gas}}}.
\newblock \emph{\bibinfo{journal}{Physical Review Letters}} \textbf{\bibinfo{volume}{124}}, \bibinfo{pages}{063601} (\bibinfo{year}{2020}).

\bibitem{sorensenEntanglementExtremeSpin2001}
\bibinfo{author}{S{\o}rensen, A.~S.} \& \bibinfo{author}{M{\o}lmer, K.}
\newblock \bibinfo{title}{Entanglement and {{Extreme Spin Squeezing}}}.
\newblock \emph{\bibinfo{journal}{Physical Review Letters}} \textbf{\bibinfo{volume}{86}}, \bibinfo{pages}{4431--4434} (\bibinfo{year}{2001}).

\bibitem{sorensenManyparticleEntanglementBose2001}
\bibinfo{author}{S{\o}rensen, A.}, \bibinfo{author}{Duan, L.-M.}, \bibinfo{author}{Cirac, J.~I.} \& \bibinfo{author}{Zoller, P.}
\newblock \bibinfo{title}{Many-particle entanglement with {{Bose}}--{{Einstein}} condensates}.
\newblock \emph{\bibinfo{journal}{Nature}} \textbf{\bibinfo{volume}{409}}, \bibinfo{pages}{63--66} (\bibinfo{year}{2001}).

\bibitem{degenQuantumSensing2017}
\bibinfo{author}{Degen, C.~L.}, \bibinfo{author}{Reinhard, F.} \& \bibinfo{author}{Cappellaro, P.}
\newblock \bibinfo{title}{Quantum sensing}.
\newblock \emph{\bibinfo{journal}{Reviews of Modern Physics}} \textbf{\bibinfo{volume}{89}}, \bibinfo{pages}{035002} (\bibinfo{year}{2017}).

\bibitem{jiangRepetitiveReadoutSingle2009}
\bibinfo{author}{Jiang, L.} \emph{et~al.}
\newblock \bibinfo{title}{Repetitive {{Readout}} of a {{Single Electronic Spin}} via {{Quantum Logic}} with {{Nuclear Spin Ancillae}}}.
\newblock \emph{\bibinfo{journal}{Science}} \textbf{\bibinfo{volume}{326}}, \bibinfo{pages}{267--272} (\bibinfo{year}{2009}).

\bibitem{robledoHighfidelityProjectiveReadout2011}
\bibinfo{author}{Robledo, L.} \emph{et~al.}
\newblock \bibinfo{title}{High-fidelity projective read-out of a solid-state spin quantum register}.
\newblock \emph{\bibinfo{journal}{Nature}} \textbf{\bibinfo{volume}{477}}, \bibinfo{pages}{574--578} (\bibinfo{year}{2011}).

\bibitem{shieldsEfficientReadoutSingle2015}
\bibinfo{author}{Shields, B.~J.}, \bibinfo{author}{Unterreithmeier, Q.~P.}, \bibinfo{author}{{de Leon}, N.~P.}, \bibinfo{author}{Park, H.} \& \bibinfo{author}{Lukin, M.~D.}
\newblock \bibinfo{title}{Efficient {{Readout}} of a {{Single Spin State}} in {{Diamond}} via {{Spin-to-Charge Conversion}}}.
\newblock \emph{\bibinfo{journal}{Physical Review Letters}} \textbf{\bibinfo{volume}{114}}, \bibinfo{pages}{136402} (\bibinfo{year}{2015}).

\bibitem{schachenmayerManyBodyQuantumSpin2015}
\bibinfo{author}{Schachenmayer, J.}, \bibinfo{author}{Pikovski, A.} \& \bibinfo{author}{Rey, A.~M.}
\newblock \bibinfo{title}{Many-{{Body Quantum Spin Dynamics}} with {{Monte Carlo Trajectories}} on a {{Discrete Phase Space}}}.
\newblock \emph{\bibinfo{journal}{Physical Review X}} \textbf{\bibinfo{volume}{5}}, \bibinfo{pages}{011022} (\bibinfo{year}{2015}).

\bibitem{braemerClusterTruncatedWigner2024}
\bibinfo{author}{Braemer, A.}, \bibinfo{author}{Vahedi, J.} \& \bibinfo{author}{G{\"a}rttner, M.}
\newblock \bibinfo{title}{Cluster truncated {{Wigner}} approximation for bond-disordered {{Heisenberg}} spin models}.
\newblock \emph{\bibinfo{journal}{Physical Review B}} \textbf{\bibinfo{volume}{110}}, \bibinfo{pages}{054204} (\bibinfo{year}{2024}).
\newblock \eprint{2407.01682}.

\bibitem{chomazDipolarPhysicsReview2022a}
\bibinfo{author}{Chomaz, L.} \emph{et~al.}
\newblock \bibinfo{title}{Dipolar physics: A review of experiments with magnetic quantum gases}.
\newblock \emph{\bibinfo{journal}{Reports on Progress in Physics}} \textbf{\bibinfo{volume}{86}}, \bibinfo{pages}{026401} (\bibinfo{year}{2022}).

\bibitem{cornishQuantumComputationQuantum2024}
\bibinfo{author}{Cornish, S.~L.}, \bibinfo{author}{Tarbutt, M.~R.} \& \bibinfo{author}{Hazzard, K. R.~A.}
\newblock \bibinfo{title}{Quantum computation and quantum simulation with ultracold molecules}.
\newblock \emph{\bibinfo{journal}{Nature Physics}} \textbf{\bibinfo{volume}{20}}, \bibinfo{pages}{730--740} (\bibinfo{year}{2024}).

\bibitem{davisProbingManybodyDynamics2023}
\bibinfo{author}{Davis, E.~J.} \emph{et~al.}
\newblock \bibinfo{title}{Probing many-body dynamics in a two-dimensional dipolar spin ensemble}.
\newblock \emph{\bibinfo{journal}{Nature Physics}} \textbf{\bibinfo{volume}{19}}, \bibinfo{pages}{836--844} (\bibinfo{year}{2023}).

\bibitem{chenLaserWritingIndividual2019}
\bibinfo{author}{Chen, Y.-C.} \emph{et~al.}
\newblock \bibinfo{title}{Laser writing of individual nitrogen-vacancy defects in diamond with near-unity yield}.
\newblock \emph{\bibinfo{journal}{Optica}} \textbf{\bibinfo{volume}{6}}, \bibinfo{pages}{662--667} (\bibinfo{year}{2019}).

\bibitem{groot-berningFabrication15NVCenters2021}
\bibinfo{author}{{Groot-Berning}, K.}, \bibinfo{author}{Jacob, G.}, \bibinfo{author}{Osterkamp, C.}, \bibinfo{author}{Jelezko, F.} \& \bibinfo{author}{{Schmidt-Kaler}, F.}
\newblock \bibinfo{title}{Fabrication of {{15NV}}- centers in diamond using a deterministic single ion implanter}.
\newblock \emph{\bibinfo{journal}{New Journal of Physics}} \textbf{\bibinfo{volume}{23}}, \bibinfo{pages}{063067} (\bibinfo{year}{2021}).

\bibitem{carrollObservationGeneralizedTJ2024}
\bibinfo{author}{Carroll, A.~N.} \emph{et~al.}
\newblock \bibinfo{title}{Observation of {{Generalized}} t-{{J Spin Dynamics}} with {{Tunable Dipolar Interactions}}} (\bibinfo{year}{2024}).
\newblock \eprint{2404.18916}.

\bibitem{signoles2021glassy}
\bibinfo{author}{Signoles, A.} \emph{et~al.}
\newblock \bibinfo{title}{Glassy dynamics in a disordered heisenberg quantum spin system}.
\newblock \emph{\bibinfo{journal}{Physical Review X}} \textbf{\bibinfo{volume}{11}}, \bibinfo{pages}{011011} (\bibinfo{year}{2021}).

\end{thebibliography}


\begin{thebibliography}{1}
\expandafter\ifx\csname url\endcsname\relax
  \def\url#1{\texttt{#1}}\fi
\expandafter\ifx\csname urlprefix\endcsname\relax\def\urlprefix{URL }\fi
\providecommand{\bibinfo}[2]{#2}
\providecommand{\eprint}[2][]{\url{#2}}

\bibitem{hughesStronglyInteractingTwodimensional2024}
\bibinfo{author}{Hughes, L.~B.} \emph{et~al.}
\newblock \bibinfo{title}{A strongly interacting, two-dimensional, dipolar spin ensemble in (111)-oriented diamond} (\bibinfo{year}{2024}).
\newblock \eprint{2404.10075}.

\bibitem{hughesTwodimensionalSpinSystems2023}
\bibinfo{author}{Hughes, L.~B.} \emph{et~al.}
\newblock \bibinfo{title}{Two-dimensional spin systems in {{PECVD-grown}} diamond with tunable density and long coherence for enhanced quantum sensing and simulation}.
\newblock \emph{\bibinfo{journal}{APL Materials}} \textbf{\bibinfo{volume}{11}}, \bibinfo{pages}{021101} (\bibinfo{year}{2023}).

\bibitem{degenQuantumSensing2017}
\bibinfo{author}{Degen, C.~L.}, \bibinfo{author}{Reinhard, F.} \& \bibinfo{author}{Cappellaro, P.}
\newblock \bibinfo{title}{Quantum sensing}.
\newblock \emph{\bibinfo{journal}{Reviews of Modern Physics}} \textbf{\bibinfo{volume}{89}}, \bibinfo{pages}{035002} (\bibinfo{year}{2017}).

\bibitem{schachenmayerManyBodyQuantumSpin2015}
\bibinfo{author}{Schachenmayer, J.}, \bibinfo{author}{Pikovski, A.} \& \bibinfo{author}{Rey, A.~M.}
\newblock \bibinfo{title}{Many-{{Body Quantum Spin Dynamics}} with {{Monte Carlo Trajectories}} on a {{Discrete Phase Space}}}.
\newblock \emph{\bibinfo{journal}{Physical Review X}} \textbf{\bibinfo{volume}{5}}, \bibinfo{pages}{011022} (\bibinfo{year}{2015}).

\bibitem{braemerClusterTruncatedWigner2024}
\bibinfo{author}{Braemer, A.}, \bibinfo{author}{Vahedi, J.} \& \bibinfo{author}{G{\"a}rttner, M.}
\newblock \bibinfo{title}{Cluster truncated {{Wigner}} approximation for bond-disordered {{Heisenberg}} spin models}.
\newblock \emph{\bibinfo{journal}{Physical Review B}} \textbf{\bibinfo{volume}{110}}, \bibinfo{pages}{054204} (\bibinfo{year}{2024}).
\newblock \eprint{2407.01682}.

\bibitem{davisProbingManybodyDynamics2023}
\bibinfo{author}{Davis, E.~J.} \emph{et~al.}
\newblock \bibinfo{title}{Probing many-body dynamics in a two-dimensional dipolar spin ensemble}.
\newblock \emph{\bibinfo{journal}{Nature Physics}} \textbf{\bibinfo{volume}{19}}, \bibinfo{pages}{836--844} (\bibinfo{year}{2023}).

\end{thebibliography}


\begin{thebibliography}{10}
\expandafter\ifx\csname url\endcsname\relax
  \def\url#1{\texttt{#1}}\fi
\expandafter\ifx\csname urlprefix\endcsname\relax\def\urlprefix{URL }\fi
\providecommand{\bibinfo}[2]{#2}
\providecommand{\eprint}[2][]{\url{#2}}

\bibitem{zuEmergentHydrodynamicsStrongly2021}
\bibinfo{author}{Zu, C.} \emph{et~al.}
\newblock \bibinfo{title}{Emergent hydrodynamics in a strongly interacting dipolar spin ensemble}.
\newblock \emph{\bibinfo{journal}{Nature}} \textbf{\bibinfo{volume}{597}}, \bibinfo{pages}{45--50} (\bibinfo{year}{2021}).

\bibitem{Krylov}
\bibinfo{author}{Liesen, J.} \& \bibinfo{author}{Strakos, Z.}
\newblock \emph{\bibinfo{title}{Krylov Subspace Methods: Principles and Analysis}} (\bibinfo{publisher}{Oxford University Press}, \bibinfo{year}{2012}).
\newblock \urlprefix\url{https://doi.org/10.1093/acprof:oso/9780199655410.001.0001}.

\bibitem{meyerKrylov2024}
\bibinfo{author}{Kahanamoku-Meyer, G.~D.} \& \bibinfo{author}{Wei, J.}
\newblock \bibinfo{title}{Gregdmeyer/dynamite: v0.4.0} (\bibinfo{year}{2024}).
\newblock \urlprefix\url{https://doi.org/10.5281/zenodo.10906046}.

\bibitem{schachenmayerManyBodyQuantumSpin2015}
\bibinfo{author}{Schachenmayer, J.}, \bibinfo{author}{Pikovski, A.} \& \bibinfo{author}{Rey, A.~M.}
\newblock \bibinfo{title}{Many-{{Body Quantum Spin Dynamics}} with {{Monte Carlo Trajectories}} on a {{Discrete Phase Space}}}.
\newblock \emph{\bibinfo{journal}{Physical Review X}} \textbf{\bibinfo{volume}{5}}, \bibinfo{pages}{011022} (\bibinfo{year}{2015}).

\bibitem{blockScalableSpinSqueezing2024}
\bibinfo{author}{Block, M.} \emph{et~al.}
\newblock \bibinfo{title}{Scalable spin squeezing from finite-temperature easy-plane magnetism}.
\newblock \emph{\bibinfo{journal}{Nature Physics}} \textbf{\bibinfo{volume}{20}}, \bibinfo{pages}{1575--1581} (\bibinfo{year}{2024}).

\bibitem{alaouiMeasuringBipartiteSpin2024}
\bibinfo{author}{Alaoui, Y.~A.} \emph{et~al.}
\newblock \bibinfo{title}{Measuring bipartite spin correlations of lattice-trapped dipolar atoms} (\bibinfo{year}{2024}).
\newblock \eprint{2404.10531}.

\bibitem{braemerClusterTruncatedWigner2024}
\bibinfo{author}{Braemer, A.}, \bibinfo{author}{Vahedi, J.} \& \bibinfo{author}{G{\"a}rttner, M.}
\newblock \bibinfo{title}{Cluster truncated {{Wigner}} approximation for bond-disordered {{Heisenberg}} spin models}.
\newblock \emph{\bibinfo{journal}{Physical Review B}} \textbf{\bibinfo{volume}{110}}, \bibinfo{pages}{054204} (\bibinfo{year}{2024}).
\newblock \eprint{2407.01682}.

\bibitem{nagaoTwodimensionalCorrelationPropagation2024}
\bibinfo{author}{Nagao, K.} \& \bibinfo{author}{Yunoki, S.}
\newblock \bibinfo{title}{Two-dimensional correlation propagation dynamics with a cluster discrete phase-space method} (\bibinfo{year}{2024}).
\newblock \eprint{2404.18594}.

\bibitem{carleoSolvingQuantumManybody2017}
\bibinfo{author}{Carleo, G.} \& \bibinfo{author}{Troyer, M.}
\newblock \bibinfo{title}{Solving the quantum many-body problem with artificial neural networks}.
\newblock \emph{\bibinfo{journal}{Science}} \textbf{\bibinfo{volume}{355}}, \bibinfo{pages}{602--606} (\bibinfo{year}{2017}).

\bibitem{haegeman2011time}
\bibinfo{author}{Haegeman, J.} \emph{et~al.}
\newblock \bibinfo{title}{Time-dependent variational principle for quantum lattices}.
\newblock \emph{\bibinfo{journal}{Physical Review Letters}} \textbf{\bibinfo{volume}{107}}, \bibinfo{pages}{070601} (\bibinfo{year}{2011}).

\bibitem{schmitt2020quantum}
\bibinfo{author}{Schmitt, M.} \& \bibinfo{author}{Heyl, M.}
\newblock \bibinfo{title}{Quantum many-body dynamics in two dimensions with artificial neural networks}.
\newblock \emph{\bibinfo{journal}{Physical Review Letters}} \textbf{\bibinfo{volume}{125}}, \bibinfo{pages}{100503} (\bibinfo{year}{2020}).

\bibitem{schmitt2022quantum}
\bibinfo{author}{Schmitt, M.}, \bibinfo{author}{Rams, M.~M.}, \bibinfo{author}{Dziarmaga, J.}, \bibinfo{author}{Heyl, M.} \& \bibinfo{author}{Zurek, W.~H.}
\newblock \bibinfo{title}{Quantum phase transition dynamics in the two-dimensional transverse-field ising model}.
\newblock \emph{\bibinfo{journal}{Science Advances}} \textbf{\bibinfo{volume}{8}}, \bibinfo{pages}{eabl6850} (\bibinfo{year}{2022}).

\bibitem{sinibaldi2023unbiasing}
\bibinfo{author}{Sinibaldi, A.}, \bibinfo{author}{Giuliani, C.}, \bibinfo{author}{Carleo, G.} \& \bibinfo{author}{Vicentini, F.}
\newblock \bibinfo{title}{Unbiasing time-dependent variational monte carlo by projected quantum evolution}.
\newblock \emph{\bibinfo{journal}{Quantum}} \textbf{\bibinfo{volume}{7}}, \bibinfo{pages}{1131} (\bibinfo{year}{2023}).

\bibitem{vicentini2022netket}
\bibinfo{author}{Vicentini, F.} \emph{et~al.}
\newblock \bibinfo{title}{Netket 3: Machine learning toolbox for many-body quantum systems}.
\newblock \emph{\bibinfo{journal}{SciPost Physics Codebases}} \bibinfo{pages}{007} (\bibinfo{year}{2022}).

\bibitem{jax2018github}
\bibinfo{author}{Bradbury, J.} \emph{et~al.}
\newblock \bibinfo{title}{{JAX}: composable transformations of {P}ython+{N}um{P}y programs} (\bibinfo{year}{2018}).
\newblock \urlprefix\url{http://github.com/google/jax}.

\bibitem{davisProbingManybodyDynamics2023}
\bibinfo{author}{Davis, E.~J.} \emph{et~al.}
\newblock \bibinfo{title}{Probing many-body dynamics in a two-dimensional dipolar spin ensemble}.
\newblock \emph{\bibinfo{journal}{Nature Physics}} \textbf{\bibinfo{volume}{19}}, \bibinfo{pages}{836--844} (\bibinfo{year}{2023}).

\end{thebibliography}

\end{document}

\title{Supplementary Information}
% \date{today}

\maketitle

\tableofcontents

\section{Derivation of Hamiltonian}

Nitrogen vacancy (NV) centers and substitutional nitrogen (P1) centers are the two primary spin defects in our diamond sample. The full Hamiltonian for all spins is given by
\begin{equation}
\begin{aligned}
    H &= \sum_j \Delta s_{z,j}^2 + \gamma_\mathrm{NV} \mathbf{s}_j \cdot \mathbf{B} 
     + A_{z,j} s_{z,j} i_{z,j} + A_\perp (s_{x,j}i_{x,j} + s_{y,j}i_{y,j})\\ &+\sum_k \gamma_\mathrm{P1} \mathbf{p}_k \cdot \mathbf{B} + H^\mathrm{HF}_\mathrm{P1} \\ 
    & + H_\mathrm{NV-NV} + H_\mathrm{NV-P1} + H_\mathrm{P1-P1} .     \label{eq:sys_H}
\end{aligned}
\end{equation}
Here, $\mathbf{s}$ is the NV electronic spin-1 operator, $\mathbf{i}$ is the nuclear spin-1/2 operator for the $^{15}$N isotope, and $\mathbf{p}$ is the electronic spin-1/2 P1 operator.
The gyromagnetic ratio of the electronic spin $\gamma_\mathrm{NV} \approx \gamma_\mathrm{P1} = (2\pi)\times2.8~$MHz/G sets interactions with external magnetic fields $\mathbf{B}$, and the zero-field splitting is $\Delta = (2\pi)\times 2.87~$GHz.
The electronic spin also interactions with the $^{15}$N nuclear spin via a hyperfine coupling with components $A_z = (2\pi)\times 3.03~$MHz and $A_\perp = (2\pi)\times 3.65~$MHz. The hyperfine interaction $H^\mathrm{HF}_\mathrm{P1}$ splits the P1 centers into four groups with different transition frequencies~\cite{zuEmergentHydrodynamicsStrongly2021}.
Finally, NV and P1 centers interact with each other $H_\text{NV-NV},~ H_\text{NV-P1},~H_\text{P1-P1}$ via magnetic dipole-dipole interactions. We are primarily interested in $H_\text{NV-NV}$ and $H_\text{NV-P1}$, whose derivation follows.

\subsection{NV-NV interaction $H_\mathrm{NV-NV}$}
The native interaction between NV centers is the magnetic dipole-dipole interaction, 
\begin{equation}
H_{\mathrm{NV-NV}}  = -\sum_{i,j}\frac{J_0}{r_{ij}^3}[3(\mathbf{s}^i\cdot\hat{\mathbf{r}}_{ij})(\mathbf{s}^j\cdot\hat{\mathbf{r}}_{ij})-\mathbf{s}^i\cdot \mathbf{s}^j],
\end{equation}
where $J_0 = 2\pi\times\SI{52}{\mega\hertz\cdot \nano\meter^3}$ is the interaction strength, $\hat{\mathbf{r}}_{ij}$ is the vector connecting spins $i$ and $j$. In the (111)-oriented diamond, we denote the out-of-the-plane direction as $\hat{\mathbf{z}}$. The vector $\hat{\mathbf{r}}_{ij}$ is perpendicular to $\hat{\mathbf{z}}$ and is parameterized by $(\cos\theta, \sin\theta, 0)$, where $\theta$ is the azimuthal angle between two spins. For the NV group quantized along $\hat{\mathbf{z}}$, the dipolar interaction can then be explicitly written as 
\begin{equation}
\begin{aligned}
    H_{\mathrm{NV-NV}}  &= -\sum_{i,j}\frac{J_0}{r_{ij}^3}[3(s^i_x\cos\theta + s^i_y\sin\theta)(s^j_x\cos\theta + s^j_y\sin\theta)  -(s^i_xs^j_x+s^i_ys^j_y+s^i_zs^j_z)] \\ 
    &= -\sum_{i,j}\frac{J}{r_{ij}^3} [(3\cos^2\theta - 1)s^i_xs^j_x+(3\sin^2\theta - 1)s^i_ys^j_y -s^i_zs^j_z  +\cos\theta\sin\theta(s^i_xs^j_y+s^i_ys^j_x)].
\end{aligned}
\end{equation}
In the rotating frame, only the secular terms survive, i.e., $s^i_xs^j_x + s^i_ys^j_y$ (flip-flop) and $s^i_zs^j_z$ (Ising). We drop the non-secular flip-flip terms ($s^i_xs^j_x - s^i_ys^j_y$) and cross-terms $(s^i_xs^j_y+s^i_ys^j_x)$. The dipolar Hamiltonian is reduced to the simple XXZ form for spin-1 operators
\begin{equation}
    H_{\mathrm{NV-NV}}  = -\sum_{i,j}\frac{J_0}{r_{ij}^3} [(s^i_xs^j_x+s^i_ys^j_y)/2 -s^i_zs^j_z].
\end{equation}
In the experiment, we work in an effective two-level subspace $\{\ket{m_s = 0}, \ket{m_s = -1}\}$ of the full spin-1 ground state manifold. It is therefore convenient to further replace the spin-1 operators $\{s_x, s_y, s_z\}$ with spin-1/2 operators $\{\tilde{s}_x, \tilde{s}_y, \tilde{s}_z\}$, i.e., 
\begin{equation}
    s_x \rightarrow \sqrt{2}\tilde{s}_x, \ \ \ s_y \rightarrow \sqrt{2}\tilde{s}_y, \ \ \ s_z \rightarrow \tilde{s}_z + \frac{I}{2} .
\end{equation}
In this case, the dipolar Hamiltonian can be rewritten as 
\begin{equation}
    H_{\mathrm{NV-NV}}  = -\sum_{i,j}\frac{J_0}{r_{ij}^3} [(\tilde{s}_x^i\tilde{s}_x^j + \tilde{s}_y^i\tilde{s}_y^j) - \tilde{s}_z^i\tilde{s}_z^j - \tilde{s}_z^i - \tilde{s}_z^j].
\end{equation}
The single-body terms $\tilde{s}_z$ can be treated as effective onsite disorder with characteristic strength $J_0$. Typically, the disorder emerging from the spin-1 to spin-1/2 reduction can be decoupled by e.g. XY-8 pulses, which are anyway needed to decouple  the disorder arising from interactions with background spins $H_\text{NV-P1}$. Finally, we arrive at the effective XXZ Hamiltonian for NV-NV interaction in the spin-1/2 subspace:
\begin{equation}
    H_{\mathrm{XXZ}} = -\sum_{i,j}\frac{J_0}{r_{ij}^3} [(\tilde{s}_x^i\tilde{s}_x^j + \tilde{s}_y^i\tilde{s}_y^j) - \tilde{s}_z^i\tilde{s}_z^j].
\end{equation}
From now on, we drop the tilde on the effective spin-1/2 operator for simplicity.

\subsection{NV-P1 interaction $H_\mathrm{NV-P1}$}
NV and P1 centers similarly interact through the magnetic dipole-dipole interaction,
\begin{equation}
H_{\mathrm{NV-P1}}  = -\sum_{i,j}\frac{J_0}{r_{ij}^3}[3(\mathbf{s}^i\cdot\hat{\mathbf{r}}_{ij})(\mathbf{p}^j\cdot\hat{\mathbf{r}}_{ij})-\mathbf{s}^i\cdot \mathbf{p}^j].
\end{equation}
Because of the zero-field splitting $\Delta$ of the NV center, spin-exchange interactions between NV and P1 centers are highly off-resonant and naturally drop out in the rotating frame. The only surviving term is the Ising coupling that conserves the total energy, i.e.,
\begin{equation}
H_{\mathrm{NV-P1}}  = \sum_{i,j}\frac{J_0}{r_{ij}^3}s^i_z p^j_z.
\end{equation}
In the experiment, we employ an XY-8 pulse sequence to decouple these Ising interactions. With NV centers mostly polarized in the $m_I = -1/2$ nuclear spin subgroup [Methods, Sec.~I.3], the only relevant term in Eq.~\ref{eq:sys_H} is $H_{\mathrm{XXZ}}$.

\section{Numerical simulations}

\subsection{Numerical methods}

To simulate the dynamics of the disordered, interacting NV centers, we use the discrete cluster TWA and neural quantum state (NQS) numerical methods, which are benchmarked against Krylov subspace methods for small system sizes.

\subsubsection{Krylov subspace method}
The Krylov method solves the Schr\"odinger equation in a so-called Krylov subspace with reduced dimensionality~\cite{Krylov}. Instead of directly diagonalizing the full Hamiltonian, the method constructs a series of smaller, more manageable Krylov subspaces. These subspaces are generated by repeatedly applying the Hamiltonian to an initial state vector, effectively capturing the most significant components of the system dynamics within a reduced basis. By projecting the Hamiltonian onto this Krylov subspace, one can compute approximate eigenvalues and eigenvectors, enabling the efficient calculation of the time evolution operator. This approach allows for the efficient simulation of quantum dynamics for systems that are otherwise too large for traditional exact diagonalization methods. In our work, we employ the Krylov subspace method using the Dynamite package \cite{meyerKrylov2024} to simulate dynamics of the NV spins. Although this method can compute the exact time evolution of the spins, it is limited to relatively small system sizes  $N\leq30$ due to memory limitations ($\sim$ 2000 GB per cluster node).

In this work, we primarily perform Krylov on a $N=20$ random spin ensemble, balancing the computational cost (runtime and memory usage) with the ability to capture the essential features of the system's time evolution. For simulations on the systems with positional disorder, we randomly sample the spatial disorder realizations and execute the calculations in parallel on the computing cluster.

\subsubsection{discrete Wigner function approximation methods}
The discrete truncated Wigner approximation (DTWA)~\cite{schachenmayerManyBodyQuantumSpin2015} is a method that approximately simulates quantum dynamics using an ensemble of classical Monte Carlo trajectories. In this approach, each trajectory involves the evolution of classical spins under the same Hamiltonian, with initial states randomly sampled from a discrete Wigner distribution corresponding to the initial quantum state. This numerical method captures the time evolution of one- and two-point correlators at the mean-field level, which is crucial for simulating the squeezing dynamics. Compared to the Krylov method, DTWA can access much larger system sizes, e.g., up to $N\sim10^5$ \cite{blockScalableSpinSqueezing2024}. However, the original method is insufficient for quantitatively studying the dynamics of disordered spin systems because it captures quantum correlations only at the mean-field level. Consequently, it fails to accurately simulate the fast dynamics between closely spaced, strongly interacting spin pairs, i.e., dimers.

To overcome this limitation, we apply a cluster generalization of the original method \cite{alaouiMeasuringBipartiteSpin2024, braemerClusterTruncatedWigner2024, nagaoTwodimensionalCorrelationPropagation2024}, known as discrete cluster truncated Wigner approximation (cluster DTWA). In this method, spins within a dimer are grouped into a cluster, and the full quantum degrees of freedom of the cluster are represented using a classical vector whose state space dimension accounts for the combined Hilbert space of the spins in the cluster.  For a dimer consisting of two spin-$1/2$ particles, this corresponds to a four-dimensional Hilbert space with 15 independent operators, reflecting the generators of the SU(4) Lie algebra. Similar to the original method, the cluster method utilizes classical Monte Carlo trajectories, but it treats the interactions within each cluster exactly in a quantum mechanical way, while employing a mean-field approximation for the interactions between clusters. The cluster method provides a much better description of the fast dynamics within dimers and is, in general, a more accurate numerical method for disordered systems compared to the original method. Note that the performance of the cluster method depends on how clusters are chosen: the numerical result is the most accurate when the most strongly interacting spins are clustered together. Here, we apply the real space renormalization group (RSRG) inspired method proposed in \cite{braemerClusterTruncatedWigner2024}, which clusters the spins based on pairwise interaction strength. 

\subsubsection{Neural quantum states}
We also utilize neural quantum states (NQS) \cite{carleoSolvingQuantumManybody2017} to simulate the dynamics. This method employs neural networks to represent and study quantum many-body systems within the framework of the time-dependent variational principle (TDVP) \cite{haegeman2011time}. Coefficients $\psi_s$ in the computational basis of the quantum state $\vert \psi \rangle = \sum_s \psi_s \vert s\rangle$ are parametrized by a neural network $\psi_{\theta}: s \mapsto \psi_{\theta} (s)$. By leveraging the TDVP, the parameters of the neural network are dynamically updated to approximate the solution of the Schrödinger equation over time \cite{carleoSolvingQuantumManybody2017}. 
Given a neural network state  $|\psi(\theta) \rangle$  at time  $t$ , we update its parameters for time  $t + \delta t$  by applying a first-order expansion of the propagator and then projecting the resulting state back onto the neural network’s parameter manifold [Fig.~\ref{fig:num_benchmark}b]. This approach leads to an ODE of the form  $S \dot{\theta} = -i F$ , where  $F = -\nabla_{\theta} \langle H \rangle$  is the gradient vector and  $S$  is the quantum Fisher matrix~\cite{schmitt2020quantum, carleoSolvingQuantumManybody2017}. Both  $F$  and  $S$  are estimated using Monte Carlo Markov Chain techniques. To find the updated neural network state  $|\psi(\theta + \delta \theta^*) \rangle$  at time  $t + \delta t$ , we solve the ODE for $\theta (t)$ using numerical integration methods such as Euler or Runge-Kutta. This framework enables simulation of real-time dynamics, including the system's response to quenches and time-dependent perturbations.

We believe that NQS has several strengths that complement other numerical approaches mentioned above. On the one hand, as compared with (essentially) exact Krylov methods, NQS can scale to much larger system sizes. On the other hand, in contrast to semiclassical DTWA methods, NQS enables a much better controlled approximation, allowing it to more accurately capture the evolution of quantum correlations~\cite{schmitt2020quantum,schmitt2022quantum,sinibaldi2023unbiasing}. The degree of approximation within NQS is mainly set by neural network hyperparameters, size of the integration timestep, and number of samples used for evaluating equations of motion for neural network parameters. Furthermore, in the ideal limit, NQS dynamics preserves conserved quantities \cite{haegeman2011time}, providing an intrinsic benchmark for checking numerical stability when comparing to an exact solution is not possible.

To arrive at optimal NQS parameters, we have conducted ablation studies. For our simulations, we use the restricted Boltzmann machine (RBM) architecture~[Fig.~\ref{fig:num_benchmark}a] with complex parameters and tunable hidden layer density $\alpha_\text{RBM}$ ranging from $0.75$ to $4$. We choose a 4th-order Runge-Kutta method with $\delta t=5\cdot10^{-3}$ for the numerical integration of the TDVP equations. We ran all simulations with $N_\mathrm{samples}=2^{16}$ run on $2^{10}$ Monte Carlo Markov chains and a quantum Fisher matrix regularization of $\epsilon_\text{RBM}=10^{-4}$. To ensure efficient parallelization and runtime, our simulations were run on NVIDIA A100 and H100 GPUs with NetKet \cite{vicentini2022netket} and JAX \cite{jax2018github}.

\subsection{Benchmark between numerical methods}
\begin{figure}[htbp]
    \includegraphics[width=0.7\linewidth]{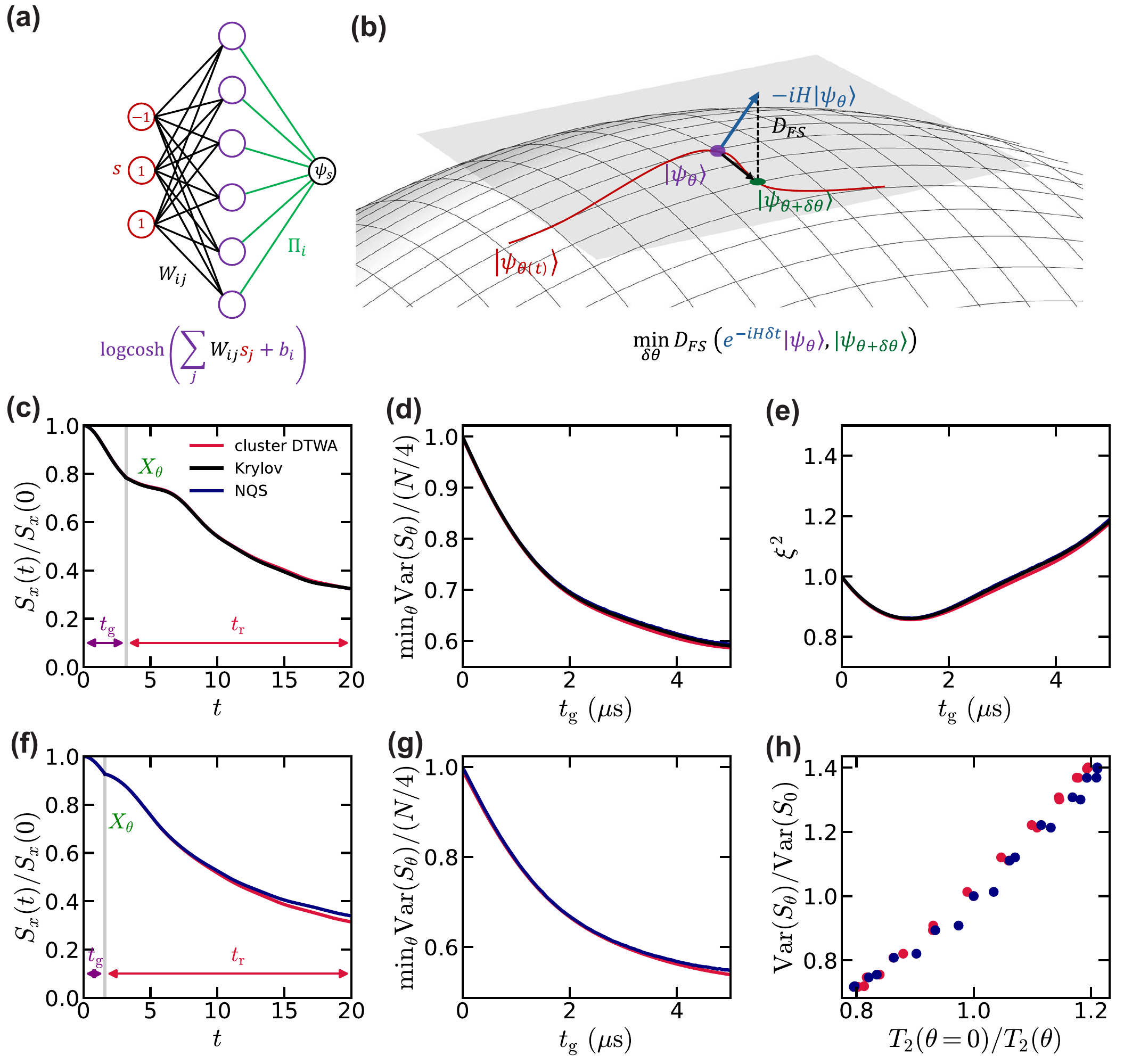}
    \caption{
    \textbf{Numerical benchmark between cluster DTWA, NQS and Krylov subspace method.}
    \textbf{(a)} Restricted Boltzmann machine neural network architecture used for NQS simulations. 
    \textbf{(b)} NQS enables the simulation of time dynamics within the TDVP framework by identifying the nearest quantum state on the NQS manifold to the Hamiltonian-evolved state, using the Fubini-Study metric $D_{FS}$ \cite{carleoSolvingQuantumManybody2017}. When expanded to first-order in $\mathcal{O(\delta \theta)}$ and $\mathcal{O}(\delta t)$, this approach results in an ODE governing the parameter evolution.
    \textbf{(c, d, e)} Numerical simulation of a small $N=20$ system with reduced disorder $r_\mathrm{shelve} = 7$ nmn and perfect initial state polarization. We simulate the $S_x(t)$ decay profile in the generation quench (up to $t_\mathrm{g} = 3.2\ \mu\mathrm{s}$) and the readout quench (up to $t_\mathrm{r} = 16.8\ \mu\mathrm{s}$) in (a). We apply a $X_\theta$ rotation ($\theta = \pi/4$) between the generation and detection quench to align certain collective spin operator $S_\theta$ into $S_z$. We simulate the time evolution of the minimal variance and squeezing parameter up to $t_\mathrm{g} = 5\ \mu\mathrm{s}$ in (d) and (e), respectively.
    \textbf{(f, g, h)} Numerical simulation of a $N = 80$ system with reduced disorder $r_\mathrm{shelve} = 7$ nm and perfect initial state polarization. We simulate the $S_x(t)$ decay profile in the generation quench (up to $t_\mathrm{g} = 1.6\ \mu\mathrm{s}$) and the readout quench (up to $t_\mathrm{r} = 18.4\ \mu\mathrm{s}$) in (a). We simulate the time evolution of the minimal variance in (g). Based on the $S_x(t_\mathrm{r})$ decay profile in the readout quench, we shows the mapping extracted in (g) following the procedure detailed in the Method. 
    }
    \label{fig:num_benchmark}
\end{figure}

We carefully benchmark the different numerical algorithms. Starting with a small system size $N = 20$, cluster DTWA and NQS are benchmarked by the Krylov subspace method, which is known to be relatively accurate. As shown in Fig.~\ref{fig:num_benchmark}(d-e), all of the three numerical methods exhibit quantitative agreement among the collective spin length, minimal variance and the squeezing parameter, indicating that both cluster DTWA and NQS capture the system dynamics to a high accuracy. 

To further benchmark the performance of cluster DTWA compared to NQS, we enlarge our system size to $N = 80$, which is used extensively in all other numerical simulations in this work. Fig.~\ref{fig:num_benchmark}(f-g), again, shows a quantitative agreement between cluster DTWA and NQS for both the collective spin length and the minimal variance. Based on the numerical simulation of the readout quench profile $S_x(t_\mathrm{r})$ in Fig.~\ref{fig:num_benchmark}f, we extract the mapping between the $T_2$ decay timescale and the variance $\mathrm{Var}(S_\theta)$ with both numerical methods [Fig.~\ref{fig:num_benchmark}h], which shows a reasonably good  agreement. By benchmarking cluster DTWA and NQS at a system size beyond the computational ability of any exact numerical methods and reaching a high consistency, we conclude that both cluster DTWA and NQS accurately simulate the time evolution for a positionally disordered system under $H_\mathrm{XXZ}$.

\section{Probing quantum variance via quench dynamics} \label{sec:probe}
In this section, we present details of the experimental protocol that probes the variance of the quantum spin projection noise by measuring the timescale of the quench dynamics under the $H_\mathrm{XXZ}$ after the preparation of a spin-squeezed state. The essence of this protocol is a relation between the twisting speed, which is proportional to the quantum variance, and the $S_x$ decay timescale. 

\subsection{Dynamics under OAT Hamiltonian}
We first present an analytical derivation of the $S_x(t_\mathrm{r})$ decay profile for the quench under $H_\mathrm{OAT}$. We start with a Gaussian initial state along $+x$ axis ($\expval{S_x} > 0$, $\expval{S_y} = \expval{S_z} = 0$), and its Wigner function is a Gaussian distribution, 
\begin{equation} \label{eq:OAT_dis}
    P(S_y, S_z) = \frac{1}{2\pi\sigma_Y\sigma_Z\sqrt{1-\rho^2}}\exp(-\frac{1}{2(1-\rho^2)}\qty[\frac{S_y^2}{\sigma_Y^2}+\frac{S_z^2}{\sigma_Z^2}-2\rho \frac{S_yS_z}{\sigma_Y\sigma_Z}]),
\end{equation}
where we slightly abuse the notation by denoting $S_\mu$ as both the quantum spin operator and the classical variable associated with the spin operator. The quantum noise is captured by $\mathrm{Var}(S_y) = \sigma_Y^2$, $\mathrm{Var}(S_z) = \sigma_Z^2$ and $\expval{S_yS_z} = \rho \sigma_Y\sigma_Z$. Under $H_\mathrm{OAT} = \chi S_z^2$, the time evolution of the Wigner quasiprobability distribution is the precession about the $z$-axis with a rate proportional to $Z$, i.e.,  
\begin{equation} \label{eq:OAT_dyn}
\begin{aligned}
    \expval{S_x(t_\mathrm{r})|S_z} &= L\cos(\phi_0S_z + 2S_z\chi t_\mathrm{r}) \\
    \expval{S_y(t_\mathrm{r})|S_z} &= L\sin(\phi_0S_z + 2S_z\chi t_\mathrm{r}),
\end{aligned}    
\end{equation}
where $L = \sqrt{\expval{S_x^2 + S_y^2}}$ is the effective spin length, and $t_\mathrm{r}$ is the evolution time. $\phi_0$ is specified by the initial condition
\begin{equation}
    L\sin(\phi_0S_z) = \expval{S_y(t_\mathrm{r}=0)|S_z} = \rho \frac{\sigma_Y}{\sigma_Z} S_z = \frac{\expval{S_yS_z}}{\sigma_Z^2}S_z,
\end{equation}
where $\expval{S_y(t_\mathrm{r}=0)|S_z} = \rho \frac{\sigma_Y}{\sigma_Z} S_z$ is a direct consequence of Eq.~\eqref{eq:OAT_dis}.  

When $\phi_0$ is small, we have $\phi_0 = \frac{\expval{S_yS_z}}{\sigma_Z^2L}$. Therefore,
\begin{equation}
    \begin{aligned}
    \expval{S_x(t_\mathrm{r})} &= \int \expval{S_x(t_\mathrm{r})|S_z} P(S_z)\dd{S_z} = L \int \cos( \qty[\frac{\expval{S_yS_z}}{\sigma_Z^2L} + 2\chi t_\mathrm{r}]S_z)P(S_z)\dd{S_z} \\
    &= L\exp(-\frac{1}{2}\sigma_Z^2\qty(\frac{\expval{S_yS_z}}{\sigma_Z^2L} + 2\chi t_\mathrm{r})^2) = L\exp(-2\chi^2\mathrm{Var}(S_z)(t_\mathrm{r}-t_\mathrm{o})^2) \\ 
    &= L\exp(-\qty(\frac{t_\mathrm{r}-t_\mathrm{o}}{T_2})^2),
    \end{aligned}
\end{equation}
where we use $P(S_z) = \frac{1}{\sqrt{2\pi} \sigma_Z} e^{-\frac{S_z^2}{2\sigma_Z^2}}$. Here, $T_2$ is the characteristic decay timescale and $t_\mathrm{o}$ is an offset time, i.e.,
\begin{equation}
    T_2 = \frac{1}{\sqrt{2}\chi\sqrt{\mathrm{Var}(S_z)}}, \ \ \ t_\mathrm{o} = -\frac{\expval{S_yS_z}}{2\chi\mathrm{Var}(S_z)L}.
\end{equation}
Note that the offset time $t_\mathrm{o}$ can be either positive and negative, depending on the sign of the correlator $\expval{S_yS_z}$.

Crucially, $T_2$ has a one-to-one correspondence to the quantum variance $\mathrm{Var}(S_z) = \sigma_Z$, which exactly reflects the fact that the quantum state with a larger $\mathrm{Var}(S_z)$ twists and wraps along the Bloch sphere faster. This correspondence enables us to extract the quantum variance by only measuring the collective observable $S_x$, which no longer requires the quantum-projection-noise-limited state readout. Furthermore, $t_\mathrm{o}$ is time when $S_x(t)$ reaches its maximum, i.e., $\max_t S_x(t) = S_x(t_\mathrm{o})$, where the quantum state is minimally wrapped around the Bloch sphere. Note that 
\begin{equation}
    \expval{S_yS_z(t)} = \int \expval{S_x(t_\mathrm{r})|S_z} S_z P(S_z)\dd{S_z} = L \sigma_z^2 \exp(-\frac{1}{2}\sigma_z^2 \qty(2\frac{\chi}{N}t+\phi_0)^2) \qty(2\frac{\chi}{N}t+\phi_0).
\end{equation}
Therefore, we have $\expval{S_yS_z(t_\mathrm{o})} = 0$, indicating that the squeezing direction of the state is either along $y$ or $z$ direction at $t = t_\mathrm{o}$, which further justify that the quantum state at $t = t_\mathrm{o}$ is minimally wrapped around the Bloch sphere. Here, a positive (negative) $t_\mathrm{o}$ means that one has to perform forward (backward) time evolution to unwrapped the initial state. 

Practically, to extract the decay time $T_2$ from the $S_x(t)$ decay profile, we first redefine the effective readout time $t_\mathrm{r}^\mathrm{eff} = t_\mathrm{r} - t_\mathrm{o}$, where $t_\mathrm{r}^\mathrm{eff} = 0$ corresponds to the minimally-wrapped state. Then, we extract the decay timescale by fitting the decay profile with a stretched exponential functional form, i.e, $S_x(t_\mathrm{r}^\mathrm{eff})\sim\exp(-(t_\mathrm{r}^\mathrm{eff}/T_2)^\alpha)$, where $\alpha = 2$ for the OAT case. This protocol can be alternatively understood as follows: 
\begin{enumerate}
    \item For an initial state with $\expval{S_yS_z} = 0$, $S_x(t_\mathrm{r})$ has a monotonic decay profile, where one can directly extract a timescale $T_2$ by fitting the decay profile to a stretched exponential functional form without any offset time $t_\mathrm{o}$.
    \item For an initial state with $\expval{S_yS_z} \neq 0$, one can forward (backward) evolution the quantum state by a time $t = -t_\mathrm{o}$ to prepare an intermediate state with $\expval{S_yS_z} = 0$. Then, one measure thes quench dynamics for the intermediate step and follow the step 1 to extract a decay timescale $T_2$, which corresponds to $\mathrm{Var}(S_z)$ of the intermediate state. Use the fact that time evolution under $H_\mathrm{OAT}$ preserve $S_z$, the initial state ($\expval{S_yS_z} \neq 0$) and the intermediate state ($\expval{S_yS_z} = 0$) has the same $\mathrm{Var}(S_z)$. Therefore, the one-to-one correspondence between $T_2$ and $\mathrm{Var}(S_z)$ for the intermediate state can be directly applied to the initial state, i.e., the measured timescale $T_2$ can be directly used to infer $\mathrm{Var}(S_z)$ for the initial state. 
\end{enumerate}
The second case above is exactly the same as redefining the effective readout time, since the preparation of the intermediate state and the quench measurement of the intermediate are all just the quench from the initial state. In practice, one can extract $t_\mathrm{o}$ from the effective twisting strength (Sec.~\ref{sec:tau}). Beyond $H_\mathrm{OAT}$, the aforementioned protocols also works for $H_\mathrm{XXZ}$ in our experiment, where the same derivation can be performed with an early-time expansion as sketched in the next section.

\subsection{Dynamics under XXZ Hamiltonian} \label{sec:tau}
After establishing the protocol under the $H_\mathrm{OAT}$, we now show that the sample procedure also applies to the disorder XXZ Hamiltonian $H_\mathrm{XXZ}$. We perform an early time expansion in the Heisenberg picture to analyze the time evolution of the collective spin operators.

The time evolution of an operator $O$ is given by
\begin{equation}
    O(t) = O + it[H, O] + \frac{1}{2} (it)^2 [H, [H, O]] + \mathcal{O}(t^3).
\end{equation}
For the dipolar interacted system, the Hamiltonian can be separated into Heisenberg part and Ising part, i.e.,
\begin{equation}
    H = -J\sum_{i,j} \frac{1}{r_{ij}^3} [(s_x^is_x^j + s_y^is_y^j + s_z^is_z^j) - 2s_z^is_z^j] = H_{\mathrm{Hei}} + H_{\mathrm{Ising}}.
\end{equation}
Because the initial coherent spin state is the eigenstate of the Heisenberg Hamiltonian $H_{\mathrm{Hei}}$, we expect that $\expval{[H_{\mathrm{Hei}}, O]} \simeq 0$ in the early time, where $O$ can be any (single or multiple body) operator. Therefore, we have
\begin{equation}
    \expval{[H, O]} = \expval{[H_{\mathrm{Hei}}, O]} + \expval{[H_{\mathrm{Ising}}, O]} = \expval{[H_{\mathrm{Ising}}, O]}
\end{equation}

Denote $J' = -2J$ in the following derivation. We have
\begin{equation}
\begin{aligned}\
    [H_{\mathrm{Ising}}, S_x] &= J'\sum_{ij} \frac{1}{r_{ij}^3} \sum_{m} [s_z^is_z^j, s_x^m]  = J'\sum_{i<j,m} \frac{1}{r_{ij}^3} i[ \delta_{im} s_y^i s_z^j + \delta_{jm} s_z^i s_y^j] = J'\sum_{i<j} \frac{1}{r_{ij}^3} i[ s_y^i s_z^j +  s_z^i s_y^j]  \\
    [H_{\mathrm{Ising}}, S_y] &= J'\sum_{ij} \frac{1}{r_{ij}^3} \sum_{m} [s_z^is_z^j, s_y^m]  = J'\sum_{i<j,m} \frac{1}{r_{ij}^3} (-i) [ \delta_{im} s_x^i s_z^j + \delta_{jm} s_z^i s_x^j]  = J'\sum_{i<j} \frac{1}{r_{ij}^3} (-i) [  s_x^i s_z^j + s_z^i s_x^j] \\
\end{aligned}
\end{equation}

\begin{equation}
    \begin{aligned}\
        [H_{\mathrm{Ising}}, [H_{\mathrm{Ising}}, S_x]] &= J'^2 \sum_{i<j; m<n} \frac{i}{r_{ij}^3r_{mn}^3}[  s_z^is_z^j,  s_y^m s_z^n +  s_z^m s_y^n] \\ 
         &= J'^2 \sum_{i<j; m<n} \frac{i}{r_{ij}^3r_{mn}^3} ([ s_z^is_z^j,  s_y^m] s_z^n +  s_z^m[s_z^is_z^j, s_y^n]) \\
         &= J'^2 \sum_{i<j; m<n} \frac{1}{r_{ij}^3r_{mn}^3} (\delta_{im}s_x^is_z^js_z^n + \delta_{jm}s_z^is_x^js_z^n + \delta_{in}s_z^ms_x^is_z^j + \delta_{jn}s_z^ms_z^is_x^j)
    \end{aligned}
\end{equation}

Here, we would like to derive Eq.\eqref{eq:OAT_dyn} for $H_\mathrm{XXZ}$, which requires the calculation of the expectation value of collective spin operators for an initial spin coherent states offsetted from the $+x$ axis by an angle $\theta$ such that $\expval{S_z} = N\sin\theta/2$. Rather than rotate the quantum state, we can rotate the reference frame by an angle of $\theta$ (redefining $s_\mu^i$ operators) such that the offset state becomes $\ket{+x}$ in the new frame. In the leading order, we get

\begin{equation}
\begin{aligned}
    \expval{[H_{\mathrm{Ising}}, S_x]} &= 0 \\
    \expval{[H_{\mathrm{Ising}}, S_y]} &= J' \sum_{i<j} \frac{1}{r_{ij}^3} (-i) 2\sin\theta\cos\theta = -2i\sin\theta\cos\theta J_{\mathrm{eff}}, \\
\end{aligned}
\end{equation}
where $J_{\mathrm{eff}} = J'\sum_{i<j}1/r_{ij}^3$. Therefore, the semiclassical evolution of the collective spin operator is 
\begin{equation} \label{eq:mf}
    \begin{aligned} 
        \expval{S_x(t)|S_z} &= S_x(0)|S_z + it\expval{[H, S_x]} + \mathcal{O}(t^2) =  \frac{N}{2}  + \mathcal{O}(t^2),\\
        \expval{S_y(t)|S_z} &= S_y(0)|S_z + it\expval{[H, S_y]} + \mathcal{O}(t^2) = 2\sin\theta \cos\theta J_{\mathrm{eff}} t + \mathcal{O}(t^2) \simeq \frac{4S_z}{N}J_{\mathrm{eff}} t + \mathcal{O}(t^2),\\
    \end{aligned}
\end{equation}
where we have used $S_z = \frac{N}{2}\sin\theta$ and assumed $\cos\theta \simeq 1$ for small $\theta$. Note that this equation is in the same form as Eq.\eqref{eq:OAT_dyn}. In the semiclassical picture, the initial coherent spin state is approximated by a Gaussian distribution $P(S_z) = \sqrt{\frac{2}{\pi \eta N}} \exp(-\frac{2S_z^2}{\eta N})$, where $\mathrm{Var}(S_z) = \expval{S_z^2} = \eta N/4$. Therefore, 
\begin{equation} 
    \begin{aligned}
        \expval{S_x(t)} &= \int \expval{S_x(t)|S_z}  P(S_z)\dd{S_z} \\
        \expval{S_yS_z(t)} &= \int \expval{S_y(t)|S_z} S_zP(S_z)\dd{S_z} = \int \qty( \frac{4S_z}{N}J_{\mathrm{eff}} t ) S_zP(S_z) \dd{S_z} = \eta J_{\mathrm{eff}} t, \\ 
        \expval{S_y^2(t)} &= \mathrm{Var}(S_y|S_z) + \int \expval{S_y(t)|S_z}^2  P(S_z)\dd{S_z} = \frac{N}{4\eta} + \frac{4 \eta J_{\mathrm{eff}}^2 t^2}{N}
    \end{aligned}
\end{equation}

As a result, we have 
\begin{equation} \label{eq:corr}
\begin{aligned}
    \frac{\expval{S_yS_z(t)}}{\expval{S_z^2}} &= \frac{4J_{\mathrm{eff}} t}{N} = \chi t, \\
    \frac{\expval{S_y^2(t)}}{\expval{S_z^2}} &= \frac{1}{\eta^2} + \frac{16 J_{\mathrm{eff}}^2 t^2}{N^2} = \frac{1}{\eta^2} + (\chi t)^2,
\end{aligned}
\end{equation}
where we denote $\chi = 4J_{\mathrm{eff}}/N$. 

Now, given that we have prepared a squeezing state by the time evolution under $H_\mathrm{XXZ}$, in order to measure $\mathrm{Var}(S_\theta)$, we apply a global rotation $\theta$ along the $x$ axis. The correlators for the state after rotation is 
\begin{equation}
    \begin{aligned}
        \expval{S_yS_z(\theta)} &= (\cos^2\theta-\sin^2\theta)\expval{S_yS_z} + \sin\theta \cos\theta(\expval{S_y^2} - \expval{S_z^2}) \\ 
        \expval{S_z^2(\theta)} &= \cos^2\theta\expval{S_z^2} + \sin^2 \theta \expval{S_y^2} + 2\sin\theta \cos\theta \expval{S_yS_z},
    \end{aligned}
\end{equation}
where $\expval{S_y^2}$, $\expval{S_z^2}$ and $\expval{S_yS_z}$ denote the correlators for the state before the $\theta$ rotation, which satisfy Eq.\eqref{eq:corr}.

Note that the offset time $t_\mathrm{o}$ is defined as the time such that $\expval{S_yS_z(t_\mathrm{o})} = 0$, i.e., the squeezing direction is either along the $+y$ or $+z$ direction. From Eq.~\eqref{eq:corr}, we have
\begin{equation}
\begin{aligned}
    \chi t_\mathrm{o} &= -\frac{\expval{S_yS_z(\theta)}}{\expval{S_z^2(\theta)}} \\ 
    &= -\frac{(\cos^2\theta-\sin^2\theta)\expval{S_yS_z} + \sin\theta \cos\theta(\expval{S_y^2} - \expval{S_z^2})}{\cos^2\theta\expval{Z^2} + \sin^2 \theta \expval{S_y^2} + 2\sin\theta \cos\theta \expval{S_yS_z}} \\ 
    &= -\frac{(1-\tan^2\theta)\frac{\expval{S_yS_z}}{\expval{S_z^2}} + \tan\theta \qty(\frac{\expval{S_y^2}}{\expval{S_z^2}} - 1)}{1 + \tan^2\theta\frac{\expval{S_y^2}}{\expval{S_z^2}} + 2\tan\theta 
 \frac{\expval{S_yS_z}}{\expval{S_z^2}}}
\end{aligned}
\end{equation}

Now, we can apply Eq.~\eqref{eq:corr} again to replace the correlators $\expval{S_y^2}$, $\expval{S_z^2}$ and $\expval{S_yS_z}$ by the preparation time, 
\begin{equation} \label{eq:data_process}
\begin{aligned}
    \chi t_\mathrm{o} &= -\frac{(1-\tan^2\theta) \chi t_\mathrm{g} + \tan\theta (\chi t_\mathrm{g})^2}{1 + \tan^2\theta(1 + (\chi t_\mathrm{g})^2) + 2\tan\theta \chi t_\mathrm{g}}. 
\end{aligned}
\end{equation} 
This is the analytical formula to extract the offset time $t_\mathrm{o}$. 

We note that the asymptotic behavior in the $\chi t_\mathrm{g} \rightarrow 0$ limit is 
\begin{equation}
\begin{aligned}
    \lim_{\chi t_\mathrm{g} \rightarrow 0} \frac{\chi t_\mathrm{o}}{\chi t_\mathrm{g}} &= -\frac{1-\tan^2\theta}{1 + \tan^2\theta}  = -\cos(2\theta),
\end{aligned}
\end{equation}
which can be used as a rough estimation if $\chi$ is unknown. However, in reality, the effective mean field interaction strength $\chi$ can be characterized by the average twisting experiment, where the initial coherent spin state is rotated along the $y$ axis by an angle $\theta$ such that $Z = \frac{N}{2}\sin\theta$. Eq.\eqref{eq:mf} provides a description for dynamics of the average motion of the coherent spin state, i.e.,
\begin{equation}
     \frac{S_y(t)}{S_x(t)} = \chi t \sin\theta,
\end{equation}
where $S_y(t)$ and $S_x(t)$ are measured in the average twisting experiment. Therefore, we finally get
\begin{equation} \label{eq:chi_MF}
    \chi = \frac{1}{\sin\theta}\lim_{t\rightarrow 0} \qty( \frac{S_y(t)/S_x(t)}{t} ),
\end{equation}  
where we take the early time limit $t\rightarrow 0$. This equation is used to extract the interaction strength $\chi$ for the average spin twisting measurement.

\section{Effect of dimers in the  disordered system} \label{sec:dimer}
In the system with positional disorder, as discussed in the main text, the existence of strongly-interacting NV dimers causes a rapid decay of the collective spin length which prohibits spin squeezing. In this section, we provide a more thorough discussion of the effect of the dimers on the mean-field twisting and variance readout measurements.

\subsection{Stretch power as a measure of lattice geometry}
The dynamics of the spin-polarized state $\ket{\mathbf{x}}$ under $H_{\mathrm{XXZ}}$ exhibit a single stretched exponential decay $S_x(t)\sim e^{-(t/\tau)^p}$. As described in the main text, the stretch power $p$ indicates the spatial geometry of the spin ensemble. Here, we provide an analytical derivation showing the crossover from the early-time Gaussian dynamics $p = 2$ to late-time disordered dynamics $p = 2/3$ in the lattice-engineered system, as discussed in Sec.~III.1 of the Methods. 

In the early time regime, the Hamiltonian can be separated into Heisenberg and Ising parts as 
\begin{equation}
    H_{\mathrm{XXZ}} = H_{\mathrm{Hei}} + H_{\mathrm{Ising}} = -J_0\sum_{i,j} \frac{1}{r_{ij}^3} (s_x^is_x^j + s_y^is_y^j + s_z^is_z^j) + 2 J_0\sum_{i,j} \frac{1}{r_{ij}^3} s_z^is_z^j;
\end{equation}
we hereafter set the interaction strength $J_0 = 1$ throughout this section for simplicity. Because $H_{\mathrm{Hei}}$ commutes with $S_x$, the quench dynamics $S_x(t)$ are dominated by the Ising interaction. Therefore, the quench profile $S_x(t)$ can be derived analytically following Eq.(S9) in Ref.~\cite{davisProbingManybodyDynamics2023}, i.e., 
\begin{equation}
\begin{aligned}
    S_x(t) &= \exp{-n\int_{r_\mathrm{min}}^\infty \dd{r} r \int \dd{\Omega} \qty(1-e^{-\frac{\chi(t)}{8r^{6}} })} \\ 
    &= \exp{-2\pi n\int_{r_\mathrm{min}}^\infty \dd{r} r \qty (1-e^{-\frac{\chi(t)}{8r^{6}} })} \\ 
    &= \exp{-2\pi n \qty[\frac{r_\mathrm{min}^2}{6} \qty( \mathrm{E}_{4/3} \qty(\frac{\chi(t)}{8r_\mathrm{min}^6})-3) - \frac{\chi(t)^{1/3}}{12}\Gamma \qty(-\frac{1}{3}) ]}.
\end{aligned}
\end{equation}
Here, $n$ is the spin density, $E_m(x) = \int_1^\infty e^{-xt}/t^m\dd{t}$ is the exponential integral function, and $\chi(t)$ encodes the response of the probe spins to the noise. In particular, $\chi(t) = t^2$ in the early-time ballistic region (a detailed definition and derivation can be found in the supplementary information of \cite{davisProbingManybodyDynamics2023}). 

In the limit $t\rightarrow 0$, by expanding the exponential integral function around $x = \chi(t)/8r_\mathrm{min}^6 = 0$, we have $E_{4/3}(x)\simeq 3 + \Gamma(-\frac{1}{3})x^{1/3} + \frac{3}{2}x + \mathcal{O}(x^2)$, thus
\begin{equation}
\begin{aligned}    
    S_x(t) &=  \exp{-2\pi n \qty[\frac{r_\mathrm{min}^2}{6} \qty(\Gamma\qty(-\frac{1}{3}) \qty(\frac{\chi(t)}{8r_\mathrm{min}^6})^{1/3} + \frac{3}{2}\qty(\frac{\chi(t)}{8r_\mathrm{min}^6}) +\mathcal{O}\qty(\qty(\frac{\chi(t)}{8r_\mathrm{min}^6})^3 ) ) - \frac{\chi(t)^{1/3}}{12}\Gamma\qty(-\frac{1}{3}) ]} \\ 
    &= \exp{-2\pi n \qty[ \Gamma\qty(-\frac{1}{3}) \frac{\chi(t)^{1/3}}{12} + \frac{\chi}{32r_\mathrm{min}^6} \mathcal{O}\qty(\frac{\chi(t)^2}{r_\mathrm{min}^{10}} )  - \frac{\chi(t)^{1/3}}{12}\Gamma\qty(-\frac{1}{3}) ]}  \\ 
    &\simeq \exp{-2\pi n \qty[ \frac{\chi(t)}{32r_\mathrm{min}^6} ]} \sim e^{-(t/\tau_i)^2},
\end{aligned}
\end{equation}
where $\tau_i$ is the characteristic decay timescale.

In the limit $t\rightarrow \infty$, by expanding the exponential integral function around $x=\infty$, we have $E_{4/3}(x) = e^{-x + \mathcal{O}(1/x^6)}(1/x + \mathcal{O}(1/x))$, thus,
\begin{equation}
\begin{aligned}    
    S_x(t) &=  \exp{-2\pi n \qty[\frac{r_\mathrm{min}^2}{6} \qty( e^{-\chi(t)/8r_\mathrm{min}^6} \qty(\frac{8r_\mathrm{min}^6}{\chi(t)} + \mathcal{O}\qty( \qty(\frac{8r_\mathrm{min}^6}{\chi(t)})^2)) -3) - \frac{\chi(t)^{1/3}}{12}\Gamma \qty(-\frac{1}{3}) ]} \\
    &\simeq  \exp{-2\pi n \qty[\frac{r_\mathrm{min}^2}{6} \qty( e^{-\chi(t)/8r_\mathrm{min}^6}\frac{8r_\mathrm{min}^6}{\chi(t)} -3) - \frac{\chi(t)^{1/3}}{12}\Gamma \qty(-\frac{1}{3}) ]} \\ 
    &\simeq \exp{-2\pi n \frac{\chi(t)^{1/3}}{12}\qty(-\Gamma\qty(-\frac{1}{3}))} \sim e^{-(t/\tau_f)^{2/3}},
\end{aligned}
\end{equation}
where we use the fact that the exponential integral function vanishes when $t$ is sufficiently large. 

To summarize, the stretch power is $p = 2$ ($\alpha = 2/3$) in the early (late) time limit. The crossover timescale is given by $t_\mathrm{c}\sim r_\mathrm{min}^3$. A larger $r_\mathrm{min}$ corresponds to the situation where more dimers are removed, such that the system behaves more like a regular lattice. As a result, the decay profile $S_x(t)$ stays in the lattice regime $\alpha = 2$ for a longer time. This derivation is consistent with our numerical simulation [Method Fig.~3]. 

\subsection{Dimer dynamics during mean-field twisting}
In the fully disordered system, both the ``typical'' NV centers and the closely-spaced NV dimers contribute to the mean-field twisting signal shown in Fig.~1 of the main text. The contribution of the typical spins resembles OAT dynamics, and leads to a linear increase of the precession angle $\varphi_p(t)$. To understand the more subtle contribution of the dimers, we study the time evolution of a single dimer under 
\begin{equation} \label{eq:H_dimer}
\begin{aligned}
    H_d &= -J (s_x^1s_x^2 + s_y^1s_y^2 - s_z^1s_z^2)= -J \qty[s_x^1s_x^2 + \frac{s_+^1s_+^2 + s_-^1s_-^2}{2}],
\end{aligned}
\end{equation}
where $s_\pm = s_y \pm i s_z$. 
For an initial state prepared on the $+x$ axis, $\ket{++}$, it is first rotated along the $+y$ axis by an angle $\varphi_o$, and the state becomes $\ket{\psi} = \cos^2(\varphi_o/2)\ket{++} + \sin^2(\varphi_o/2)\ket{--} - \sin(\varphi_o/2)\cos(\varphi_o/2)(\ket{+-}+\ket{-+})$. After evolving under $H_\mathrm{XXZ}$ for a time $t$,  we have 
\begin{equation}
\begin{aligned}
    \ket{\psi(t)} =& ~e^{iJ t/4}\qty[\cos^2\qty(\frac{\varphi_o}{2})\cos(\frac{J_t}{2}) - i\sin^2\qty(\frac{\theta}{2})\sin(\frac{Jt}{2})]\ket{++}  \\ 
    &\ + e^{iJt/4}\qty[-i\cos^2\qty(\frac{\theta}{2})\sin(\frac{Jt}{2}) + \sin^2\qty(\frac{\varphi_o}{2})\cos(\frac{Jt}{2})]\ket{--}  \\
    & \ - e^{-iJt/4}\sin(\frac{\varphi_o}{2})\cos(\frac{\varphi_o}{2})(\ket{+-}+\ket{-+}).
\end{aligned}
\end{equation}
As a result, we have 
\begin{equation} \label{eq:dimr_MF}
\begin{aligned}
    \expval{S_x(t)} &= \cos\varphi_o \cos(Jt), \\ 
    \expval{S_y(t)} &= -\sin\varphi_o\cos\varphi_o \sin(Jt).
\end{aligned}
\end{equation}
For a single spin, the probability distribution of the distance $r_\mathrm{NN}$  to its nearest spin is
\begin{equation} \label{eq:r_dis}
    P(r_\mathrm{NN}) = n\cdot 4\pi r_\mathrm{NN}^2\exp(-\frac{4}{3}\pi r_\mathrm{NN}^3 n),
\end{equation}
where $n$ is the density of spins. 

\begin{figure}[htbp]
    \includegraphics[width=0.7\linewidth]{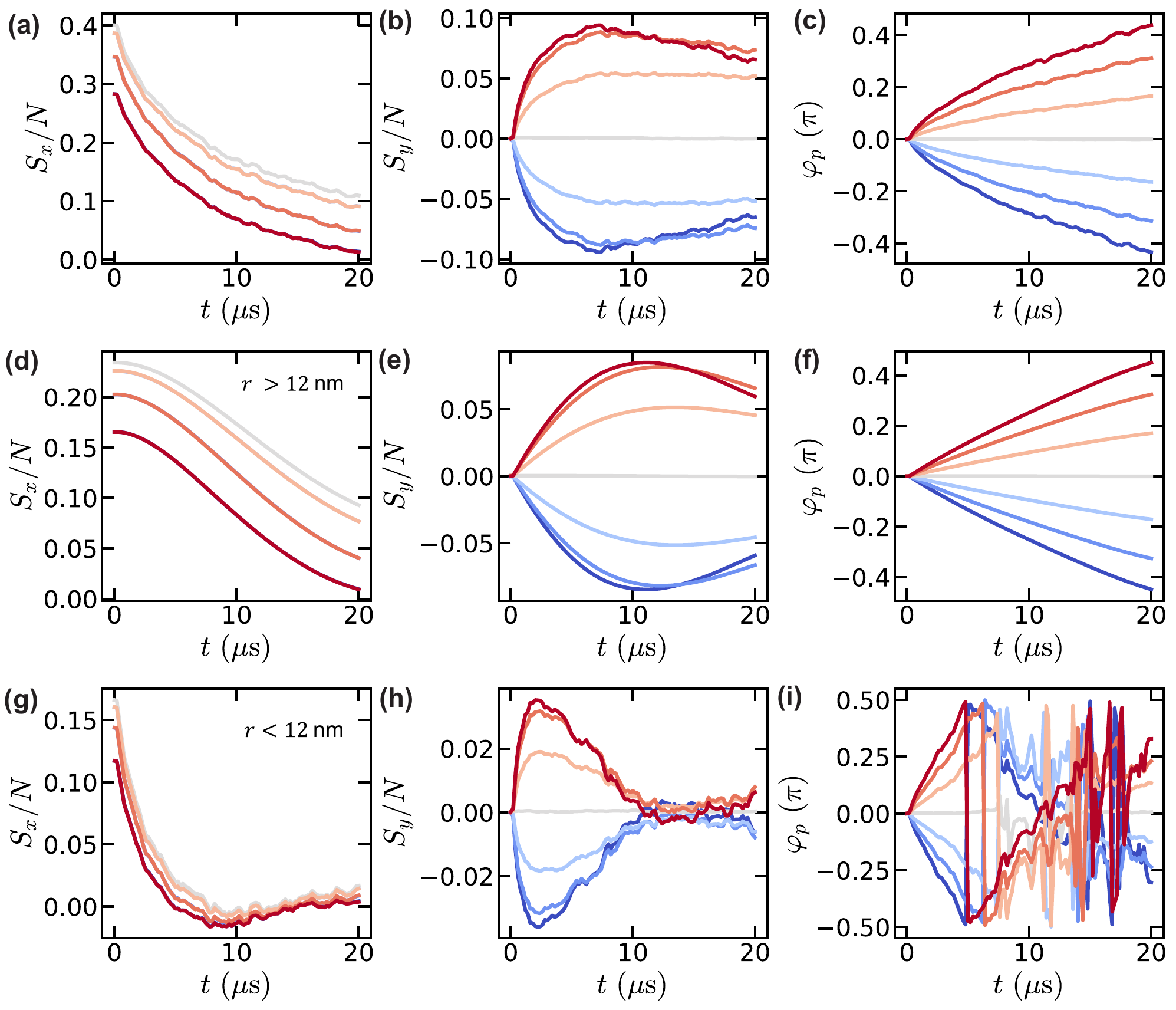}
    \caption{
    \textbf{Numerical simulation of average twisting dynamics.} We simulate the average twisting dynamics on a fully disordered system with $N = 80$ spins by the cluster DTWA method. We plot the evolution of $S_x$, $S_y$ and $\varphi_p = \arctan (S_y/S_x)$ for the full system (\textbf{a-c}), the subsystem consisting of the spins whose distance to the closest neighboring spin is larger than $12$ nm (\textbf{d-f}) and smaller than $12$ nm (\textbf{g-i}). The initial polarization is set as $80\%$. In all plots, different colors correspond to different initial rotation angles $\varphi_o$, ranging from $\varphi_o = -\pi/4$ (red) to $\varphi_o = \pi/4$ (blue).
    }
    \label{fig:MF_cut}
\end{figure}

Eq.~\eqref{eq:dimr_MF} reveals that the early time behavior of the dimers is similar to the twisting dynamics of ``typical'' spins. Specifically, the precession angle satisfies $\varphi_p \simeq \expval{S_y(t)}/\expval{S_x(t)} \simeq -\tan(Jt)\sin\theta \simeq -J_{12}t\sin\varphi_o$, which is proportional to both the evolution time $t$ and $\expval{S_z}\sim \sin\varphi_o$. These dynamics are plotted in Figure~\ref{fig:MF_cut}. The first row (a-c) shows the mean-field twisting for a fully disordered system, which is a combination of typical NV and dimer dynamics. In the second and third rows, we cut the system in two and plot the ``typical" (d-f) and dimer (g-i) contributions separately. The cut is chosen to correspond to the optimal shelving radius, $r_\text{shelve}=12~$nm, used to produce the light pink data in Fig.~4(a-c) of the main text. 

\subsection{Dimer dynamics during quantum variance readout}
Here, we analyze the effect of the dimers dynamics on the readout of the variance.  
Recall that the experiment protocol consists of two quenches: generation quench (with an evolution time $t_\mathrm{g}$) and readout quench (with an evolution time $t_\mathrm{r}$), with a $X_\theta$ rotation between the two quenches. 
Our discussion of one-axis twisting dynamics in the main text draws from the OAT picture, which maps well onto ordered spins in a lattice. In this picture, the twisting rate acts as a measurement of the variance of the spins.

Under the same pulse sequence, dimers behave distinctly but still contribute to the measured $\avg{S_x(t)}$ signal. The primary contribution of the dimers in our measurement is to alter the offset time $t_\mathrm{o}$ compared with the OAT case. As discussed in Sec.~\ref{sec:probe}, the decay of the spin length $\langle S_x(t_\mathrm{r}) \rangle$ originates from the twisting of the spin projection noise around the Bloch sphere. Naively, when $\expval{S_yS_z} = 0$, the spin length $S_x(t)$ reaches its maximum. While this is true for spins on a lattice, in the disordered system the time when $\expval{S_yS_z} = 0$ does not coincide with the time at which $\expval{S_x(t)}$ is maximized, due to dimer dynamics.

The dimers always at least partially recover their polarization at $t_\mathrm{r} = t_\mathrm{g}$.
Using Eq.~\eqref{eq:H_dimer}, we can restrict the dynamics to the $\{\ket{++}, \ket{--}\}$ subspace, such that $H_\mathrm{XXZ} = -J\qty[\frac{1}{4} - \frac{\Sigma_x}{2}]$, where $\Sigma_\mu\ (\mu = x, y, z)$ is the Pauli operator in this subspace. For dimers initially prepared in $\ket{\psi(t_\mathrm{g} = 0)} = \ket{++}$, the generation quench yields $\ket{\psi(t_\mathrm{g})} = \cos(J t_\mathrm{g}/2)\ket{++} - i\sin(Jt_\mathrm{g}/2)\ket{--}$. After the rotation $X_\theta$, the state becomes $\ket{\psi(t_\mathrm{g}, t_\mathrm{r} = 0)} = \cos(Jt_\mathrm{g}/2) e^{i\theta}\ket{++} - i\sin(Jt_\mathrm{g}/2)e^{-i\theta}\ket{--}$. Finally, we evolve this quantum state under $H_\mathrm{XXZ}$ again for $t_\mathrm{r}$, i.e.,
\begin{equation}
\begin{aligned}
    \ket{\psi(t_\mathrm{g}, t_\mathrm{r})} &= \cos(\frac{t_\mathrm{g}}{2}) e^{i\theta} \qty[\cos(\frac{Jt_\mathrm{r}}{2})\ket{++} - i\sin(\frac{Jt_\mathrm{r}}{2})\ket{--}] \\ 
    & \ - i\sin(\frac{Jt_\mathrm{g}}{2})e^{-i\theta}\qty[-i\sin(\frac{Jt_\mathrm{r}}{2})\ket{++} + \cos(\frac{Jt_\mathrm{r}}{2})\ket{--}] \\ 
    &= [\cos(\frac{Jt_\mathrm{g}}{2})\cos(\frac{Jt_\mathrm{r}}{2})e^{i\theta} - \sin(\frac{Jt_\mathrm{g}}{2})\sin(\frac{Jt_\mathrm{r}}{2})e^{-i\theta}]\ket{++} \\ 
    & \ +  \qty[-ie^{i\theta}\cos(\frac{Jt_\mathrm{g}}{2})\sin(\frac{Jt_\mathrm{r}}{2}) - ie^{-i\theta} \sin(\frac{Jt_\mathrm{g}}{2})\cos(\frac{Jt_\mathrm{r}}{2})]\ket{--}.
\end{aligned}
\end{equation}
As a result, the dimer contribution to the spin length is 
\begin{equation}
\begin{aligned}
    \expval{S_x(t_\mathrm{g}, t_\mathrm{r})} &= \cos(Jt_\mathrm{g}) \cos(Jt_\mathrm{r}) - \sin(Jt_\mathrm{g}) \sin(Jt_\mathrm{r})\cos(2\theta) \\ 
    &= \sin^2\theta \cos(J(t_\mathrm{g} - t_\mathrm{r})) + \cos^2\theta\cos(J(t_\mathrm{g} + t_\mathrm{r})).
\end{aligned}
\end{equation}

In a system with positional disorder, dimers interact with different couplings $J = J_0/r_{12}^3$, where the inter-spin distance $r_{12}$ can be randomly sampled from Eq.~\eqref{eq:r_dis}. Most of the time ($t_\mathrm{g}\neq t_\mathrm{r}$), due to the random distribution of $J$, the dimers with different inter-spin distances oscillate between $\ket{++}$ and $\ket{--}$ with different frequencies, and $\expval{S_x(t_\mathrm{g}, t_\mathrm{r})}$ averages to zero. However, when $t_\mathrm{g} = t_\mathrm{r}$, the first term in the equation above is no longer vanishing, i.e., $\expval{S_x(t_\mathrm{g}, t_\mathrm{g})} = \sin^2\theta$ while the second term still averages to zero. Therefore, regardless of $\theta$, the dimers always partially recover their polarization at $t_\mathrm{r} = t_\mathrm{g}$. When $\theta = \pi/2$, we have $\expval{S_x(t_\mathrm{g}, t_\mathrm{g})} = 1$, indicating that the dimers fully recover their polarization. Indeed, we observe this feature in the experiment (see main text Fig.~2(e), inset). This feature fully originates from dimers, and is unrelated to spin projection noise of the collective spin wrapping around the Bloch sphere. While the offset time $t_\mathrm{o}$ is the time when the $\expval{S_x(t_\mathrm{r})}$ reaches its maximum for the OAT and lattice systems, we cannot use this as the procedure to identify $t_\mathrm{o}$ in the system with positional disorder. Instead, to determine $t_\mathrm{o}$, we use the condition that the state at $t_\mathrm{r} = t_\mathrm{o}$ should minimally wrap around the Bloch sphere, i.e., $\expval{S_yS_z} = 0$. 

\bibliographystyle{naturemag}
\bibliography{supplementalInformation_arXiv.bib}

\title{Methods and Extended Data}

\maketitle

% \tableofcontents

\section{Experimental setup and sample properties}

\subsection{Sample preparation}\label{sec:sample}
Our sample \cite{hughesStronglyInteractingTwodimensional2024} is grown via plasma-enhanced chemical vapor deposition (PECVD) with nitrogen delta-doping. The growth substrate is sliced from an electronic grade $(001)$-oriented CVD substrate (Element Six Ltd.) along the $(111)$ plane and then polished by Applied Diamond, Inc. to a surface roughness less than $500$ pm. Prior to growth, the miscut angle is measured using x-ray diffractometry (XRD) rocking curves about the (111) omega peak and surface roughness is verified with AFM. PECVD growth is performed in a SEKI SDS6300 reactor at $770\ ^\circ\mathrm{C}$, which finally produces a $99.998\%$ $^{12}$C isotopically purified epilayer. In the nitrogen delta-doping process, $^{15}$N gas is introduced into the PECVD chamber for $1$ minute. After growth, secondary ion mass spectroscopy (SIMS) is performed in-house with conditions as outlined previously in Refs.~\cite{hughesTwodimensionalSpinSystems2023, hughesStronglyInteractingTwodimensional2024} to estimate the isotopic purity, epilayer thickness, and the properties of the delta-doped layer [Fig.~\ref{fig:supp_main}a.i-iii].

Because the growth substrate is a diamond with natural carbon abundance while the epilayer is isotopically purified, the epilayer thickness is given by the low $^{13}$C density layer, i.e., $\sim 270$ nm [Fig.~\ref{fig:supp_main}a.iii]. 
A significantly higher $^{15}$N density is observed in the delta-doped layer $125$ nm below the diamond surface compared to the background within the epilayer~[Fig.~\ref{fig:supp_main}a.ii]. 
The thickness of the delta-doped layer extracted from SIMS is estimated from the FWHM of the peak to be $7^{+11}_{-7}$ nm. The error is calculated from the $^{13}$C SIMS step function used to calculate the growth rate of the epitaxial layer, and it represents the data broadening due to surface roughness. Layer thicknesses extracted from SIMS should only be taken as an estimate, and represent an upper bound due to the roughness-induced broadening.

To further increase the density of NV centers, the sample is then irradiated with electrons at 200 keV from a transmission electron microscope (TEM, ThermoFisher Talos F200X G2 TEM). The irradiation dosage is $1.29\times 10^{20}\ \mathrm{e/cm}^2$ for the spot used in this experiment. The sample is then annealed at $400^\circ$C for 2 hours and $850^\circ$C for 4 hours in a vacuum furnace ($4\times 10^{-8}$ torr) to promote vacancy diffusion and NV formation. Finally, the sample is further cleaned by boiling in a tri-acid solution (1:1:1 \ce{H2SO4}:\ce{HNO3}:\ce{HClO4}) and annealed at $450^\circ$C in air for 4 hours to remove surface contaminants and form oxygen termination for the diamond surface. Further growth details can be found in Ref.~\cite{hughesStronglyInteractingTwodimensional2024}.

\subsection{Apparatus}
The delta-doped sample is measured with a home-built scanning confocal microscope at room temperature. For optical initialization and readout, $75\ \mu$W of 532~nm light is sent through an air objective (Nikon CFI Plan Apo 60X, NA 0.95). The NV fluorescence is separated
from the green excitation light with a dichroic filter and collected on fiber-coupled single-photon counters (Excelitas SPCM-AQRH-12-FC). A typical scan of our irradiated spot is shown in Figure~\ref{fig:supp_main}b. A magnetic field is generated by three sets of orthogonal electromagnets with tunable current sources. The field is aligned to the out-of-the-plane direction of the sample and set to 393~Gauss to maximize the nuclear spin hyperpolarization. Microwave pulses at 1.742~GHz ($\ket{m_s=0}\leftrightarrow \ket{m_s=-1}$ transition) and 3.998~GHz ($\ket{m_s=0}\leftrightarrow \ket{m_s=1}$ transition) at 393~G are generated by an arbitrary waveform generator (AWG, Tektronix AWG7122C), amplified by a broadband microwave amplifier (Mini-Circuits ZHL-50W-63+) and delivered through a stripline to achieve a typical Rabi frequency $\sim (2\pi)\times 25$ MHz. Various microwave filters and circulators are inserted into the signal chain to suppress reflections and improve the microwave pulse quality. To generate strong and weak pulses simultaneously, we use two channels of the AWG and combine the signals with a resistive power combiner (Pulsar RP2-05-411). The amplitude of the microwave pulse is further tuned with a programmable attenuator (Mini-Circuits RUDAT-6000-90).

\subsection{Nuclear spin polarization}
At the magnetic field $B = 393$~Gauss used in our experiment, the nuclear spin polarization is $89(1)\%$, measured via relative amplitudes of the optically detected magnetic resonances (ODMR)~[Fig.~\ref{fig:supp_main}e]. In our experiments, we work with the majority $m_I = +1/2$ nuclear spin subgroup and shelve the residual $m_I = -1/2$ population in the state $\ket{m_s = 1}$ to avoid off-resonant driving of these spins. The Rabi frequency of the shelving pulse is chosen to be $\sim (2\pi) \times 600$ kHz, which is smaller than the hyperfine splitting and larger than the inhomogeneous broadening. The ODMR spectra before and after shelving are shown in Fig.~\ref{fig:supp_main}e.

\subsection{NV and P1 densities}\label{sec:NVdensity}
\begin{figure}[h!]
    \centering
    \includegraphics[width=0.8\linewidth]{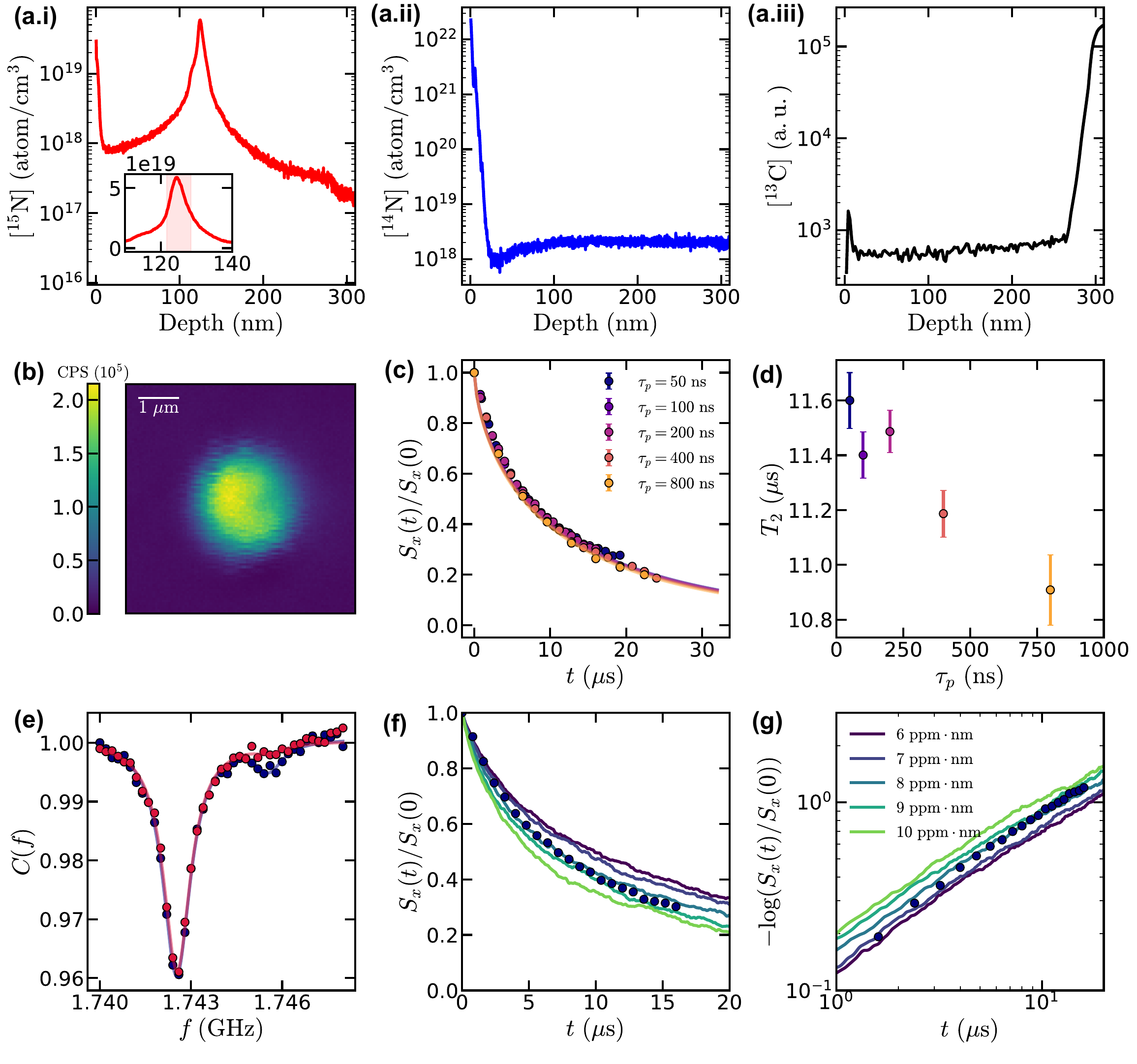}
    \caption{\textbf{Sample characterization.}
    \textbf{(a)} Secondary ion mass spectroscopy (SIMS) measurements of the sample showing the densities of $^{15}$N (i), $^{14}$N (ii), and $^{13}$C atoms (iii) as a function of distance from the surface of the diamond.
    \textbf{(b)} A typical confocal scan of the irradiated spot used in our mesaurements. 
    \textbf{(c)} Normalized quench decay profile $S_x(t)/S_x(0)$ measured by XY-8 dynamical decoupling sequence with different interpulse spacings $\tau_p$. We fit each curve to a stretched exponential  $e^{-(t/T_2)^{2/3}}$ to extract $T_2$.
    \textbf{(d)} We observe a saturation of $T_2$ as the interpulse spacing $\tau_p$ is decreased, indicating that NV-P1 interactions are sufficiently decoupled. 
    \textbf{(e)} Nuclear spin polarization as measured via ODMR. At 393 Gauss, the nuclear spin exhibits a hyperpolarization of $89(1)\%$ from a double-Lorentzian fit to the blue data. Red data shows the ODMR spectrum after shelving the less populated $m_I = -1/2$ nuclear spin subgroup in the $\ket{m_s = 1}$ state.
    \textbf{(f, g)} Measurement of NV density. We compare the XY-8 decoherence profile on both a linear (f) and log (g) scale with cluster DTWA numerics for NV densities ranging from $5.0$ to $11.0\ \mathrm{ppm}\cdot\mathrm{nm}$, and extract an NV density of 8 $\mathrm{ppm}\cdot\mathrm{nm}$. 
    } 
    \label{fig:supp_main}
\end{figure}

Substitutional nitrogen (P1) centers are the densest defects in the delta-doped layer. We estimate the P1 density from ODMR spectroscopy, using the linewidth of the NV $\ket{m_s = 0} \rightarrow \ket{m_s = -1}$ transition [Fig.~\ref{fig:supp_main}e]. From the FWHM, the NV-P1 interaction strength is $\sim (2\pi)\times 360~$kHz (setting $\hbar$ to 1 throughout), leading to an estimated P1 density $\sim 75$ ppm $\cdot$ nm.

To characterize the NV areal density in the irradiation spot used for our experiments, we measure the NV decoherence timescale under an XY-8 pulse sequence. To show that this timescale is limited by interactions between NV centers rather than by the stronger interactions between NV and P1 centers, we measure  $T_2^\text{XY8}$ as a function of the interpulse spacing $\tau_p$ [Fig.~\ref{fig:supp_main}(c, d)]. Because smaller interpulse spacings result in improved decoupling of the NV-P1 interactions~\cite{degenQuantumSensing2017}, we observe an increase in the  decoherence time when the interpulse spacing $\tau_p$ decreases. We choose an optimal interpulse spacing $\tau_p = 50~$ns for our measurements.

We then extract the NV density by comparing cluster DTWA numerical simulations~\cite{schachenmayerManyBodyQuantumSpin2015, braemerClusterTruncatedWigner2024} with the measured decoherence [Fig.~\ref{fig:supp_main}(f, g)], from which we infer an NV density of $8\ \mathrm{ppm}\cdot\mathrm{nm}$ for the out-of-the-plane NV group. The total NV density in the irradiation spot is thus $32\ \mathrm{ppm}\cdot\mathrm{nm}$. The NV-NV interaction strength in the out-of-plane group, with which we work exclusively, is then $\sim (2\pi)\times 80~$kHz.

\section{Pulse Sequences}\label{sec:pulseSequences}
As shown in Figure~\ref{fig:seq}, the full experiment pulse sequence consists of lattice engineering, and squeezing generation and readout steps. Each is discussed in detail in Secs.~\ref{sec:statePrep}, \ref{sec:generation}, \ref{sec:readout}, respectively, below.

\subsection{NV cluster removal via lattice engineering}\label{sec:statePrep}

\subsubsection{Overview} % Interaction spectrum
The first step in the experiment is to prepare a ``lattice-like'' geometry with relatively homogeneous spin-spin couplings by removing closely-spaced NV clusters from the system. Removal of these clusters is accomplished by harnessing their most prominent feature, i.e. the large interaction energy induced by the small distance between spins. The Ising component of this interaction shifts the bare splitting of the $\ket{m_s = 0}\rightarrow\ket{m_s = \pm1}$ transition for a spin $i$ by $\pm J_i$, defined as 
\begin{equation}
   J_i = \sum_{j} \frac{J_0}{r_{ij}^3}.
\end{equation}
For sufficiently large system sizes, the mean-field coupling $J_i$ can be treated as a continuous random variable $J$ with probability $P(J)$.
For disordered spins in 2D, clusters -- in particular, dimers formed by nearby pairs of spins -- give rise to the heavy tail in the distribution $P(J)$~[Fig.~3a, main text]. Because the dimer energy is larger than the average mean field interaction $ \overline{J} = \int P(J) J dJ$, dimers can be isolated from the remainder of the system using a frequency-selective control pulse in either the $z$ [Sec.~\ref{sec:shelving}] or $x$ [Sec.~\ref{sec:depol}] basis as described below.

\subsubsection{Dimer removal method I: frequency-resolved shelving}\label{sec:shelving}
Our first method for removing dimers from the system is frequency-resolved shelving~[Fig.~\ref{fig:seq}b.i]. After optically initializing the spins in the state $\ket{m_s=0}$, we apply a weak $\pi$-pulse whose amplitude $\Omega$ is set to the average NV coupling strength $\overline{J}$. This pulse drives ``typical'' NV centers to the $\ket{m_s = -1}$ state while dimers remain in $\ket{m_s=0}$. Because the relevant NV-NV interactions are dominated by larger NV-P1 interactions, we decouple the latter by interleaving XY-8 pulses with the weak control pulse. After the weak $\pi$-pulse, we apply two strong $\pi$-pulses: the first shelves the dimers in the $\ket{m_s = 1}$ state, and the second transfers the target spins back to the $\ket{m_s=0}$ state. At the end of the state preparation sequence, the dimers are shelved in $\ket{m_s = 1}$ and the remaining spins are prepared in $\ket{m_s = 0}$.

Because of the large difference between the amplitudes of the XY-8 and frequency-selective control pulses, we use two channels of the AWG and attenuate the microwaves separately before sending them to the microwave amplifier. We calibrate the relative phase between the microwaves generated by the two channels such that the $+x$ axis defined by the XY-8 pulses coincides with the one defined by the weak continuous drive.

In principle, the dimers shelved in $\ket{m_s = 1}$ can still interact with the rest of the system. However, the XY-8 pulse sequence is always applied during the subsequent quench dynamics~[Secs.~\ref{sec:generation} and \ref{sec:readout}], so the Ising interaction between the dimers and the rest of the system is decoupled. Furthermore, the NV-P1 Ising interactions suppress NV spin-exchange within the $\{\ket{m_s = 0}, \ket{m_s=1}\}$ subspace. As a result, the shelved spins do not participate in the squeezing or readout dynamics.

Several effects can limit the performance of this protocol, including XY-8 pulse imperfections and dimer dynamics during the slow pulse. In practice, we optimize the shelving parameters empirically to maximize the average twisting rate compared to the initial decay of the collective spin $\langle S_x \rangle $.

\subsubsection{Dimer removal method II: adiabatic depolarization}\label{sec:depol}
Our second method for removing NV dimers from the system is adiabatic depolarization~[Fig.~\ref{fig:seq}b.ii].
After optical initialization, we prepare the spin-polarized state $\ket{\mathbf{x}}$ with a $Y_{\frac{\pi}{2}}$ pulse. We then apply a weak, time-dependent transverse field $h_x(t)$ while simultaneously performing XY-8 decoupling pulses. 
The initial field $h_x(0)$ is much larger than NV dimer interactions so the spin-polarized state $\ket{\mathbf{x}}$ is an eigenstate. Then, we ramp  the field down to a final value $h_x^f\sim\overline{J}$ to depolarize dimers via two complementary mechanisms. First, spins with $J \gg \hx$ undergo fast decoherence, while spins with $J \ll \hx$ remain spin-locked along the $+x$ quantization axis. Second, the temporal profile of the field $\hx(t)$ improves the efficiency of depolarization by constraining the spins to adiabatically follow the eigenstates of the Hamiltonian. As a result of these two effects, this protocol removes the $S_x$ polarization for the dimers without significantly depolarizing the other spins.

We determine the ramp-down profile of the transverse field $h_x(t)$ by optimizing for adiabaticity. Specifically, for a two-body XXZ interaction with a strong transverse field $h_x\gg J$,
\begin{equation}\label{eq:Hdimers}
    H_d = -J_0(s_x^1s_x^2 + s_y^1s_y^2 - s_z^1s_z^2) - h_x(t)(s_x^1 + s_x^2).
\end{equation}
The dimer eigenstate to first order in perturbation theory in $J$ is 
\begin{equation}
    \ket{\boldsymbol{\epsilon}} = \ket{\mathbf{x}} - \frac{J_0}{4\hx(t)}\ket{\mathbf{-x}}. \label{eq:adia_eigen}
\end{equation}
Using Eqn.~\eqref{eq:adia_eigen}, we can rewrite the adiabatic condition $|\langle-\mathbf{x}|\dot{\boldsymbol{\epsilon}} \rangle|/|E_\mathbf{-x}-E_{\boldsymbol{\epsilon}}|\ll 1$ in terms of the transverse field.
The optimal ramp-down profile $h_x(t)$ then obeys the differential equation $\dot{h_x} = - kh_x^3$, whose solution is given by
\begin{equation}
    \hx(t) = \frac{\hx(0)}{\sqrt{1 + \hx(0)^2 kt}},
\end{equation}
where $k$ is the parameter controlling the adiabaticity. Note that we assume that $h_x(t)$ is much larger than $J$ in this derivation, but the same ramp-down profile still holds even when  $h_x(t)$ is no longer large compared to $h_x(t)$ as verified by the numerical diagonalization of the Hamiltonian. We empirically optimize the parameter $k$ and the total ramp duration by maximizing the twisting rate $\chi_\mathrm{eff}$ compared to the initial decay of the spin during the squeezing dynamics. After the state preparation protocol, we rotate the remaining polarized spins back to $\ket{m_s=0}$ and wait for a time $\tw \gg T_2^*$ to fully decohere the dimers.

\subsection{Generation of spin squeezing}\label{sec:generation}
After lattice engineering, we apply a $Y_{\frac{\pi}{2}}$ pulse to prepare the spin-polarized state $\ket{\mathbf{x}}$ for the remaining system spins. To generate the squeezed state, we allow the system to evolve under the native NV-NV interactions $\Hxxz$ for a time $t_\mathrm{g}$ by applying an XY-8 pulse sequence with interpulse spacing $\tau_p = 50$ ns to suppress the larger NV-P1 interactions~[Fig.~\ref{fig:seq}c].

\subsection{Readout of quantum spin projection noise}\label{sec:readout}
\begin{figure}[h!]
    \includegraphics[width=0.9\linewidth]{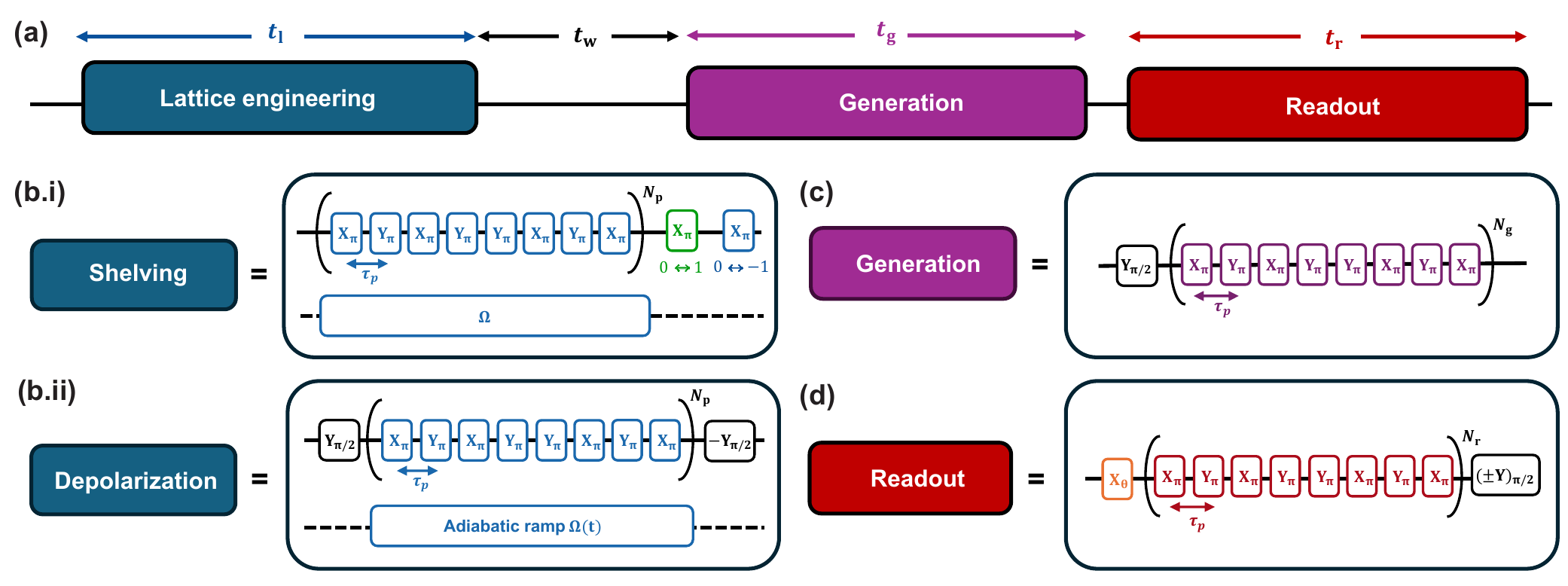}
    \caption{\textbf{Full experiment sequence.}
    (a) The experiment consists of three parts: lattice engineering, squeezing generation, and squeezing readout. (b) In the lattice engineering step, we prepare an effective lattice from the disordered system by removing NV dimers with either the shelving method (i) or the depolarization method (ii). (c) To generate squeezing, we evolve the system under its native dipolar interactions $\Hxxz$ for a time $\tg$. (d) In the readout step, we first apply a pulse with variable rotation angle $\theta$ along the $x$ axis (yellow) to map the spin operator $S_\theta$ into $S_z$. Then we probe the variance $\mathrm{Var}(S_\theta)$ by measuring the quench dynamics of the rotated squeezed state for a time $t_r$. The initial and final $\pi/2$ pulses that initialize and readout the state on $+x$ axis are shown in black. Throughout the time evolution, we apply XY-8 pulses to decouple the Ising interaction between NV and P1 centers. We conclude the experiment by a fluorescence readout while shining the 532 nm laser (not shown). 
    }
    \label{fig:seq}
\end{figure}

In order to measure the variance of the 
operator $S_\theta = S_x\cos\theta + S_y \sin\theta$, we perform a $X_\theta$ pulse to rotate the squeezed state about the $+x$ axis by an angle $\theta$, such that the $S_\theta$ operator for the pre-rotated state becomes the $S_z$ operator for the post-rotated state $X_\theta^\dagger S_\theta X_\theta = S_z$. We then measure the decay profile $S_x(t_r)$ for the post-rotated state, while continuously decoupling NV-P1 interactions with an XY-8 pulse sequence with interpulse spacing $\tau_p = 50$ ns~[Fig.~\ref{fig:seq}d].

\section{Data analysis}

\subsection{Stretch power as a measure of spatial geometry}\label{sec:crossover}

A spin-polarized state $\ket{\mathbf{x}}$ evolving under $\Hxxz$ decays as a stretched exponential $\expval{S_x(t)}/\expval{S_x(0)} \sim e^{-(t/\tau)^p}$. In particular, a disordered dipolar ensemble in 2D decays with stretch power $p = 2/3$, whereas spins on a regular lattice decay with $p = 2$~\cite{davisProbingManybodyDynamics2023}. The stretch power $p$ thus characterizes the spatial distribution of the spins and reveals the efficacy of our dimer removal protocols~[Sec.~\ref{sec:statePrep}]. The removal of dimers imposes a minimum distance $r_\mathrm{min}$ between the remaining spins, such that the local geometry is lattice-like while the global geometry is still disordered. One thus expects Gaussian decay ($p=2$) at early times, crossing over to disordered dynamics ($p = 2/3$) when the effective interaction length scale is much larger than the minimum spacing $r_\mathrm{min}$.

\begin{figure}[h!]
    \centering
    \includegraphics[width=0.6\linewidth]{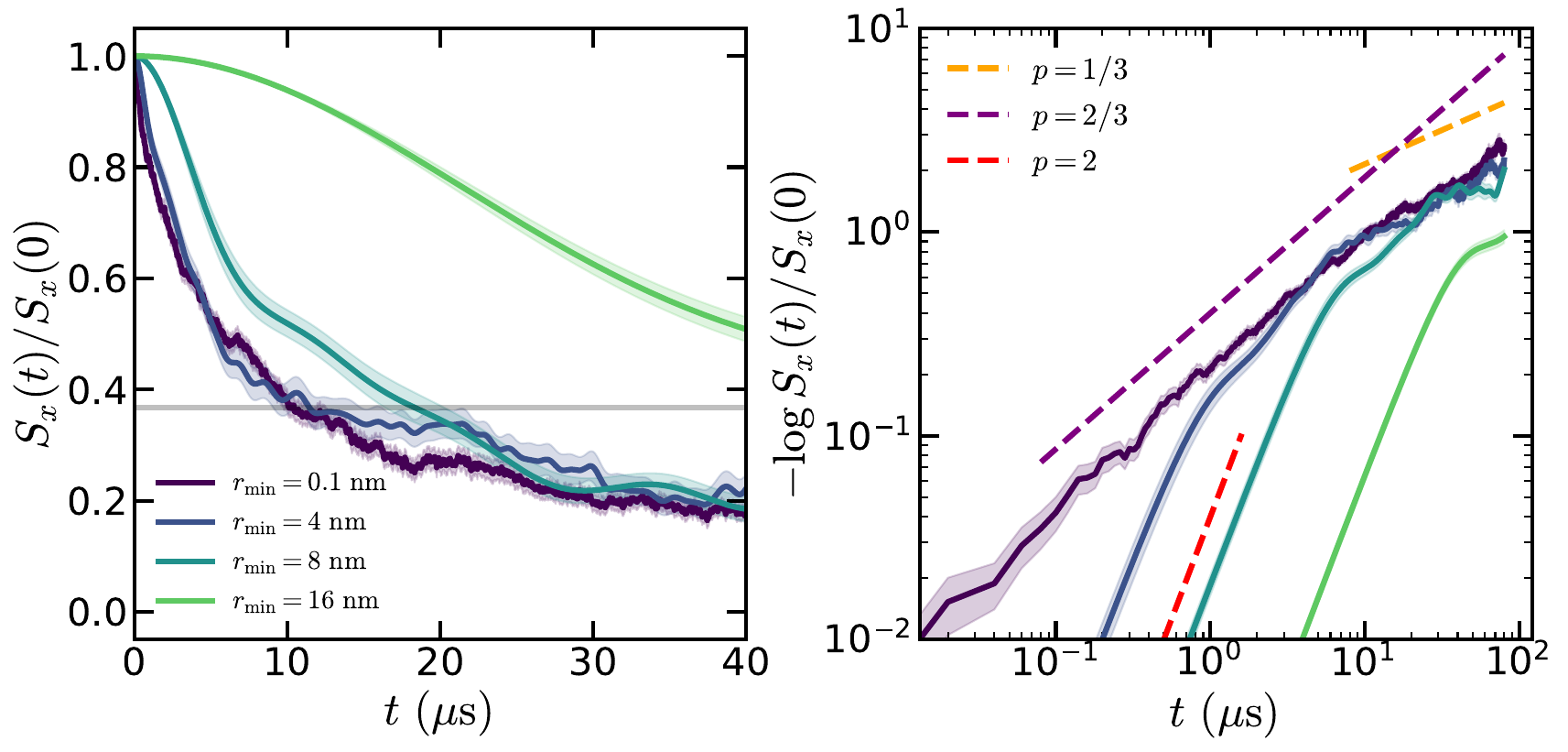}
    \caption{\textbf{Stretched power as a measure of spatial geometry.} Krylov subspace numerics averaged over 40 realizations of positional disorder for $N = 15$ spins. Data shows quench dynamics of spins evolving under $\Hxxz$ with minimum spin spacing $r_\mathrm{min} = 0.1, 4, 8, 16~$nm on linear (left) and log (right) scales. 
    The shaded area indicates the standard deviation of the positional disorder averaging. The deviation from $p = 2/3$ at long times can be explained by the crossover from the early-time ballistic regime into the late-time random-walk regime ($p=1/3$)~\cite{davisProbingManybodyDynamics2023}.}
    \label{fig:dim_crossover}
\end{figure}

To verify the above intuition, we simulate the decoherence profile as a function of $r_\mathrm{min}$~[Fig.~\ref{fig:dim_crossover}]. Beginning with spins randomly distributed in a 2D plane with areal density $8~\mathrm{ppm}\cdot\mathrm{nm}$, we then remove spins that have a neighbor within $r_\mathrm{min}$. 
When no spins are removed from the system, the stretch power of the decay profile is $p = 2/3$ for three decades in time~[Fig.~\ref{fig:dim_crossover}, purple]. As the minimum distance $r_\mathrm{min}$ is increased [Fig.~\ref{fig:dim_crossover}, blue to green curves], we observe the predicted crossover of the stretch power from $p = 2$ (early time lattice regime) to $p = 2/3$ (late time disordered regime). Furthermore, the crossover timescale $t_\mathrm{c}$ increases as $r_\mathrm{min}$ increases, which is consistent with the analytical prediction $t_\mathrm{c}\sim r_\mathrm{min}^3$ (see Supplemental Information).% (Supplementary information).
 
In the analysis below, we describe how we extract an effective minimum radius imposed by the shelving and depolarization protocols from XY-8 decoherence timescales. These models for spectral tailoring of the spatial geometry are then used to compute the input state for the squeezing and readout numerics described in Sec.~\ref{sec:dict}.

\subsubsection{Shelving}
To quantitatively model dimer shelving, we first calculate the distribution $P(J)$ for NV centers in a 2D plane. With the sequence in Sec.~\ref{sec:shelving}, the probability that the frequency-selective control $\pi$-pulse drives a spin to the $\ket{m_s = -1}$ state is 
\begin{equation}
    p_{-1}(J) = \frac{\Omega^2}{\Omega^2+J^2}\sin^2 \qty(\frac{\pi}{2}\sqrt{\qty (\frac{J}{\Omega})^2 + 1}).
\end{equation}
The shelving probability for a dimer is then $p_\mathrm{shelve} = 1 - p_{-1}(J)$. 

To understand shelving in terms of an effective cutoff radius $r_\mathrm{shelve}$, we can express the Rabi frequency of the shelving pulse as $\Omega = J_0/r_\mathrm{shelve}^3$. Similarly, we can rewrite any coupling $J = J_0/(r_\mathrm{NN})^3$, where $r_\mathrm{NN}$ is the effective nearest-neighbor distance. 
After plotting the shelving probability $p_\mathrm{shelve}$ as a function of $r_\mathrm{NN}$ [Fig.~\ref{fig:LE}a.i], we observe that $r_\mathrm{shelve}$ faithfully captures the dimer removal criteria: spins with $r_\mathrm{NN} < r_\mathrm{shelve}$ ($r_\mathrm{NN} > r_\mathrm{shelve}$) are removed (preserved). 

As shown in Figure~\ref{fig:LE}a.ii-iii, we measure the XY-8 decoherence profile after the shelving protocol, and compare the experimental result (black points) with cluster DTWA numerical simulations (colored curves) assuming the above shelving probability and areal NV density $8\ \mathrm{ppm}\cdot\mathrm{nm}$. Different curves correspond to simulated decoherence profiles assuming different shelving radii $r_\mathrm{shelve}$. We thus extract a dimer removal radius $r_\mathrm{shelve} = 7\ \mathrm{nm}$ ($r_\mathrm{shelve} = 6\ \mathrm{nm}$) for the shelving protocol for the pink (purple) data shown in Figure~4c of the main text.

\subsubsection{Depolarization}
\begin{figure}[h!]
    \centering
    \includegraphics[width=0.8\linewidth]{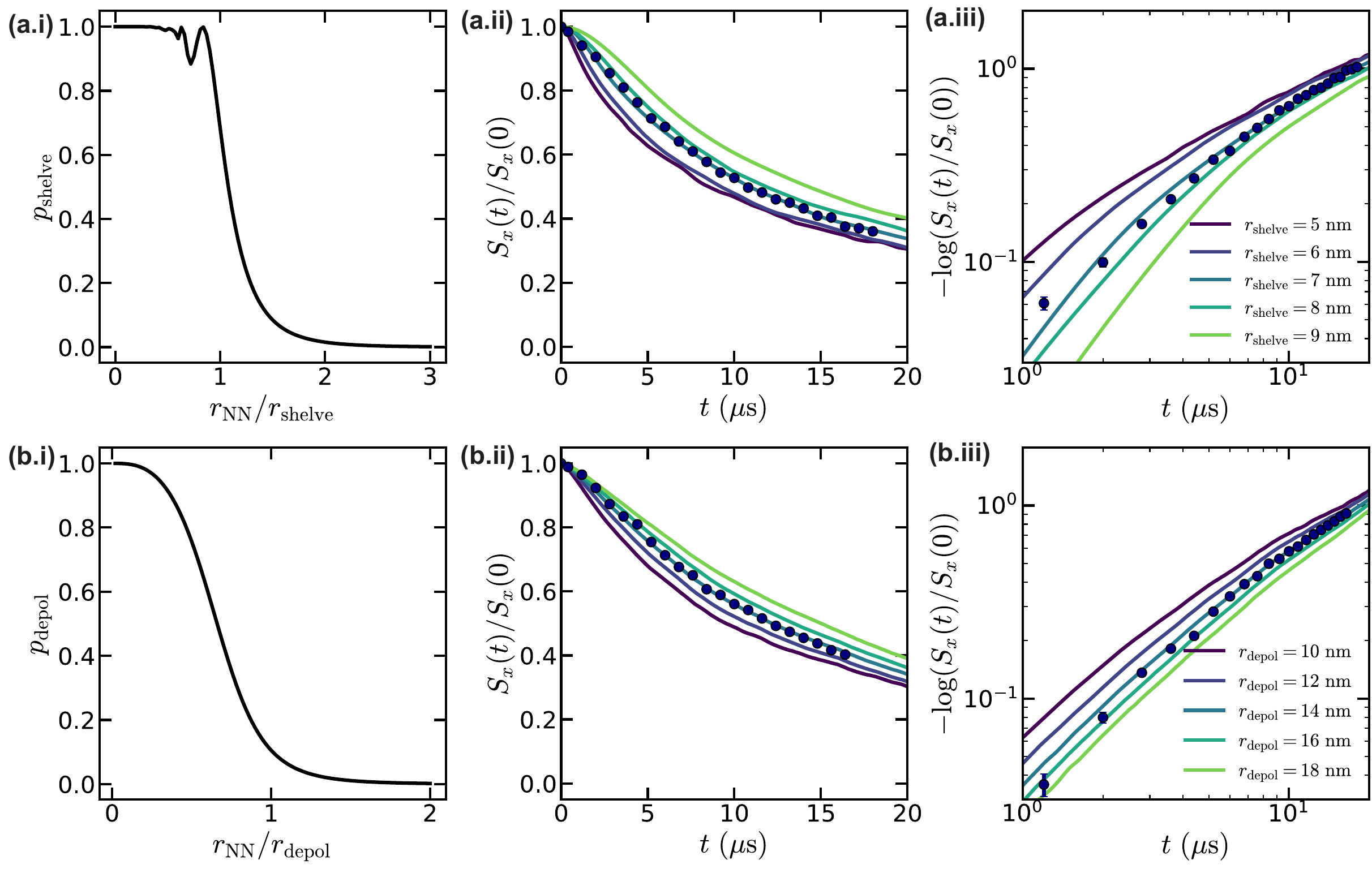}
    \caption{\textbf{Minimum spin spacing after lattice engineering.}
    \textbf{(a, b).i}. The shelving probability $p_\mathrm{shelve}$ (a) and the dimer depolarization probability $p_\mathrm{depol}$ (b) as a function of nearest-neighbor distance $r_\mathrm{NN}$.
    \textbf{(a, b).ii-iii} We measure the XY-8 decoherence profile after the shelving (a) and depolarization (b) protocols, and compare the experimental result with the cluster DTWA numerical simulation on linear (ii) and log (iii) scales. The curves are the decoherence profiles assuming different cut radii $r_\mathrm{shelve}$ or $r_\mathrm{depol}$. The NV density is fixed at $8\ \mathrm{ppm}\cdot\mathrm{nm}$. We extract a dimer removal radius $r_\mathrm{shelve} = 7\ \mathrm{nm}$ for the shelving protocol and  $r_\mathrm{depol} = 14\ \mathrm{nm}$ for the adiabatic depolarization protocol.
    }
    \label{fig:LE}
\end{figure}

To quantitatively model dimer depolarization, we estimate the depolarization rate for dimers with spacings $r_\mathrm{NN}$ by calculating the expectation value $\expval{S_x}$ for the dimer eigenstate. Assuming a transverse field strength $\hx = J_0/r_\mathrm{depol}^3$, the dimers whose inter-spin distance $r_\mathrm{NN}<r_\mathrm{depol}$ are mostly removed. Here, $r_\mathrm{depol}$ parametrizes the dimer removal radius for the depolarization protocol. Rewriting the dimer Hamiltonian [Eqn.~\ref{eq:Hdimers}] as 
\begin{equation}
\begin{aligned}
     H_d &= \frac{J}{r_\mathrm{NN}^3}(s_x^1s_x^2 + s_y^1s_y^2 - s_z^1s_z^2) + \frac{J_0}{r_\mathrm{depol}^3} (s_x^1 + s_x^2) \\ 
     &\sim(s_x^1s_x^2 + s_y^1s_y^2 - s_z^1s_z^2) + \qty(\frac{r_\mathrm{NN}}{r_\mathrm{depol}})^3 (s_x^1 + s_x^2),
\end{aligned}
\end{equation}
we obtain an expectation value $\expval{S_x}$ for the dimer eigenstate  
\begin{equation}
    \expval{s_x^1+s_x^2} = \frac{4(r_\mathrm{NN}/r_\mathrm{depol})^3(\sqrt{1 + 4(r_\mathrm{NN}/r_\mathrm{depol})^6} - 1)}{4(r_\mathrm{NN}/r_\mathrm{depol})^6 + (\sqrt{1+4(r_\mathrm{NN}/r_\mathrm{depol})^6} - 1)^2}.
\end{equation}
We have $\expval{s_x^1+s_x^2} \rightarrow 1$ when $r_\mathrm{depol} \rightarrow 0$, and  $\expval{s_x^1+s_x^2} \rightarrow 0$ when $r_\mathrm{depol}\rightarrow \infty$. Therefore,  dimers with an inter-spin distance $r_\mathrm{NN}$ will be depolarized with a probability $p_\mathrm{depol} = 1 - \expval{s_x^1+s_x^2}$ (Fig.~\ref{fig:LE}b.i). We again observe that $r_\mathrm{depol}$ faithfully quantifies the dimer removal criteria: the spins with $r_\mathrm{NN} < r_\mathrm{depol}$ ($r_\mathrm{NN} > r_\mathrm{depol}$) are removed (preserved). However, the curve is less steep than in the shelving protocol. 

As shown in Figure~\ref{fig:LE}b.ii-iii, we measure the XY-8 decoherence profile after the depolarization protocol, and compare the experimental result with the cluster DTWA numerical simulation assuming the above depolarization probability. The curves are the decoherence profiles assuming different cut radius $r_\mathrm{depol}$. The NV density is fixed at $8\ \mathrm{ppm}\cdot\mathrm{nm}$. We extract a dimer removal radius $r_\mathrm{depol} = 14\ \mathrm{nm}$ for the depolarization protocol corresponding to the data in Figure~4f of the main text.

\subsection{Probing spin projection noise via quench dynamics}

\subsubsection{Overview}
Briefly, we outline our procedure for extracting the variance $\mathrm{Var}(S_\theta)$ from readout quench dynamics $\langle S_x(\tr) \rangle$; each step is then detailed individually in Secs.~\ref{sec:offsetTime}-\ref{sec:dict}. After evolving the input state for a time $t_\mathrm{g}$ to generate squeezing, we measure the spin length decay $\expval{S_x(t_r)}$ as a function of the readout time $t_r$ for a set of rotation angles $\theta$. We then shift each measured curve by an offset time $\to$ %(\theta, t_\mathrm{g})
to obtain  $\langle S_x(t_r - t_0) \rangle \equiv \langle S_x(t_r^\text{eff}) \rangle$~[Sec.~\ref{sec:offsetTime}]. After shifting, we fit the set of curves $\langle S_x(t_r^\text{eff}) \rangle $ with a stretched exponential decay $e^{-(t_r^\text{eff}/T_2)^p}$ within a window $t_\text{min} \leq r_r^\text{eff} \leq t_\text{max}$. The timescale $T_2(\theta, \tg)$ is fit for each curve individually and the stretch power $p$ is globally optimized over all curves~[Sec.~\ref{sec:curveFitting}]. Next, we calculate a one-to-one mapping $T_2(\theta, \tg) \rightarrow \text{Var}(S_\theta(\tg))$ numerically~[Sec.~\ref{sec:dict}]. We finally use this mapping to calculate the squeezing parameter $\xi^2(\tg)$ for our data~[Sec.~\ref{sec:xi}].

\subsubsection{Offset time}\label{sec:offsetTime}
In order to systematically extract the decay times 
across a set of measurements $\langle S_x(t_r) \rangle$ for different rotation angles $\theta$, we must compare curves that satisfy the same fiducial initial condition. Specifically, we require that the spin state has no YZ correlations, $\expval{S_y S_z} = 0$, at the beginning of readout. More intuitively, this condition implies (i) that only the variance $\mathrm{Var}(S_z)$ differs between states rotated by different angles $\theta$, and (ii) that (for spins on a lattice) the total spin $S_x$ is maximized. 

\begin{figure}[h!]
    \centering
    \includegraphics[width=0.8\linewidth]{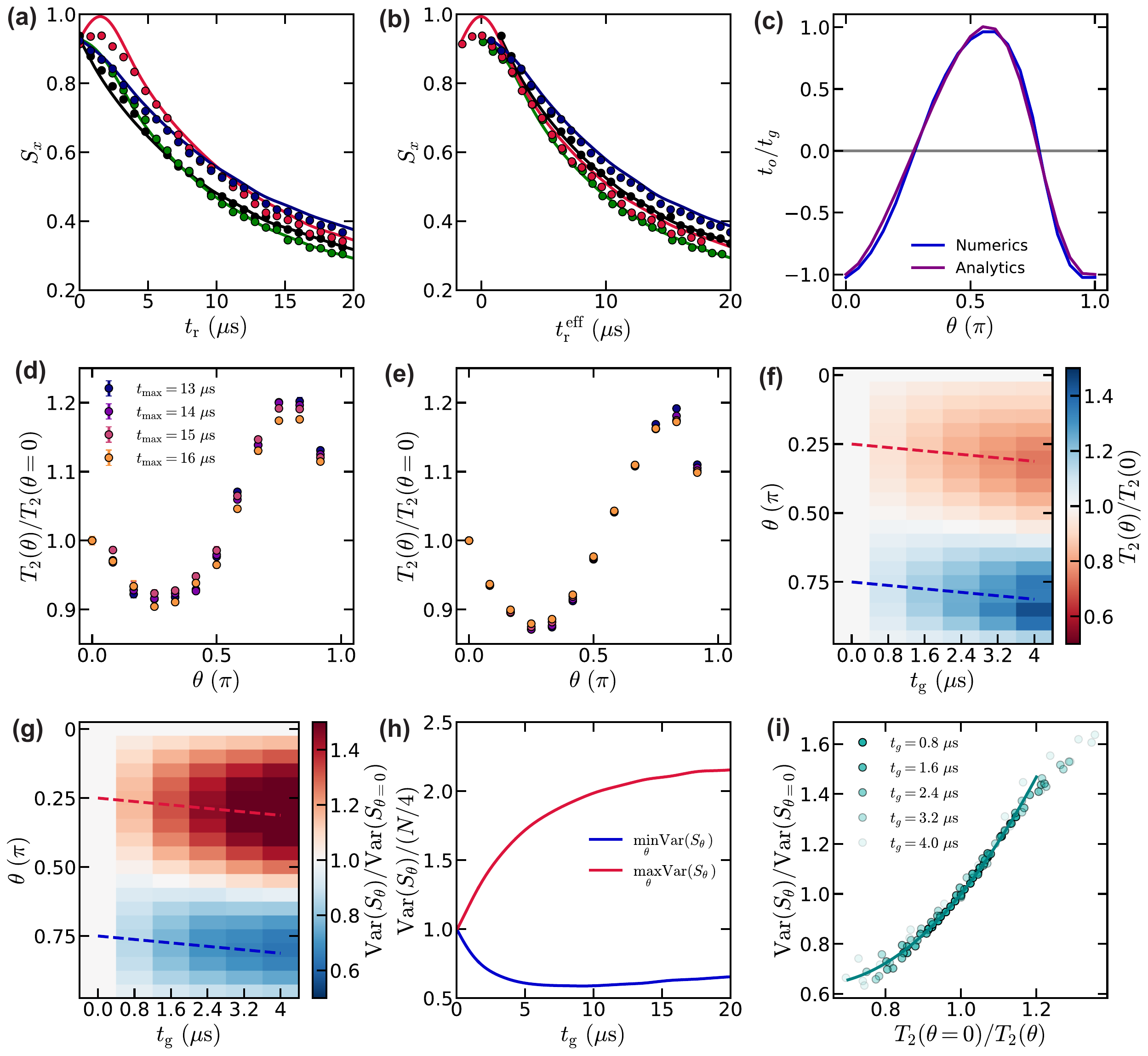}
    \caption{\textbf{Data analysis} 
    \textbf{(a)} Sample decay curves $\langle S_x(t_r)\rangle$ comparing experimental and numerical data. Here, shelving has been used to prepare the initial state which is then squeezed for a time $\tg = 1.6~\mu$s. 
    \textbf{(b)} The same data plotted in (a) after shifting each curve by $\to(\tg, \theta)$. 
    \textbf{(c)} Comparison of the offset time calculated using the OAT model (purple) with direct numerical extraction of the offset time for these data (blue). 
    \textbf{(d, e)} Comparison between the fitted decay timescales for experimental (d) and numerical (e) data. Here we also show the change in extracted timescale if the upper bound on the fitting window $t_\mathrm{max}$ is adjusted between 12 and 16~$\mu$s.
    \textbf{(f, g)} We plot the extracted $T_2$ decay timescale for the numerical data (f) in the full parameter space $\{\tg, \theta\}$. Maximum and minimum $T_2$ are indicated with the dashed blue and red lines, respectively. The same plot after mapping $T_2$ to Var($S_\theta$) is shown in (g).
    \textbf{(h)} Red and blue curves show the maximum and minimum variances Var($S_\theta$) as a function of $\tg$, corresponding to the dashed lines in (g). 
    \textbf{(i)} Final mapping obtained by matching the timescale data plotted in (f) to the variance data plotted in (g). We observe that the dictionaries at different squeezing generation times $\tg$ (solid to transparent green) overlap, as expected. 
    }
    \label{fig:dataAnalysis}
\end{figure}

In practice, each fiducial state is prepared at some offset time $t_r=\to(\theta,t_g)$ during the readout quench dynamics, which satisfies one of the following inequalities:
\begin{enumerate}[label=\roman*.]%\alph*]
    \item $ 0 < \to \leq \tg$: For angles $\pi/4 \lesssim \theta \lesssim 3\pi/4$, the state first ``untwists'' after the $X_\theta$ rotation, reaches the fiducial state satisfying $\langle S_y S_z \rangle = 0$, then ``retwists''~[Fig.~\ref{fig:dataAnalysis}c], e.g. for $\theta = \pi/2$, $\to\sim t_g$~[Fig.~\ref{fig:dataAnalysis}(a, b), red]. 
    \item $-\tg\leq \to<0$: For rotation angles $0\leq\theta \lesssim \pi/4$ and $3\pi/4 \lesssim \theta \leq \pi$, reaching a state with $\langle S_z S_y\rangle = 0$ would require backward time  evolution~[Fig.~\ref{fig:dataAnalysis}c], e.g. when $\theta = 0$, the offset $\to = -\tg$~[Fig.~\ref{fig:dataAnalysis}(a, b), black]. %For $\theta = 0$, $\to = -t_g$ as indicated in Fig.~\ref{fig:dataAnalysis}c.
    \item $\to=0$: No shift is required if the state is prepared with $\langle S_z S_y \rangle = 0$ by the rotation $X_\theta$, which occurs only at the two angles $\theta$ which orient the squeezed and anti-squeezed quadratures along $z$~[Fig.~\ref{fig:dataAnalysis}(a, b), blue and green]. 
\end{enumerate}
The offset time $\to(\theta, \tg)$ depends on both the squeezing duration $\tg$ and the rotation $X_\theta$. We determine the offset for each measurement $\langle S_x(\tr)\rangle$ analytically from the OAT model (see Supplementary Information) and benchmark against numerics for the disordered system, finding good agreement~[Fig~\ref{fig:dataAnalysis}c]. We then shift each measurement $\langle S_x(t_r-\to) \rangle$ to obtain $\langle S_x(t_r^\text{eff}) \rangle $.

\subsubsection{Curve Fitting}\label{sec:curveFitting}
We fit each shifted decoherence profile $\langle S_x(t_r^\text{eff})\rangle$ to the functional form $A(\theta) e^{-(t_r^\text{eff}/T_2(\theta))^p}$, where $T_2(\theta), A(\theta)$ are local fit parameters and the stretch power $p$ is optimized globally across all measurements $\{\theta, \tg \}$. We fix a fitting window $t_r^\mathrm{eff}\in[t_\mathrm{min}, t_\mathrm{max}]$. We choose the lower limit of the fitting window $t_\mathrm{min}$ by ensuring that the decay profile $S_x(t_r^\mathrm{eff})$ for each rotation angle $\theta$ has the same number of data points in the fitting window. In the experiment, data is only available for the detection quench time $t_r > 0$, i.e., $t_r^\mathrm{eff} = t_r - t_0 > -t_0$. Because $t_0$ can be negative, we set the lower limit of the fitting window $t_\mathrm{min} = \max_\theta(-t_0(\theta)) = \tg$, where the maximum is reached at $\theta = 0$. 

We choose the upper limit of the fitting window by evaluating how well the decay profile can be effectively described as a stretched exponential decay with a single stretch power $p$. Since the $\langle S_x(t_r^\text{eff}) \rangle$ decay profile has a crossover behavior for the stretch exponent~[Sec.~\ref{sec:crossover}],  the crossover timescale sets $t_\mathrm{max}$. We perform the fits for a range of reasonable fitting windows with $13\ \mu\mathrm{s}\leq t_\mathrm{max} \leq 16~\mu$s [Fig.~\ref{fig:dataAnalysis}(d, e)] and include the resulting fluctuations in the errorbars of Figure~4 in the main text.

\subsubsection{Mapping between decay time scale and quantum variance}\label{sec:dict}
In order to map a measured decay time $T_2$ to $\mathrm{Var}(S_z)$, we establish a one-to-one mapping between the extracted decay timescale and the quantum variance. A quantitative study for the mapping is provided as follows. 

To build the mapping, we place $N=80$ spins in a 2D plane and remove dimers according to $r_\mathrm{shelve}$ or $r_\mathrm{depol}$ as detailed in Sec.~\ref{sec:crossover}. We perform cluster DTWA numerics with $50$ realizations of positional disorder to simulate quench dynamics under $\Hxxz$: generation of squeezing for a time $\tg$, rotation $X_\theta$, and subsequent decay $\langle S_x(t_r) \rangle$. The input parameters for our simulations are determined from independent measurements~[Table~\ref{tab:dictionary}]. Thus, the numerics used to generate the mapping contains no free parameters. 

\begin{table}[h]
    \centering
    \begin{tabular}{|c|c|c|c|c|c|c}
        \hline Configuration & Initial NV polarization & NV density (ppm$\cdot$nm) & layer thickness (nm) &  Geometry  \\ \hline\hline
        Full positional disorder & 80\% & 8 & 7 & N/A \\ \hline
        Optimal shelving & 80\%  & 8 & 7 & $r_\mathrm{shelve} = 7\ $nm  \\ \hline
        Intermediate shelving & 80\%   & 8 & 7 & $r_\mathrm{shelve} = 6\ $nm  \\ \hline
        Optimal adiabatic depol & 80\%  & 8 & 7 & $r_\mathrm{depol} = 14\ $nm  \\ \hline
    \end{tabular}
    \caption{Parameters used in our numerical simulations of squeezing and readout quench dynamics. NV density is obtained from $T_2^\text{XY8}$~[Sec.~\ref{sec:NVdensity}]. Layer thickness from SIMS~[Sec.~\ref{sec:sample}]. The effective radii $r_\text{shelve}$ and $r_\text{depol}$ are determined by fitting to the models described in Sec.~\ref{sec:shelving} and \ref{sec:depol}.} 
    \label{tab:dictionary}
\end{table}
 
A comparison between the resulting raw numerical and experimental data is shown in Figure~\ref{fig:dataAnalysis}a. We note that the solid curves $\langle S_x(t_r) \rangle$ are directly output from our numerics, i.e. there are no fit parameters. We shift and fit the numerical and experimental data following the procedures in Secs.~\ref{sec:offsetTime}, \ref{sec:curveFitting}. We can thus compare the timescales $T_2$ obtained via the numerics and the experiment, as shown in Figure~\ref{fig:dataAnalysis}(d-f). From the same numerical data, we compute the variance $\mathrm{Var}(S_\theta)$ as a function of $\tg$ and $\theta$. Finally, plotting $\mathrm{Var}(S_\theta)$ vs $T_2$ exhibits a clear one-to-one correspondence, with consistent results across the measured squeezing durations from 0 to 4 microseconds. From the plot in Figure~\ref{fig:dataAnalysis}i, we can thus measure the variance $\mathrm{Var}(S_\theta)$ via the decay timescale $T_2$.

\subsection{Extraction of squeezing parameter}\label{sec:xi}
The squeezing parameter for a system with $N$ spins is defined as 
\begin{equation}
    \xi^2(\tg) = N\frac{\min_\theta\mathrm{Var}(S_\theta(\tg))}{S_x^2(\tg)},
\end{equation}
where $\xi^2=1$ for an unentangled spin-polarized state. The squeezing parameter can be expressed as the ratio  
\begin{equation}
    \xi^2(\tg) = \frac{\xi^2(\tg)}{\xi^2(0)} = \frac{\min_\theta\mathrm{Var}(S_\theta(\tg))}{\min_\theta\mathrm{Var}(S_\theta(0))} \qty(\frac{S_x(0)}{S_x(\tg)})^2. 
\end{equation}
Because the $S_z$ operator is conserved under $\Hxxz$, the variance $\mathrm{Var}(S_{\theta=0}(t))$ remains a constant during the squeezing generation, i.e. $\mathrm{Var}(S_{\theta=0}(0)) = \mathrm{Var}(S_{\theta=0}(\tg))$. Furthermore, the initial state is symmetric under rotations $X_\theta$, i.e., $\mathrm{Var}(S_{\theta}(0))$ is identical for all $\theta$, such that $\min_\theta\mathrm{Var}(S_{\theta=0}(0)) = \mathrm{Var}(S_{\theta=0}(0))$. Therefore, we have $\min_\theta\mathrm{Var}(S_{\theta=0}(0)) = \mathrm{Var}(S_{\theta=0}(\tg))$, and the squeezing parameter becomes
\begin{equation}
    \xi^2(\tg) = \min_\theta\qty(\frac{\mathrm{Var}(S_\theta(\tg))}{\mathrm{Var}(S_{\theta=0}(\tg))}) \qty(\frac{S_x(0)}{S_x(\tg)})^2.
\end{equation}

With the protocol explained in Sec.~\ref{sec:dict}, we extract the relative variance $\mathrm{Var}(S_\theta(\tg))/\mathrm{Var}(S_{\theta=0}(\tg))$ by measuring the $S_x$ decay in the readout quench. Meanwhile, we directly measure $S_x(\tg)/S_x(0)$ in the generation quench. In practice, the squeezing parameter here is the ``error corrected'' squeezing parameter which captures the squeezing dynamics under $\Hxxz$ while the imperfect state preparation, e.g., finite initial polarization, is normalized out.

\bibliographystyle{naturemag}
\bibliography{method_arXiv.bib}

\newpage

\newpage

\newpage

\newpage 

\newpage

\newpage